\documentclass{aa}
\usepackage{graphicx}
\begin{document}

   \title{Oxygen and nitrogen abundances in nearby galaxies}
   \subtitle{Correlations between oxygen abundance and macroscopic properties}

\author{L.S.~Pilyugin \inst{1,2},
        Jos\'{e} M. V\'{\i}lchez \inst{2},
        Thierry Contini \inst{3} }

  \offprints{L.S. Pilyugin }

   \institute{   Main Astronomical Observatory
                 of National Academy of Sciences of Ukraine,
                 27 Zabolotnogo str., 03680 Kiev, Ukraine
                 (pilyugin@mao.kiev.ua)
                 \and
                 Instituto de Astrof\'{\i}sica de Andaluc\'{\i}a,
                 CSIC, Apdo, 3004, 18080 Granada, Spain
                 (jvm@iaa.es)
                 \and
                 Laboratoire d'Astrophysique de l'Observatoire Midi--Pyr\'{e}n\'{e}es -- UMR 5572,
                 14 avenue E. Belin, 31400 Toulouse, France
                 (contini@ast.obs-mip.fr)           }

\date{Received 16 October 2003 / accepted 16 June 2004}

\abstract{
We performed a compilation of more than 1000 published spectra of H\,{\sc ii}
regions in spiral galaxies. The oxygen and nitrogen abundances in each
H\,{\sc ii} region were recomputed in a homogeneous way, using the P--method.
The radial distributions of oxygen and nitrogen abundances were derived.
The correlations between oxygen abundance and macroscopic properties  are examined.
We found that the oxygen abundance in spiral galaxies  correlates with its
luminosity, rotation velocity, and morphological type: the correlation 
with the rotation velocity may be slightly tighter. 
There is a significant difference between the luminosity -- metallicity relationship
obtained here and that based on the oxygen abundances determined through the
$R_{\rm 23}$--calibrations. The oxygen abundance of NGC 5457 recently determined
using direct measurements of T$_e$ (Kennicutt, Bresolin \& Garnett 2003)
agrees with the luminosity -- metallicity
relationship derived in this paper, but is in conflict with the luminosity --
metallicity relationship derived with the $R_{\rm 23}$-based oxygen abundances.
The obtained luminosity -- metallicity relation for spiral galaxies is compared
to that for irregular galaxies. Our sample of galaxies shows evidence that
the slope of the O/H  -- M$_B$ relationship for spirals
(--0.079 $\pm$ 0.018) is slightly more shallow
than that for irregular galaxies (--0.139 $\pm$ 0.011).
The effective oxygen yields were estimated for spiral and irregular galaxies.
The effective oxygen yield increases with increasing luminosity from $M_{\rm B} \sim -11$
to $M_{\rm B} \sim -18$ (or with increasing rotation velocity from
$V_{\rm rot} \sim 10$ km s$^{-1}$ to $V_{\rm rot} \sim 100$ km s$^{-1}$) and 
then remains approximately constant.
Irregular galaxies from our sample have effective oxygen yields
lowered by a factor of 3 at maximum, i.e. irregular galaxies usually keep 
at least 1/3 of the oxygen they manufactured during their evolution.
}

\titlerunning{Oxygen abundance vs. macroscopic properties in nearby galaxies}

\authorrunning{L.S.Pilyugin et al.}

\maketitle

\section{Introduction}

Investigating the macroscopic properties of galaxies that could drive their
chemical evolution is very important in understanding their global
evolution, which has been the goal of many studies over the past twenty years.
It has been found, for example, that the properties
of H\,{\sc ii} regions in late-type galaxies are linked to macroscopic
characteristics of galaxies such as luminosity or Hubble type. Smith (1975) 
concluded that excitation differences among the H\,{\sc ii} regions of
Sbc--Scd--Irr galaxies can be best understood in terms of an abundance sequence
which progresses from higher to lower heavy-element enrichment as one progresses
from earlier to later type galaxies. He also noted that his results show no
apparent correlation between the average heavy-element abundance and galaxy mass.
The correlation between oxygen abundance and the morphological type of galaxy
was later confirmed by Vila-Costas \& Edmunds (1992) and Zaritsky, Kennicutt  
\& Huchra (1994).

Lequeux et al. (1979) revealed that the oxygen abundance correlates with total
galaxy mass for irregular galaxies, in the sense that the higher the total mass,
the higher the heavy element content. Since the galaxy mass is a poorly known
parameter, the metallicity -- luminosity relation instead of the
mass -- metallicity relation is usually considered (Skillman, Kennicutt \& 
Hodge (1989);  Richer \& McCall (1995); Hunter \& Hoffman (2000); Pilyugin 
(2001c); Melbourne \& Salzer (2002), among others).
Garnett \& Shields (1987) found that spiral disk abundance also correlates
very well with galaxy luminosity. They concluded that the metallicity of
galaxies correlates better with galaxy luminosity than with morphological type.
Zaritsky, Kennicutt \& Huchra (1994) found that the characteristic
gas-phase abundances and luminosities of spiral galaxies are strongly correlated,
and this relationship maps almost directly onto the luminosity -- metallicity
relationship of irregular galaxies.

The origin of this correlation is open to debate. It is widely suggested that
there are two mechanisms which can be responsible for a luminosity -- metallicity
relation for spirals and irregulars: higher astration level and decreasing
efficiency of heavy-element loss with increasing luminosity.
The mass exchange between a galaxy and its environment can alter the relation
between oxygen abundance and gas mass fraction; it mimics the variation in the
oxygen yield. To investigate the possibility of a varying yield, 
Edmunds (1990) and Vila-Costa \&
Edmunds (1992) have suggested to use the ``effective'' oxygen yield,
$y_{\rm eff}$, as the yield that would be deduced if a system was assumed to
behave as in the simplest model of chemical evolution. The variation of the value
of the effective oxygen yield from galaxy to galaxy can be considered as indicative
of the efficiency of mass exchange between galaxies and their environments.
A similar approach, the concept of the oxygen abundance deficiency in the
galaxy which is introduced as a deficiency of the oxygen abundance observed in
the galaxy in comparison with the oxygen abundance predicted by the closed-box
model for the same gas mass fraction, has been used by Pilyugin \& Ferrini
(1998; 2000a,b). The effective oxygen yields for a set of spiral and irregular
galaxies were derived recently by Garnett (2002). He found that the
value of effective oxygen yield is approximately constant, $y_{\rm eff}$ =
0.0112, for $V_{\rm rot} > 150$ km s$^{-1}$.
The effective oxygen yield decreases by a factor of 10 -- 20 from
$V_{\rm rot} \sim 300$ km s$^{-1}$ to 5. This means that low-mass galaxies have 
lost the bulk of their manufactured oxygen, up to 90 -- 95\%.

Accurate abundance determinations are mandatory for such investigations.
In H\,{\sc ii} regions, they can be derived from measurements
of temperature-sensitive line ratios, such as
[OIII]$\lambda \lambda$4959,5007/[OIII]$\lambda$4363.
Unfortunately, in oxygen-rich H\,{\sc ii} regions, the temperature-sensitive
lines such as [OIII]$\lambda$4363 are often too weak to be detected. For such
H\,{\sc ii} regions, abundance indicators based on more readily observable
lines were suggested (Pagel et al. 1979; Alloin et al. 1979).
The oxygen abundance indicator
$R_{\rm 23}$ = ([OII]$\lambda \lambda$3727,3729 + [OIII]$\lambda \lambda$4959,5007)/H$_{\beta}$,
suggested by Pagel et al. (1979), has found widespread acceptance and use for
the oxygen abundance determination in H\,{\sc ii} regions where the
temperature-sensitive lines are undetectable. Grids of photoionization models
are often used to establish the relation between strong oxygen line intensities
and oxygen abundances (Edmunds \& Pagel 1984; McCall et al. 1985; Dopita \&
Evans 1986; Kobulnicky et al. 1999; Kewley \& Dopita 2002; among others).
Radial distributions of oxygen abundance determined with theoretical (or model)
calibrations have been obtained for large samples of spiral galaxies by
Vila-Costas \& Edmunds (1992), Zaritsky et al. (1994), van Zee et al. (1998), Consid\`ere
et al. (2000), among others.

The early calibrations were one-dimensional (Edmunds \& Pagel 1984; McCall
et al. 1985; Dopita \& Evans 1986; Zaritsky et al. 1994), i.e. a relation of
the type $O/H = f(R_{\rm 23})$ was used. It has been shown (Pilyugin 2000; 2001a,b)
that the error in the oxygen abundance derived with the one-dimensional
calibrations involves a systematic error. The origin of this systematic error
is evident. In a general case, the intensity of oxygen emission lines in
spectra of H\,{\sc ii} regions depends not only on the oxygen abundance but also on the
physical conditions (hardness of the ionizing radiation and geometrical factors) in the 
ionized gas.
Thus, when one estimates the oxygen abundance from emission line intensities, the
physical conditions in H\,{\sc ii} regions should be taken into account. In the T$_e$ --
method this is done via the electron temperature T$_{e}$. In one-dimensional 
calibrations the physical conditions in H\,{\sc ii} regions are ignored. 
Starting from the idea of McGaugh (1991)
that the strong oxygen lines contain the necessary information for determinining 
accurate abundances in (low-metallicity) H\,{\sc ii} regions, it has been shown
(Pilyugin 2000; 2001a,b) that the physical conditions in H\,{\sc ii} regions can be
estimated and taken into account via the excitation parameter $P$. A two-dimensional or
parametric calibration (the so-called ``P -- method'') has been suggested. 
A more general relation
of the type $O/H = f(P, R_{\rm 23})$ is used in the P -- method, compared with the
relation of the type $O/H = f(R_{\rm 23})$ used in one-dimensional calibrations.
Thus, the one-dimensial $R_{\rm 23}$ -- calibration provides more
or less realistic oxygen abundances in high-excitation H\,{\sc ii} regions, but
yields overestimated oxygen abundances in low-excitation H\,{\sc ii} regions.
This is in agreement with the results of Kinkel \& Rosa (1994), Castellanos et al. (2002),
Kennicutt, Bresolin \& Garnett (2003), Garnett, Kennicutt \& Bresolin (2004), 
who found that the $R_{\rm 23}$ -- method
yields overestimated oxygen abundances in high-metallicity H\,{\sc ii} regions.

It should be stressed that "strong lines -- oxygen abundance" calibrations do not
form an uniform family. The calibrations of the first type are the empirical
calibrations, established on the basis of H\,{\sc ii} regions in which the
oxygen abundances are determined through the T$_e$ -- method. Two-dimensional
empirical calibrations both at low and high metallicities were recently
derived by Pilyugin (2000; 2001a,c). The calibrations of the second type are
the theoretical (or model) calibrations, established on the basis of the grids
of photoionization models of H\,{\sc ii} regions. The two-dimensional
theoretical calibrations were recently proposed by Kobulnicky et al. (1999)
and Kewley \& Dopita (2002).
It has been shown (Pilyugin 2003b) that there is a discrepancy between the oxygen
abundances in high-metallicity H\,{\sc ii} regions determined with the
T$_e$ -- method (and/or with the corresponding
"strong lines -- oxygen abundance" calibration) and that determined with the
model fitting (and/or with the corresponding "strong lines -- oxygen
abundance" calibration).
Thus, so far, there actually exist two scales of oxygen abundances
in H\,{\sc ii} regions. The first (empirical) scale corresponds to the oxygen abundances
derived with the T$_e$ -- method or with empirical calibrations (the P --
method). The second (theoretical or model) scale corresponds to the oxygen
abundances derived through model fitting or with theoretical (model) calibrations.

Pilyugin (2003b) suggested to use the interstellar oxygen abundance in the solar
vicinity, derived with very high precision from the high-resolution observations
of the weak interstellar OI$\lambda$1356 absorption line towards the stars, as
a ``Rosetta Stone'' to make a choice between "theoretical" and "empirical" scales
of oxygen abundances in high-metallicity H\,{\sc ii} regions.
The agreement between the value of the oxygen
abundance at the solar galactocentric distance traced by the abundances derived
in H\,{\sc ii} regions through the T$_e$ -- method and that derived from the
interstellar absorption lines towards the stars is strong evidence that the
classical T$_e$ -- method provides accurate oxygen abundances in H\,{\sc ii} regions,
i.e. the "empirical" scale of oxygen abundances in high-metallicity H\,{\sc ii}
regions is correct.  Therefore, at high metallicities the "strong lines -- oxygen
abundance" calibrations must be based on the H\,{\sc ii} regions with the
oxygen abundances derived through the T$_e$ -- method but not on the existing
grids of models for H\,{\sc ii} regions.

The usual quantities: the oxygen and nitrogen abundance distributions in
galaxies (Sect. 2), the correlations between oxygen abundance and
macroscopic properties of galaxies (Sect. 3 and Sect. 4), and the estimation of
effective oxygen yields and their variations among galaxies (Sect. 5) will
be considered in this paper. 

What is the novelty of our study? All previous investigations
of these problems were carried out with the $R_{\rm 23}$-based oxygen abundances.
Let us refer to the recent analysis and conclusion of B. Pagel who is
the foundator of the $R_{\rm 23}$ -- calibrations: ``Starting from around 1980, the
notorious $R_{\rm 23}$ method has been the most widely used method for the oxygen
abundance determination. Investigations
since then have shown that our calibrations have overestimated oxygen abundances
near solar. There is some promise in newly developed strong-line indices and
especially in refinements of the $R_{\rm 23}$ method (P -- calibration). Until more
of these refinements have been applied, abundances near and above solar have
to be taken with a grain of salt'' (Pagel 2003).
For the present study, the oxygen and nitrogen abundances for a large sample
of H\,{\sc ii} regions in late-type galaxies will be derived using a homogeneous 
method: the two-dimensional calibration (P -- method).
In Sect. 6, the oxygen abundances obtained here will be compared with the
recently derived T$_{e}$-based oxygen abundances in the disk of the galaxy
NGC 5457 (Kennicutt, Bresolin \& Garnett 2003). The O/H -- M$_B$ relationship
obtained here with the P-based oxygen abundances will be compared with
the O/H -- M$_B$ relathionship obtained recently by Garnett (2002) on the basis
of oxygen abundances derived through the $R_{\rm 23}$ -- calibrations. The comparison
between P-based, T$_e$-based, and $R_{\rm 23}$-based data provides
an additional check of the P-based abundances and shows the difference
between the P-based and $R_{\rm 23}$-based data.

\section{The chemical abundances}

\begin{table*}
\caption[]{\label{table:oxygen}
The adopted and derived parameters of the radial oxygen abundance distributions in galaxies}
{\footnotesize
\begin{center}
\begin{tabular}{lrccccl}  \\ \hline \hline
galaxy       & $R _{\rm 25}$   &  12+log(O/H)$_0$  &   gradient      & $\Delta$log(O/H)  &  number of  &  references                \\
             & arcmin      &                   &  dex/R$ _{25}$  &   dex             & H\,{\sc ii} regions  &                            \\  \hline
NGC0224      & 102.07      &    8.76           &    -0.52        &   0.09            &   22     & BKC82, BKG99, DK81         \\
NGC0253      &  14.09      &    8.71           &    -0.54        &   0.10            &    9     & WS83                       \\
NGC0300      &  11.19      &    8.49           &    -0.40        &   0.13            &   45     & CPG97, DCL88, dORW83, \\
             &             &                   &                 &                   &          & EP84, PEB79, WS83         \\
NGC0598      &  37.06      &    8.57           &    -0.20        &   0.08            &   41     & BKG99, DTP87, GOS92, KA81,  \\
             &             &                   &                 &                   &          & MRS85, P70, S71, S75, VPD98  \\
NGC0628      &   5.36      &    8.68           &    -0.45        &   0.08            &   37     & BKG99, FGW98, MRS85, vZSH98   \\
NGC0753      &   1.32      &    8.82           &    -0.21        &   0.08            &   12     & HBC96                      \\
NGC0925      &   5.48      &    8.51           &    -0.48        &   0.10            &   20     & vZSH98, ZKH94              \\
NGC1058      &   1.62      &    8.71           &    -0.32        &   0.04            &    6     & FGW98                      \\
NGC1068      &   3.54      &    8.83           &    -0.26        &   0.03            &    9     & OK93, vZSH98               \\
NGC1232      &   3.71      &    8.73           &    -0.58        &   0.09            &   15     & vZSH98                     \\
NGC1365      &   5.61      &    8.74           &    -0.70        &   0.14            &   53     & AEL81, PEB79, RW97         \\
NGC1637      &   2.04      &    8.67           &     0.02        &   0.11            &   14     & vZSH98                     \\
NGC2403      &  11.45      &    8.52           &    -0.35        &   0.11            &   56     & BKG99, GSP99, GSS97, FTP86, \\
             &             &                   &                 &                   &          & MRS85, S75, vZSH98         \\
NGC2442      &   3.23      &    8.70           &    -0.17        &   0.05            &    8     & R95                        \\
NGC2541      &   3.30      &    8.23           &     0.14        &   0.14            &   19     & ZKH94                      \\
NGC2805      &   3.30      &    8.44           &    -0.26        &   0.11            &   17     & vZSH98                     \\
NGC2835      &   3.62      &    8.31           &    -0.07        &   0.12            &   17     & R95                        \\
NGC2841      &   4.06      &    9.12           &    -0.78        &   0.01            &    5     & BKG99, OK93                \\
NGC2903      &   6.29      &    8.94           &    -0.71        &   0.09            &   31     & MRS85, vZSH98, ZKH94       \\
NGC2997      &   5.00      &    8.66           &    -0.39        &   0.08            &    6     & EP84, MRS85                \\
NGC3031      &  13.77      &    8.69           &    -0.43        &   0.09            &   36     & BKG99, GS87, OK93, SB84    \\
NGC3184      &   3.71      &    8.97           &    -0.63        &   0.06            &   30     & MRS85, vZSH98, ZKH94       \\
NGC3198      &   4.26      &    8.69           &    -0.64        &   0.07            &   14     & ZKH94                      \\
NGC3344      &   3.54      &    8.63           &    -0.48        &   0.09            &   15     & MRS85, VEP88, ZKH94        \\
NGC3351      &   3.79      &    8.90           &    -0.26        &   0.05            &   19     & BK02, BKG99, MRS85, OK93   \\
NGC3521      &   5.48      &    8.86           &    -0.93        &   0.06            &    9     & ZKH94                      \\
NGC3621      &   6.74      &    8.55           &    -0.44        &   0.12            &   26     & R95, ZKH94                 \\
NGC4254      &   2.81      &    8.94           &    -0.65        &   0.06            &   19     & HPC94, MRS85, SSK91        \\
NGC4258      &   9.31      &    8.57           &    -0.20        &   0.08            &   33     & BKG99, C00, OK93, ZKH94    \\
NGC4303      &   3.23      &    8.84           &    -0.72        &   0.10            &   22     & HPL92, SSK91               \\
NGC4321      &   3.79      &    8.86           &    -0.37        &   0.06            &   10     & MRS85, SSK91               \\
NGC4395      &   6.59      &    8.27           &    -0.02        &   0.08            &   16     & MRS85, vZSH98              \\
NGC4501      &   3.54      &    8.99           &    -0.52        &   0.07            &    5     & SKS96                      \\
NGC4559      &   5.48      &    8.48           &    -0.38        &   0.13            &   20     & ZKH94                      \\
NGC4571      &   1.86      &    8.90           &    -0.20        &   0.04            &    4     & SKS96, SSK91               \\
NGC4651      &   2.04      &    8.72           &    -0.64        &   0.06            &    6     & SKS96                      \\
NGC4654      &   2.45      &    8.85           &    -0.77        &   0.06            &    7     & SKS96                      \\
NGC4689      &   2.18      &    8.89           &    -0.41        &   0.08            &    5     & SKS96, SSK91               \\
NGC4713      &   1.35      &    8.71           &    -0.73        &   0.05            &    4     & SKS96                      \\
NGC4725      &   5.34      &    9.01           &    -0.88        &   0.13            &    8     & ZKH94                      \\
NGC4736      &   5.61      &    8.60           &    -0.26        &   0.05            &   16     & BKG99, MRS85, OK93         \\
NGC5033      &   5.36      &    9.06           &    -1.78        &   0.06            &    5     & ZKH94                      \\
NGC5055      &   6.29      &    9.01           &    -0.83        &   0.07            &    5     & MRS85                      \\
NGC5068      &   3.88      &    8.32           &    +0.08        &   0.12            &   23     & MRS85, R95                 \\
NGC5194      &   5.61      &    8.92           &    -0.40        &   0.06            &   21     & BKG99, DTV91, MRS85        \\
NGC5236      &   6.59      &    8.79           &    -0.28        &   0.07            &   27     & BK02, DTJ80, WS83          \\
NGC5457      &  14.42      &    8.80           &    -0.88        &   0.09            &   65     & GSP99, KG96, KR94, MRS85, RPT82,  \\
             &             &                   &                 &                   &          & S71, S75, SS78, TPF89, vZSH98        \\
NGC6384      &   3.38      &    8.90           &    -0.62        &   0.05            &   16     & BK02, BKG99, OK93          \\
NGC6744      &  10.21      &    9.00           &    -0.89        &   0.04            &   16     & R95                        \\
NGC6946      &   8.30      &    8.70           &    -0.41        &   0.06            &    9     & FGW98, MRS85               \\
NGC7331      &   5.74      &    8.68           &    -0.48        &   0.04            &   12     & BKG99, OK93, ZKH94         \\
NGC7793      &   4.78      &    8.54           &    -0.50        &   0.07            &   22     & EP84, MRS85, WS83          \\
IC0342       &  22.33      &    8.85           &    -0.90        &   0.12            &    5     & MRS85                      \\
IC5201       &   4.26      &    8.27           &    +0.09        &   0.14            &   11     & R95                        \\  \hline
\end{tabular}
\end{center}
}
\end{table*}

\subsection{The algorithm for oxygen and nitrogen abundances determination}

The oxygen and nitrogen abundances in H\,{\sc ii} regions are derived in
the following way. We adopt a two-zone model for the temperature structure
within  H\,{\sc ii} regions.

As the first step, the (O/H)$_{P}$ oxygen abundance in H\,{\sc ii} regions is
determined with the expression suggested in Pilyugin (2001a)
\begin{equation}
12+log(O/H)_{P} = \frac{R_{23} + 54.2  + 59.45 P + 7.31 P^{2}}
                       {6.07  + 6.71 P + 0.37 P^{2} + 0.243 R_{23}}  ,
\label{equation:ohp}
\end{equation}
where $R_{\rm 23}$ =$R_{2}$ + $R_{3}$,
$R_{\rm 2}$ = $I_{[OII] \lambda 3727+ \lambda 3729} /I_{H\beta }$,
$R_{\rm 3}$ = $I_{[OIII] \lambda 4959+ \lambda 5007} /I_{H\beta }$,
and P = $R_{\rm 3}$/$R_{\rm 23}$.

Then, the electron temperatures within the [OIII] and [OII] zones are derived.
For this purpose the expressions for the oxygen abundance determination from Pagel et al.
(1992) and the T$_{e}$([OII]) -- T$_{e}$([OIII]) relation from Garnett (1992) are used,
\begin{equation}
\frac{O}{H} = \frac{O^+}{H^+} + \frac{O^{++}}{H^+}                ,
\label{equation:otot}
\end{equation}
\begin{eqnarray}
12+ \log (O^{++}/H^+) = \log \frac{I_{[OIII] \lambda 4959 + \lambda 5007}}
{I_{H_{\beta}}} +   \nonumber  \\
6.174 + \frac{1.251}{t_3}  - 0.55 \log t_3 ,
\label{equation:oplus2}
\end{eqnarray}
\begin{eqnarray}
12+ \log (O^{+}/H^+) = \log \frac{I_{[OII] \lambda 3726 + \lambda 3729}}
{I_{H_{\beta}}} + 5.890 +  \nonumber  \\
\frac{1.676}{t_2}  - 0.40 \log t_2 + \log (1+1.35x)  ,
\label{equation:oplus}
\end{eqnarray}
\begin{equation}
x= 10^{-4} n_e t_2^{-1/2},
\end{equation}
where $n_e$ is the electron density in cm$^{-3}$ (we assume a constant value of
$n_e = 100$ cm$^{-3}$), $t_3$ = $t_{[OIII]}$
is the electron temperature within the [OIII] zone in units of 10$^4$K, $t_2$ =
$t_{[OII]}$ is the electron temperature within the [OII] zone in units of 10$^4$K.
The $t_2$ value is determined from the equation (Garnett 1992)
\begin{equation}
t_2 =  0.7 \, t_3 + 0.3 .
\label{equation:t2t3}
\end{equation}
Using the value of O/H derived from Eq.(\ref{equation:ohp}) and the measured emission
line ratios, Eqs.(\ref{equation:otot}) - (\ref{equation:t2t3}) can be solved for
$t_{[OII]}$ and $t_{[OIII]}$.

Assuming $t_2=t_{[NII]}=t_{[OII]}$, the N/O abundance ratio in H\,{\sc ii} regions 
is determined from the expression (Pagel et al. 1992)
\begin{eqnarray}
\log (N/O) = \log (N^+/O^+) = \log \frac{I_{[NII] \lambda 6548 + \lambda 6584}}
{I_{[OII] \lambda 3726 + \lambda 3729}} +   \nonumber  \\
0.307 - \frac{0.726}{t_2}
 - 0.02 \log t_2  .
\label{equation:no}
\end{eqnarray}

Alternatively, the value of $t_{3}$ can be derived from the following expression 
for $t_P$ = $t_3$ (Pilyugin 2001a)
\begin{equation}
t_{P} = \frac{R_{23} + 3.09  + 7.05 P + 2.87 P^{2}}
                       {9.90  + 11.86 P + 7.05 P^{2} - 0.583 R_{23}}. 
\label{equation:tp}
\end{equation}
Then the oxygen (and nitrogen) abundances (O/H)$_{t_P}$ can be determined from
Eqs.(\ref{equation:otot})-(\ref{equation:no}) with t$_3$=t$_P$.  The oxygen 
abundance (O/H)$_{t_P}$ is in good agreement with the value of (O/H)$_P$ 
(the differences between (O/H)$_{t_P}$ and (O/H)$_P$ are usually less than 0.03 
dex, see Fig. 12 in Pilyugin 2001a). 
However, for H\,{\sc ii} regions in which most of the oxygen is in the O$^+$ 
stage, the value of (O/H)$_{t_P}$ is less reliable than the value of (O/H)$_P$.

\begin{table}
\caption[]{\label{table:refero}
List of references to Table \ref{table:oxygen}.
}
\begin{center}
\begin{tabular}{ll} \hline \hline
abrrev       & source                                       \\
\hline
AEL81        &  Alloin, Edmunds, Lindblad, et al.  1981     \\
BK02         &  Bresolin \& Kennicutt   2002               \\
BKC82        &  Blair, Kirshner \& Chevalier  1982         \\
BKG99        &  Bresolin, Kennicutt \& Garnett  1999       \\
C00          &  Castellanos  2000                           \\
CPG97        &  Christensen, Petersen \& Gammelgaard  1997 \\
DCL88        &  Deharveng, Caplan, Lequeux, et al.  1988    \\
DK81         &  Dennefeld \& Kunth  1981                   \\
dORW83       &  d'Odorico, Rosa \& Wampler  1983           \\
DTJ80        &  Dufour, Talbot, Jensen, et al.  1980        \\
DTP87        &  Diaz, Terlevich, Pagel, et al.  1987        \\
DTV91        &  Diaz, Terlevich, V\'{\i}lchez, et al.  1991 \\
EP84         &  Edmunds  \& Pagel  1984                     \\
FGW98        &  Ferguson, Gallagher \& Wyse  1998          \\
FTP86        &  Fierro, Torres-Peimbert \& Peimbert  1986  \\
GOS92        &  Garnett, Odewanh, \& Skillman  1992         \\
GS87         &  Garnett \& Shields  1987                   \\
GSP99        &  Garnett, Shields, Peimbert, et al.  1999    \\
GSS97        &  Garnett, Shields, Skillman, et al.  1997    \\
HBC96        &  Henry, Balkowski, Cayatte, et al.  1996     \\
HPC94        &  Henry, Pagel \& Chincarini  1994           \\
HPL92        &  Henry, Pagel, Lasseter, et al.  1992        \\
KA81         &  Kwitter \& Aller  1981                     \\
KG96         &  Kennicutt \& Garnett  1996                 \\
KR94         &  Kinkel \& Rosa  1994                       \\
MRS85        &  McCall, Rybsky \& Shields  1985            \\
OK93         &  Oey \& Kennicutt  1993                     \\
P70          &  Peimbert  1970                              \\
PEB79        &  Pagel, Edmunds, Blackwell, et al.  1979     \\
R95          &  Ryder  1995                                 \\
RPT82        &  Rayo, Peimbert \& Torres-Peimbert  1982    \\
RW97         &  Roy \& Walsh  1997                         \\
S71          &  Searle  1971                                \\
S75          &  Smith  1975                                 \\
SB84         &  Stauffer \& Bothun  1984                   \\
SKS96        &  Skillman, Kennicutt, Shields, et al.   1996 \\
SS78         &  Shields \& Searle  1978                    \\
SSK91        &  Shields, Skillman \& Kennicutt  1991       \\
TPF89        &  Torres-Peimbert, Peimbert \& Fierro  1989  \\
VEP88        &  V\'{\i}lchez, Edmunds \& Pagel  1988       \\
VPD88        &  V\'{\i}lchez, Pagel, Diaz, et al.  1988     \\
vZSH98       &  van Zee, Salzer, Haynes, et al.  1998       \\
WS83         &  Webster \& Smith  1983                     \\
ZKH94        &  Zaritsky, Kennicutt \& Huchra  1994        \\
 \hline
\end{tabular}
\end{center}
\end{table}

\subsection{Central abundances and radial gradients}

\begin{figure*}
\resizebox{0.98\hsize}{!}{\includegraphics[angle=0]{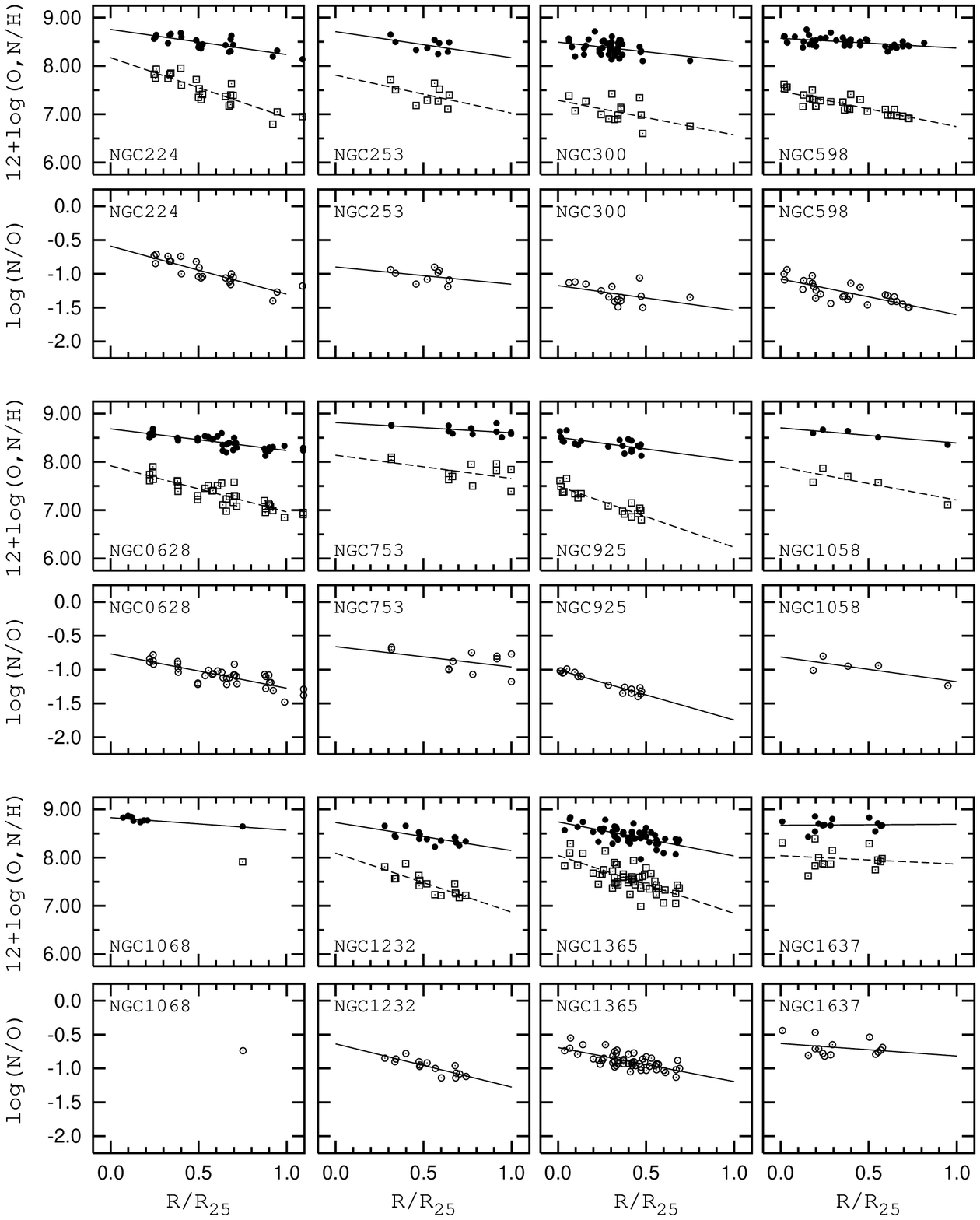}}
\caption{ Oxygen and nitrogen abundances, and nitrogen-to-oxygen abundance
ratios versus galactocentric distance for late-type galaxies. The oxygen
abundances are shown by filled circles, the linear least-squares fits to these data
are presented by solid lines. The nitrogen abundances are shown by open squares, 
the linear least-squares fits to these data are presented by dashed lines. The nitrogen-to-oxygen 
abundance ratios are shown by open circles, the linear least-squares fits to these data are 
presented by solid lines. The galactocentric distances are normalized to the 
isophotal radius.}
\label{figure:sample1}
\end{figure*}

\begin{figure*}
\resizebox{0.98\hsize}{!}{\includegraphics[angle=0]{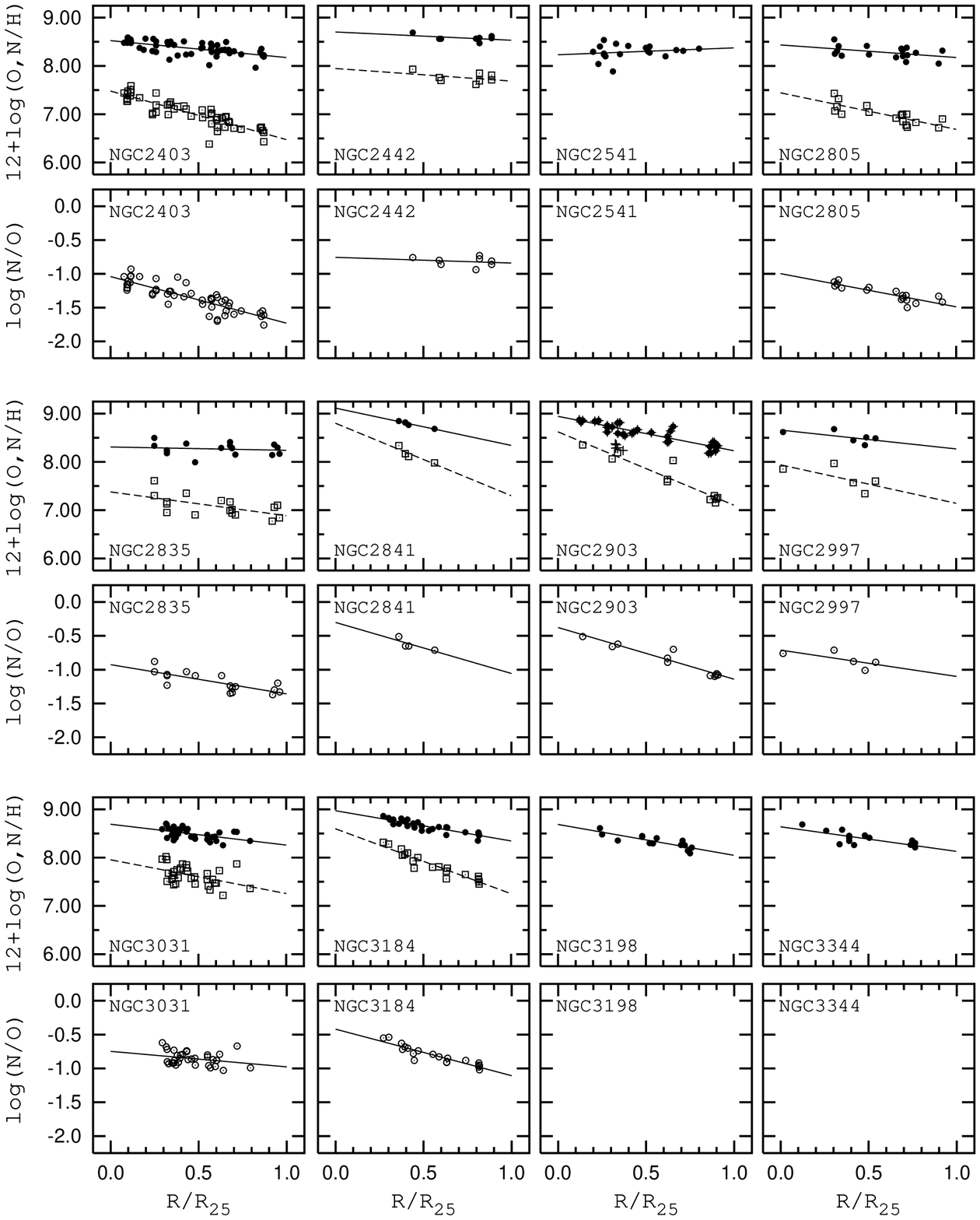}}
\caption{ Oxygen and nitrogen abundances and nitrogen-to-oxygen abundance
ratio versus galactocentric distance for late-type galaxies. The oxygen
abundances are shown by filled circles, the linear least-squares fits to these data
are presented by solid lines. The nitrogen abundances are shown by open squares,
 the linear least-squares  fits to these data are presented by dashed lines.
The nitrogen-to-oxygen abundance ratios  are shown by open circles,
the linear least-squares  fits to these data are presented by solid lines.
The galactocentric distances are normalized to the isophotal radius.
The H\,{\sc ii} regions with depleted oxygen abundances in NGC 2903 are
shown by pluses.}
\label{figure:sample2}
\end{figure*}

\begin{figure*}
\resizebox{0.98\hsize}{!}{\includegraphics[angle=0]{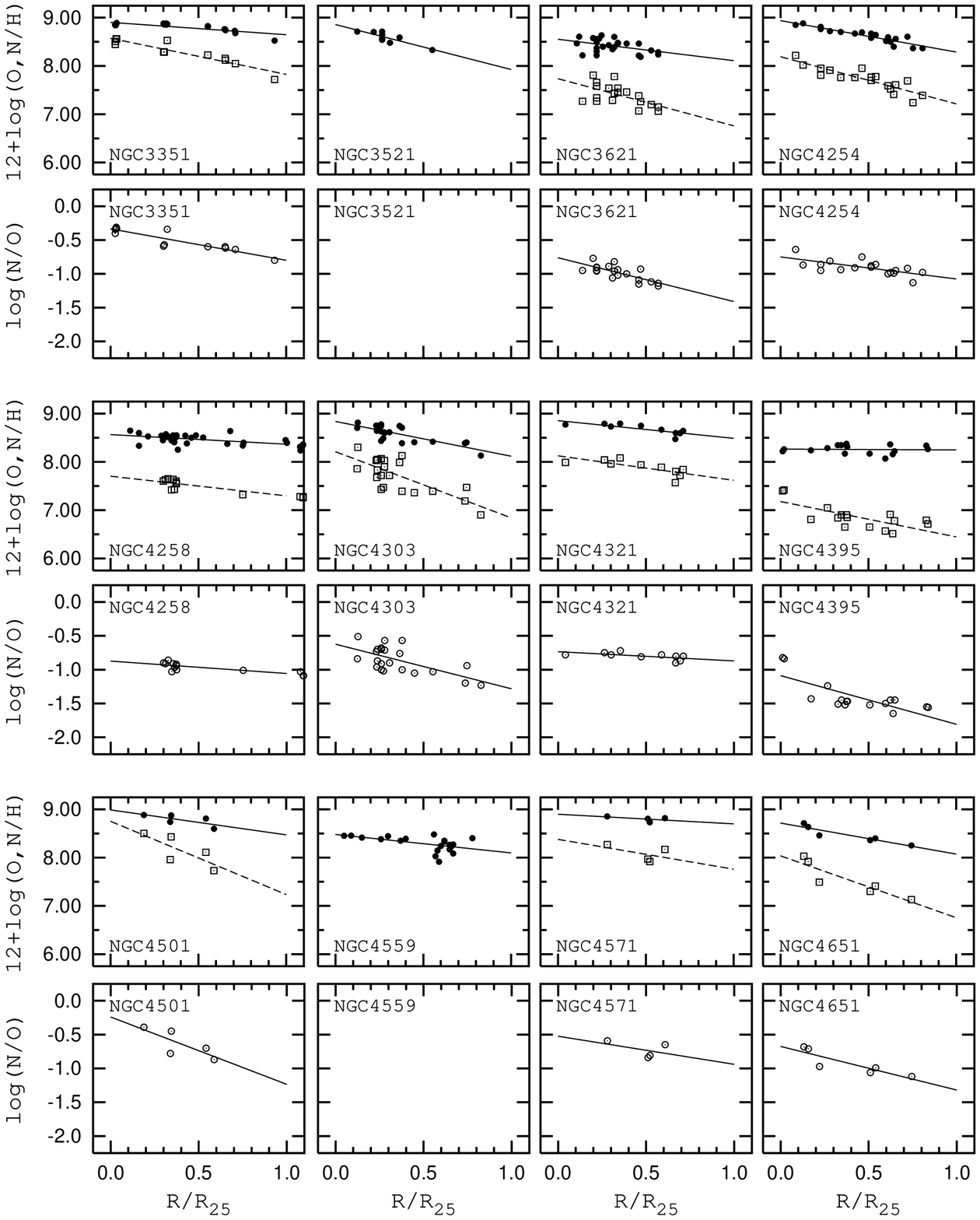}}
\caption{ Oxygen and nitrogen abundances and nitrogen-to-oxygen abundance
ratio versus galactocentric distance for late-type galaxies. The oxygen
abundances are shown by filled circles, the linear least-squares fits to these data
are presented by solid lines. The nitrogen abundances are shown by open squares,
 the linear least-squares fits to these data are presented by dashed lines.
The nitrogen-to-oxygen abundance ratios are shown by open circles,
the linear least-squares fits to these data are presented by solid lines.
The galactocentric distances are normalized to the isophotal radius.}
\label{figure:sample3}
\end{figure*}

\begin{figure*}
\resizebox{0.98\hsize}{!}{\includegraphics[angle=0]{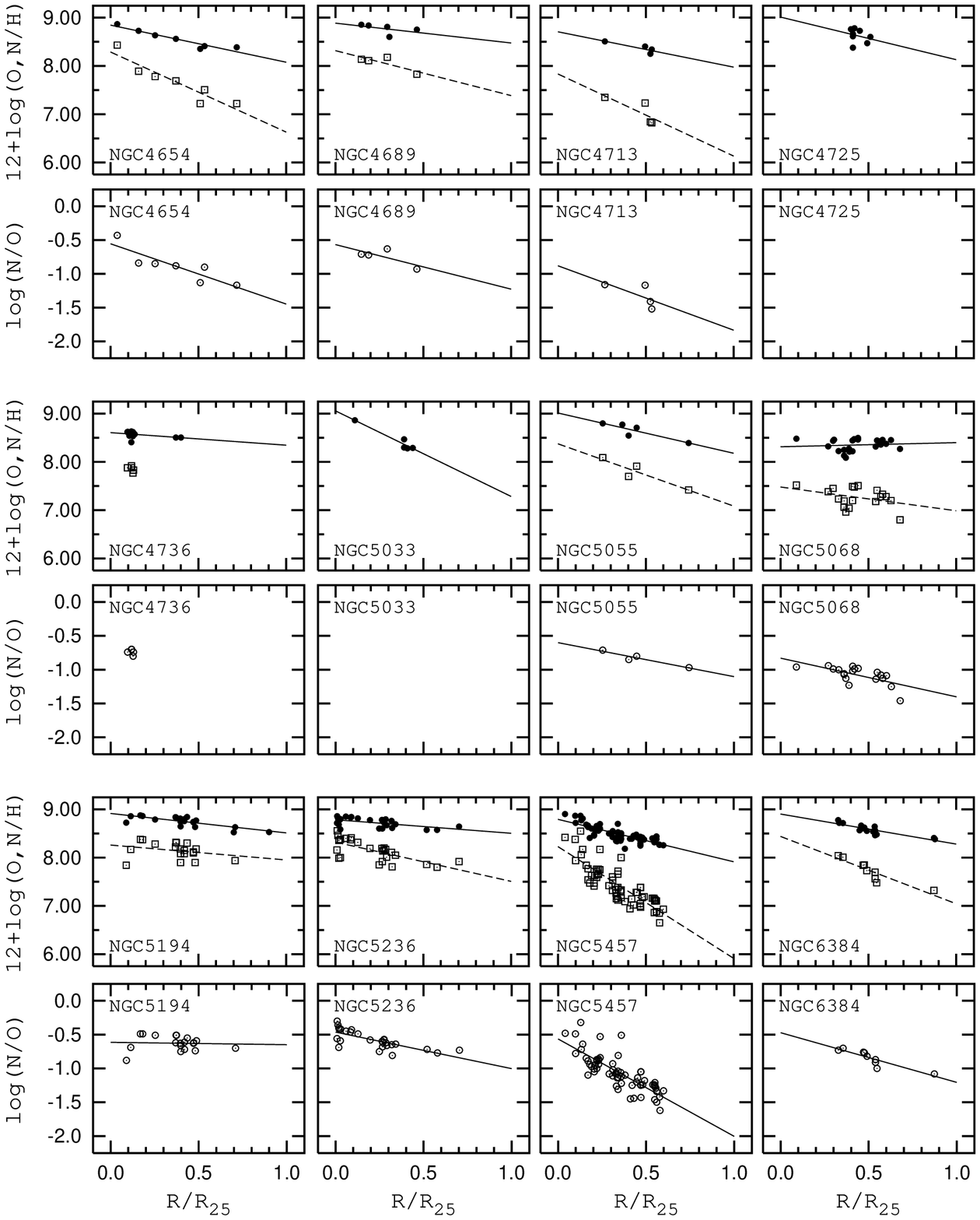}}
\caption{ Oxygen and nitrogen abundances and nitrogen-to-oxygen abundance
ratio versus galactocentric distance for late-type galaxies. The oxygen
abundances are shown by filled circles, the linear least-squares fits to these data
are presented by solid lines. The nitrogen abundances are shown by open squares,
 the linear least-squares fits to these data are presented by dashed lines.
The nitrogen-to-oxygen abundance ratios  are shown by open circles,
the linear least-squares fits to these data are presented by solid lines.
The galactocentric distances are normalized to the isophotal radius.}
\label{figure:sample4}
\end{figure*}

\begin{figure*}
\resizebox{0.98\hsize}{!}{\includegraphics[angle=0]{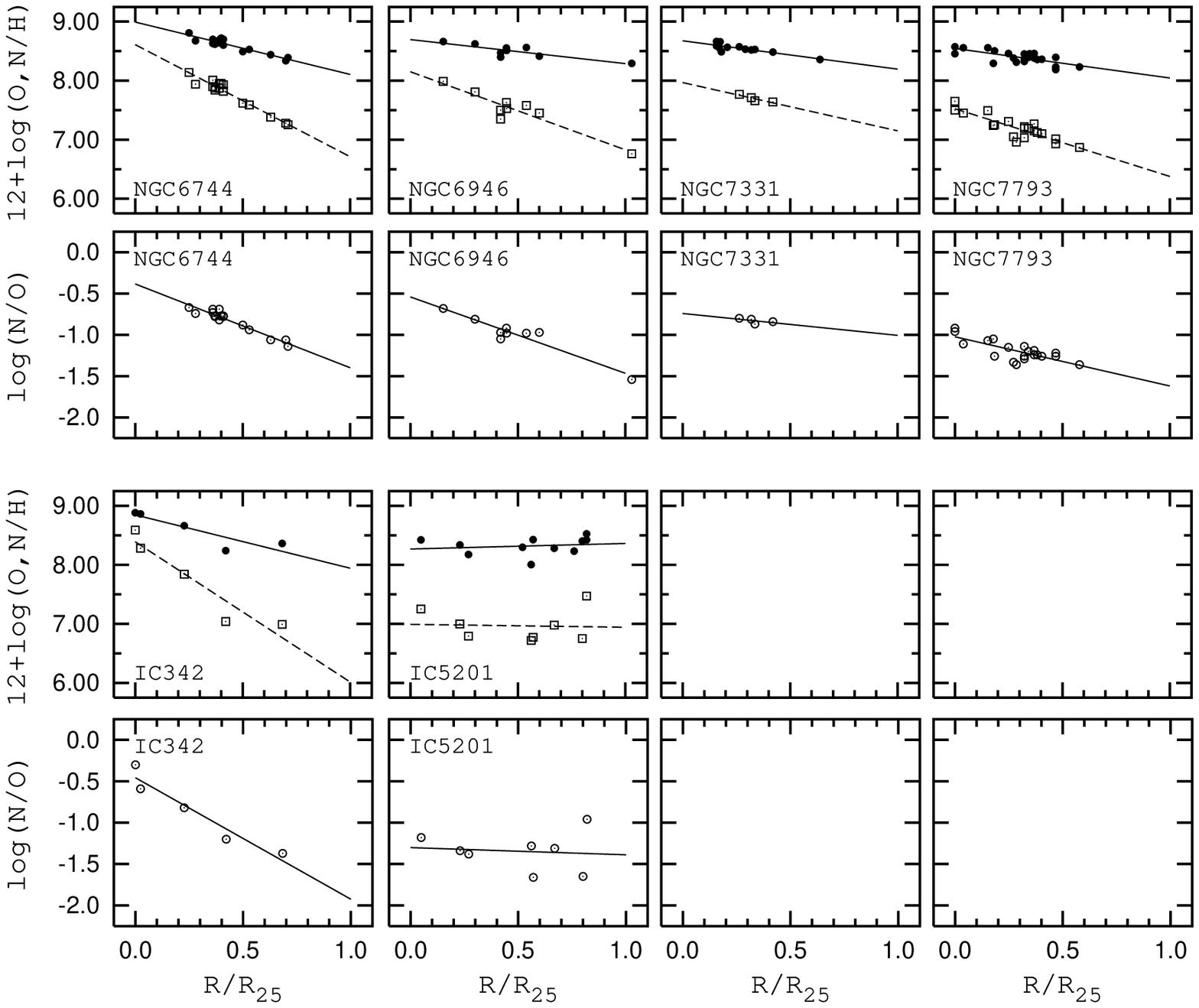}}
\caption{ Oxygen and nitrogen abundances and nitrogen-to-oxygen abundance
ratio versus galactocentric distance for late-type galaxies. The oxygen
abundances are shown by filled circles, the linear least-squares  fits to these data
are presented by solid lines. The nitrogen abundances are shown by open squares,
 the linear least-squares  fits to these data are presented by dashed lines.
The nitrogen-to-oxygen abundance ratios  are shown by open circles,
the linear least-squares fits to these data are presented by solid lines.
The galactocentric distances are normalized to the isophotal radius.}
\label{figure:sample5}
\end{figure*}

A compilation of published spectra of H\,{\sc ii} regions in late-type galaxies has been carried out.
Our list contains more than 1000 individual spectra of H\,{\sc ii} regions in 54 late-type galaxies.
We performed an extensive compilation of spectra of H\,{\sc ii} regions from the literature but our
collection does not pretend to be exhaustive. Only the galaxies with available spectra for at least
four H\,{\sc ii} regions were taken into consideration.
Using these spectrophotometric data and the algorithm described in the previous section, the oxygen and
nitrogen abundances were derived for our sample of H\,{\sc ii} regions.

In investigations of the relationships between the oxygen abundances and the
macroscopic properties of spiral galaxies, the concept of the characteristic
oxygen abundance has been introduced: it is defined as the oxygen abundance in
the disk at a predetermined galactocentric distance.
Due to the presence of radial abundance gradients in the disks of spiral
galaxies, the choice of the characteristic (or representative) value of the oxygen
abundance in a galaxy is not trivial.
The value of the oxygen abundance at the $B$-band effective (half-light) radius
of the disk (Garnett \& Shields 1987; Garnett 2002), the value of the central
oxygen abundance extrapolated to zero radius from the radial abundance gradient
(Vila-Costas \& Edmunds 1992), the value of
the oxygen abundance at $r=0.4R_{\rm 25}$, where $R_{\rm 25}$ is the isophotal
(or photometric, or Holmberg) radius, (Zaritsky, Kennicutt \& Huchra 1994),
and the value of the oxygen abundance at one disk scale length from the
nucleus (Garnett et al. 1997), have been used as the characteristic oxygen
abundance in a galaxy. To estimate the characteristic oxygen abundance in
spiral galaxies, the radial distribution of oxygen abundances within the galaxies
should be established.

The radial oxygen abundance distribution in every galaxy is well fitted by the
following equation:
\begin{equation}
12+\log(O/H)  = 12+\log(O/H)_0 + C_{O/H} \times (R/R_{25}) ,
\label{equation:grado}
\end{equation} where 12 + log(O/H)$_{0}$ is the extrapolated central oxygen abundance,
C$_{O/H}$ is the slope of the oxygen abundance gradient expressed in terms of dex/$R_{\rm 25}$,
and $R$/$R_{\rm 25}$ is the fractional radius (the galactocentric distance normalized to the disk isophotal radius).

The derived parameters of the oxygen abundance distributions are presented in Table~\ref{table:oxygen}.
The name of the galaxy is listed in column 1.
The isophotal radius $R_{\rm 25}$ (in arcmin) taken from the Third Reference Catalog of Bright Galaxies
(de Vaucouleurs et al. 1991) is given in column 2.
The extrapolated central 12 + log(O/H)$_{0}$ oxygen abundance and the gradient
(the coefficient C$_{O/H}$ in Eq.(\ref{equation:grado})) expressed in terms of dex/$R_{\rm 25}$
    are listed in columns 3 and 4.
The scatter of oxygen abundances around the general radial oxygen abundance trend is reported in column 5.
The number of available individual spectra of H\,{\sc ii} regions in the galaxy is listed in column 6.
The source(s) for the emission line flux measurements in the H\,{\sc ii} regions is(are) reported in column 7.
The list of references to Table \ref{table:oxygen} is given in Table \ref{table:refero}.

The derived radial distributions of the oxygen abundance in galaxies are
presented in Figs. 1 to 5. The oxygen
abundances for individual H\,{\sc ii} regions are shown by the filled circles.
The linear best fits (derived via the least squares method) to these points are
presented by solid lines. The galactocentric distances are normalized
to the isophotal radius.

\begin{table*}
\caption[]{\label{table:nitrogen}
The derived parameters of the radial distributions of the nitrogen abundance and
nitrogen-to-oxygen abundance ratios for our sample of galaxies}
{\small
\begin{center}
\begin{tabular}{lccccccc}  \\ \hline \hline
             &                 &               &                  &         &              &              &                  \\
galaxy       & 12+log(N/H)$_0$ &  gradient     & $\Delta$log(N/H) &  number of  & log(N/O)$_0$ &  gradient    & $\Delta$log(N/O) \\
             &                 & dex/$R _{\rm 25}$ &   dex            & H\,{\sc ii} regions &              & dex/$R_{\rm 25}$ &                  \\  \hline
             &                 &               &                  &         &              &              &                  \\
NGC0224      &    8.17         &   -1.24       &     0.15         &   22    &  -0.59       & -0.71        &   0.09           \\
NGC0253      &    7.81         &   -0.79       &     0.17         &    9    &  -0.90       & -0.26        &   0.09           \\
NGC0300      &    7.29         &   -0.72       &     0.19         &   15    &  -1.17       & -0.37        &   0.12           \\
NGC0598      &    7.48         &   -0.74       &     0.11         &   29    &  -1.08       & -0.53        &   0.10           \\
NGC0628      &    7.92         &   -0.95       &     0.13         &   36    &  -0.77       & -0.51        &   0.09           \\
NGC0753      &    8.14         &   -0.48       &     0.18         &   11    &  -0.66       & -0.30        &   0.14           \\
NGC0925      &    7.50         &   -1.27       &     0.10         &   17    &  -1.00       & -0.74        &   0.04           \\
NGC1058      &    7.89         &   -0.68       &     0.12         &    6    &  -0.81       & -0.36        &   0.08           \\
NGC1068      &                 &               &                  &         &              &              &                  \\
NGC1232      &    8.09         &   -1.22       &     0.12         &   15    &  -0.64       & -0.64        &   0.06           \\
NGC1365      &    8.05         &   -1.20       &     0.19         &   53    &  -0.69       & -0.50        &   0.08           \\
NGC1637      &    8.04         &   -0.18       &     0.21         &   14    &  -0.63       & -0.18        &   0.12           \\
NGC2403      &    7.48         &   -1.01       &     0.14         &   50    &  -1.04       & -0.69        &   0.11           \\
NGC2442      &    7.95         &   -0.26       &     0.08         &    8    &  -0.75       & -0.09        &   0.06           \\
NGC2541      &                 &               &                  &         &              &              &                  \\
NGC2805      &    7.44         &   -0.75       &     0.12         &   17    &  -1.00       & -0.49        &   0.06           \\
NGC2835      &    7.38         &   -0.50       &     0.16         &   17    &  -0.93       & -0.43        &   0.09           \\
NGC2841      &    8.80         &   -1.50       &     0.05         &    4    &  -0.30       & -0.75        &   0.04           \\
NGC2903      &    8.62         &   -1.52       &     0.14         &   11    &  -0.38       & -0.76        &   0.06           \\
NGC2997      &    7.94         &   -0.80       &     0.17         &    5    &  -0.71       & -0.39        &   0.08           \\
NGC3031      &    7.96         &   -0.70       &     0.17         &   33    &  -0.75       & -0.23        &   0.10           \\
NGC3184      &    8.60         &   -1.35       &     0.08         &   20    &  -0.42       & -0.69        &   0.05           \\
NGC3198      &                 &               &                  &         &              &              &                  \\
NGC3344      &                 &               &                  &         &              &              &                  \\
NGC3351      &    8.57         &   -0.75       &     0.08         &   13    &  -0.33       & -0.47        &   0.06           \\
NGC3521      &                 &               &                  &         &              &              &                  \\
NGC3621      &    7.74         &   -0.98       &     0.17         &   19    &  -0.76       & -0.65        &   0.07           \\
NGC4254      &    8.19         &   -0.97       &     0.11         &   18    &  -0.75       & -0.33        &   0.08           \\
NGC4258      &    7.70         &   -0.41       &     0.07         &   13    &  -0.87       & -0.18        &   0.04           \\
NGC4303      &    8.21         &   -1.37       &     0.23         &   22    &  -0.62       & -0.66        &   0.15           \\
NGC4321      &    8.12         &   -0.51       &     0.10         &   10    &  -0.73       & -0.14        &   0.04           \\
NGC4395      &    7.18         &   -0.73       &     0.17         &   16    &  -1.09       & -0.72        &   0.15           \\
NGC4501      &    8.75         &   -1.52       &     0.18         &    5    &  -0.24       & -1.00        &   0.12           \\
NGC4559      &                 &               &                  &         &              &              &                  \\
NGC4571      &    8.38         &   -0.62       &     0.12         &    4    &  -0.52       & -0.41        &   0.09           \\
NGC4651      &    8.04         &   -1.28       &     0.14         &    6    &  -0.67       & -0.64        &   0.08           \\
NGC4654      &    8.29         &   -1.66       &     0.14         &    7    &  -0.56       & -0.89        &   0.11           \\
NGC4689      &    8.32         &   -0.93       &     0.08         &    4    &  -0.57       & -0.66        &   0.08           \\
NGC4713      &    7.84         &   -1.70       &     0.14         &    4    &  -0.88       & -0.95        &   0.12           \\
NGC4725      &                 &               &                  &         &              &              &                  \\
NGC4736      &                 &               &                  &         &              &              &                  \\
NGC5033      &                 &               &                  &         &              &              &                  \\
NGC5055      &    8.38         &   -1.29       &     0.10         &    4    &  -0.60       & -0.50        &   0.03           \\
NGC5068      &    7.48         &   -0.49       &     0.18         &   19    &  -0.03       & -0.57        &   0.10           \\
NGC5194      &    8.26         &   -0.31       &     0.15         &   19    &  -0.61       & -0.04        &   0.10           \\
NGC5236      &    8.32         &   -0.82       &     0.15         &   27    &  -0.46       & -0.54        &   0.09           \\
NGC5457      &    8.23         &   -2.32       &     0.22         &   63    &  -0.56       & -1.44        &   0.16           \\
NGC6384      &    8.44         &   -1.40       &     0.10         &    9    &  -0.47       & -0.73        &   0.05           \\
NGC6744      &    8.61         &   -1.90       &     0.06         &   16    &  -0.38       & -1.01        &   0.04           \\
NGC6946      &    8.15         &   -1.32       &     0.11         &    9    &  -0.54       & -0.92        &   0.07           \\
NGC7331      &    7.97         &   -0.82       &     0.02         &    4    &  -0.74       & -0.27        &   0.02           \\
NGC7793      &    7.53         &   -1.15       &     0.10         &   20    &  -1.02       & -0.60        &   0.08           \\
IC0342       &    8.39         &   -2.38       &     0.21         &    5    &  -0.46       & -1.46        &   0.11           \\
IC5201       &    6.99         &   -0.05       &     0.25         &    8    &  -1.30       & -0.09        &   0.22           \\ \\ \hline
\end{tabular}
\end{center}
}
\end{table*}

As in the case of the oxygen abundance, the radial nitrogen abundance distribution in every galaxy
is well fitted by the following equation:
\begin{equation}
12+\log(N/H)  = 12+\log(N/H)_0 + C_{N/H} \times (R/R_{25}) ,
\label{equation:gradn}
\end{equation}
and the radial distribution of the nitrogen to oxygen abundance ratios
is well fitted by the single equation of the type:
\begin{equation}
\log(N/O)  = \log(N/O)_0 + C_{N/O} \times (R/R_{25}) .
\label{equation:gradno}
\end{equation}
The derived parameters of the nitrogen abundance distributions are presented in Table \ref{table:nitrogen}.
The name of the galaxy is listed in column 1.
The extrapolated central 12 + log(N/H)$_{0}$ nitrogen abundance and the gradient
(the coefficient C$_{N/H}$ in Eq.(\ref{equation:gradn})) expressed in terms of dex/$R_{\rm 25}$
    are listed in columns 2 and 3.
The scatter of nitrogen abundances around the general radial nitrogen abundance trend is reported in column 4.
The number of available individual spectra of H\,{\sc ii} regions in the galaxy with measured nitrogen
emission line fluxes is listed in column 5.
The extrapolated central value of the nitrogen-to-oxygen ratio log(N/O)$_{0}$ is listed in column 6.
The radial gradient of the nitrogen-to-oxygen abundance ratio
(the coefficient C$_{N/O}$ in Eq.(\ref{equation:gradno})) expressed in terms of dex/$R_{\rm 25}$
    is reported in column 7.
The scatter of log(N/O) values around the general radial trend is presented in column 8.
The nitrogen emission line fluxes in the H\,{\sc ii} regions were taken from the same sources
as the oxygen emission line fluxes (see references in Table \ref{table:oxygen}).

The derived radial distributions of the nitrogen abundances and nitrogen-to-oxygen abundance 
ratios in galaxies are shown in Figs.\ref{figure:sample1} to \ref{figure:sample5}.
The nitrogen abundances for individual H\,{\sc ii} regions are shown by the open squares.
The linear best fits to these points are presented by dashed lines.
The nitrogen-to-oxygen abundance ratios for individual H\,{\sc ii} regions are shown
by the open circles. The linear best fits to these points are presented by solid lines.

\begin{table}
\caption[]{\label{table:izlom}
 Galaxies with the observed H\,{\sc ii} regions
which are expected to not belong to the upper branch
of the $R_{\rm 23}$ -- O/H diagram and the radius R$^*$ where the oxygen
abundance decreases to around 12+log(O/H) = 8.2.
}
\begin{center}
\begin{tabular}{lc|lc} \hline \hline
galaxy       & $R^{*}$/$R_{\rm 25}$ & galaxy       & $R^{*}$/$R_{\rm 25}$     \\
 \hline
NGC300       & 0.8              & NGC4651      &   0.8                \\

NGC925       & 0.5              & NGC5033      &   0.5                \\

NGC1365      & 0.7              & NGC5457      &   0.6                \\

NGC2805      & 0.9              & NGC7793      &   0.6                \\

NGC3198      & 0.8              & IC342        &   0.7                \\  \hline
\end{tabular}
\end{center}
\end{table}

For the majority of galaxies, all the H\,{\sc ii} regions with available
oxygen and nitrogen emission line measurements were used in the analysis of the
abundance gradients. For a few galaxies, however, some H\,{\sc ii} regions were
rejected for the following reason. The relationship between oxygen abundance
and strong line intensities is double-valued with two distincts parts usually
known as the ``lower'' and ``upper'' branches of the $R_{\rm 23}$ -- O/H relationship.
Thus, one has to know apriori on which of the two branches the H\,{\sc ii} region
lies. The above expression for the oxygen abundance determination in H\,{\sc ii}
regions, Eq.(\ref{equation:ohp}), is valid only for H\,{\sc ii} regions which belong
to the upper branch, with 12+log(O/H) higher than $\sim 8.2$.
It has been known for a long time (Searle 1971; Smith 1975) that disks of spiral
galaxies show radial oxygen abundance gradients, in the sense that the oxygen
abundance is higher in the central part of the disk and decreases with
galactocentric distance. We thus start from the H\,{\sc ii} regions in the central
part of disks and move outward until the radius $R^*$ where the oxygen abundance
decreases to 12+log(O/H) $\sim$ 8.2. It should be noted that it is difficult to
establish the exact value of $R^*$ due to the scatter in oxygen abundance values at
any fixed radius. An unjustified use of Eq. (\ref{equation:ohp}) in the determination
of the oxygen abundance in low-metallicity H\,{\sc ii} regions beyond $R^*$ would result
in overestimated oxygen abundances, and would produce a false bend in the slope of
abundance gradients (Pilyugin 2003a). Therefore, H\,{\sc ii} regions with galactocentric
distances larger than $R^*$, those with 12+log(O/H) less than 8.2 were rejected.
The list of such galaxies is given in Table \ref{table:izlom}.

The credibility of the radial oxygen abundance gradients (as well as gradients
of the nitrogen abundance and gradients of the nitrogen-to-oxygen abundance
ratios) is defined not only by the large number of H\,{\sc ii} regions and
their small dispersion but also by the distribution of these H\,{\sc ii} regions
along the galactic radius. For example, the six H\,{\sc ii} regions in the galaxy NGC 4651
(Fig. 3) give a much more reliable value of abundance gradients than
the 8 H\,{\sc ii} regions in the galaxy NGC 4725 (Fig. 4). The estimated
values of the radial oxygen abundance gradient in galaxies NGC 1068, NGC 1637, NGC
2841, NGC 3521, NGC 4571, NGC 4713, NGC 4725, NGC 5033, and NGC 5055 are not beyond question. 

\section{Characteristic oxygen abundance in spirals as a function
of $M_{\rm B}$, $V_{\rm rot}$, and Hubble type}

\begin{table*}
\caption[]{\label{table:Vrot}
The characteristic oxygen abundance (at $r=0.4R_{\rm 25}$), gas (atomic and molecular) mass fraction $\mu$,
atomic and molecular hydrogen contents, and rotation velocity for spiral galaxies from our sample.
The oxygen abundance is given as 12+log(O/H).}
{\footnotesize
\begin{center}
\begin{tabular}{lccccrl} \\ \hline \hline
galaxy & O/H    & $\mu$  & M$_{H_I}/L_B$ & M$_{H_2}$/L$_B$  & V$_{rot}$    &             references                                         \\
NGC    &        &        & M$_{\odot}$/L$_{\odot}$&M$_{\odot}$/L$_{\odot}$  & km/s       & (M$_{H_I}$ ; M${H_2}$ ;     V$_{rot}$)           \\
\hline
224    & 8.55   & 0.09   &    0.15      &    0.004         & 269          & DD, HR                ; KDI                 ; H                \\
253    & 8.50   & 0.12   &    0.15      &    0.057         & 215          & SD, HR, HS            ; YXT, SNK            ; PCG, H           \\
300    & 8.33   & 0.38   &    0.93      &    ....          & 95           & RCC, HR               ;  -                  ; PCB, H, RCC      \\
598    & 8.49   & 0.24   &    0.46      &    0.004         & 116          & DD, HR                ; YXT                 ; H, CS            \\
628    & 8.51   & 0.35   &    0.76      &    0.052         & 153          & WKA, R, SD, DD        ; NN, YXT, S, HTR     ; NNK              \\
753    & 8.73   & 0.18   &    0.33      &    ....          & 215          & Sh, BSC               ;    -                ; RFT, AMB         \\
925    & 8.32   & 0.28   &    0.58      &    0.013         & 127          & WKA, R, DD, HHG       ; S                   ; KS               \\
1058   & 8.58   & 0.35   &    0.79      &    0.017         & ....         & SD, vKS               ; S                   ; -                \\
1068   & 8.73   & 0.10   &    0.04      &    0.117         & 220          & HS, HBS               ; YXT                 ; HB               \\
1232   & 8.50   & 0.19   &    0.35      &    ....          & 229          & SD, RMG, ZB, BMR      ; -                   ; ZB, BMR          \\
1365   & 8.46   & 0.21   &    0.27      &    0.130         & 285          & OH, RMG, JM           ;  SJL                ; OH               \\
2403   & 8.39   & 0.31   &    0.67      &    0.004         & 135          & WKA, R, HR            ; YXT, S              ; H, BBS           \\
2442   & 8.64   & 0.11   &    0.18      &    0.013         & ....         & BM, RMG, RKS          ; BWR                 ; -                \\
2541   & 8.29   & 0.43   &    1.16      &    ....          & ....         & R, BR, HS, HHM        ;    -                ;   -              \\
2805   & 8.33   & 0.29   &    0.63      &    ....          & 78           & Sh, Re, BCH           ;    -                ; Re               \\
2835   & 8.28   & 0.19   &    0.35      &    ....          & 120          & SD, RMG, BMR          ;       -             ; BMR              \\
2841   & 8.81   & 0.16   &    0.23      &    0.052         & 318          & R, B, HS              ; YXT                 ; BBS, B           \\
2903   & 8.66   & 0.15   &    0.22      &    0.043         & 223          & WKA, HR, HS, HHG, HHM ; NN, YXT, S, HTR     ; NNK, BBS         \\
2997   & 8.51   & 0.13   &    0.22      &    ....          & 195          & SD, RMG               ;       -             ; P, MM            \\
3031   & 8.52   & 0.10   &    0.16      &    0.004         & 240          & ADS, HR               ; S                   ; H, Ro            \\
3184   & 8.72   & 0.21   &    0.34      &    0.052         & ....         & Sh, HS                ; NN, YXT, S          ;    -             \\
3198   & 8.43   & 0.28   &    0.57      &    ....          & 154          & WKA, R, SD            ;       -             ; ABB, HRG, BBS    \\
3344   & 8.43   & 0.29   &    0.60      &    0.030         & ....         & R, HHG                ; YXT, S              ;    -             \\
3351   & 8.80   & 0.10   &    0.14      &    0.035         & 236          & Sc, HS                ; YXT, S, HTR         ; Bu               \\
3521   & 8.49   & 0.22   &    0.33      &    0.091         & 235          & R                     ; NN, YXT, S, HTR     ; NNK, CG          \\
3621   & 8.38   & 0.34   &    0.78      &    ....          & 163          & RMG, HR, BMR          ;   -                 ; BMR              \\
4254   & 8.68   & 0.18   &    0.20      &    0.126         & 264          & SD, Wb, HRb           ; YXT, SKB ,  SEC, KY ; NNK, GGK, Wb     \\
4258   & 8.49   & 0.14   &    0.23      &    0.022         & 200          & WKA, R, HS            ; YXT, HTR, CD        ; AS               \\
4303   & 8.55   & 0.17   &    0.20      &    0.104         & 178          & Sh, Wb, HS, HRb       ; YXT, SKB,  HTR, KY  ; NNK, GGK, Wb     \\
4321   & 8.71   & 0.15   &    0.12      &    0.139         & 236          & Sh, Wa, HRb           ; YXT, SKB, HTR, KY   ; GGK, Wa          \\
4395   & 8.26   & 0.43   &    1.07      &    0.083         & 83           & WKA, R, HR, HS, HHM   ; SEC                 ; BB               \\
4501   & 8.78   & 0.09   &    0.07      &    0.083         & 295          & Wa, HRb               ; YXT, SKB, KY        ; NNK, GGK, Wa, KS \\
4559   & 8.33   & 0.27   &    0.54      &    0.017         & 129          & BR, Sh, HS            ; S                   ; KS               \\
4571   & 8.82   & 0.12   &    0.14      &    0.078         & 165          & Wb, HR, HRb           ; YXT, KY             ; Wb               \\
4651   & 8.46   & 0.19   &    0.33      &    0.035         & 250          & Wa, HHG, HRb          ; YXT, SKB, KY        ; Wa               \\
4654   & 8.54   & 0.15   &    0.20      &    0.070         & 195          & Sh, Wb, HRb           ; YXT, SKB, KY        ; GGK, Wb          \\
4689   & 8.72   & 0.10   &    0.05      &    0.104         & 182          & Wb, HRb               ; YXT, SKB, SEC, KY   ; GGK, Wb          \\
4713   & 8.42   & 0.29   &    0.58      &    0.017         & 137          & Sh, Wb, HRb, HHM      ; YXT, KY             ; Wb               \\
4725   & 8.66   & 0.14   &    0.18      &    0.065         & 249          & WKA, GDH, WAD         ; YXT   -             ; Bu, WAD          \\
4736   & 8.50   & 0.05   &    0.05      &    0.022         & 198          & R,BHS, HR             ; NN, YXT, S, HTR     ; NNK, Bu, BHS     \\
5033   & 8.35   & 0.28   &    0.50      &    0.096         & 228          & WKA, R, SD,B          ; YXT, SEC, HTR       ; NNK, HAS, B      \\
5055   & 8.68   & 0.24   &    0.42      &    0.061         & 210          & WKA, R, B             ; YXT, S, HTR         ; TM, B            \\
5068   & 8.35   & 0.18   &    0.34      &    ....          & ....         & RMG, HR               ;        -            ;     -            \\
5194   & 8.75   & 0.14   &    0.14      &    0.109         & 242          & WG, DD                ; YXT, S, HTR         ; KN, WG           \\
5236   & 8.68   & 0.31   &    0.57      &    0.104         & 205          & SD, HR, DD            ; YXT                 ; VPD              \\
5457   & 8.44   & 0.32   &    0.69      &    0.026         & 180          & R, DD, HR             ; S, HTR              ; GW               \\
6384   & 8.65   & 0.19   &    0.29      &    0.070         & 200          & Sh, HHM               ; YXT                 ; SCR              \\
6744   & 8.64   & 0.31   &    0.67      &    ....          & ....         & RMG                   ;    -                ;      -           \\
6946   & 8.53   & 0.15   &    0.22      &    0.048         & 179          & R, DD, HR             ; YXT, S, HTR         ; NNK, H, CCB      \\
7331   & 8.48   & 0.18   &    0.24      &    0.078         & 262          & R, Sh, B              ; NN, YXT, HTR, HHM   ; NNK,  BBS, B, KS \\
7793   & 8.34   & 0.18   &    0.31      &    0.013         & 117          & CP, RMG, HR, BMR      ; SEC                 ; CP, BMR          \\
IC0342 & 8.49   & 0.16   &    0.26      &    0.026         & 192          & R, HR                 ; YXT, S              ; H                \\
IC5201 & 8.31   & 0.36   &    0.86      &    ....          & 97           & BM, RMG, BMR          ;    -                ; BMR              \\  \hline
\end{tabular}
\end{center}
}
\end{table*}

\begin{table*}
\caption[]{\label{table:referV}
List of references to Table \ref{table:Vrot}}
{\small
\begin{center}
\begin{tabular}{ll|ll} \hline \hline
Abbr         & Reference                                         & Abbr         & Reference                                         \\
 \hline
ABB          &  van Albada, Bahcall, Begeman, et al. 1985        & KN           &  Kuno   \& Nakai 1997                             \\
ADS          &  Appleton, Davies \& Stephenson 1981              & KS           &  Krumm  \& Salpeter 1979                           \\
AMB          &  Amram, Marcelin, Balkowski, et al. 1994          & KY           &  Kenney \& Young 1988                              \\
AS           &  van Albada  \& Shane 1975                        & MM           &  Milliard  \& Marcelin 1981                       \\
B            &  Bosma 1981                                       & NN           &  Nishiyama  \& Nakai 2001                         \\
Bu           &  Buta 1988                                        & NNK          &  Nishiyama, Nakai  \& Kuno 2001                    \\
BB           &  de Blok  \& Bosma 2002                           & OH           &  Ondrechen  \& van der Hulst 1989                  \\
BBS          &  Begeman, Broeils  \& Sanders 1991                & P            &  Peterson 1978                                     \\
BCH          &  Bosma, Casini, Heidmann, et al. 1980             & PCB          &  Puche, Carignan  \& Bosma 1990                   \\
BHS          &  Bosma, van der Hulst  \& Sullivan 1977           & PCG          &  Puche, Carignan  \& van Gorkom 1991               \\
BM           &  Bajaja   \& Martin  1985                         & R            &  Rots 1980                                        \\
BMR          &  Becker, Mebold, Reif, et al. 1988                & RCC          &  Rogstad, Crutcher  \& Chu 1979                    \\
BR           &  Broeils  \& Rhee 1997                            & Re           &  Reakes 1979                                       \\
BSC          &  Bravo-Alfaro, Szomoru, Cayatte, et al. 1997      & Ro           &  Rots 1975                                         \\
BWR          &  Bajaja, Wielebinski, Reuter, et al. 1995         & RFT          &  Rubin, Ford  \& Thonnard 1980                     \\
CCB          &  Carignan, Charbonneau, Boulanger, et al. 1990    & RKS          &  Ryder, Koribalski, Staveley-Smith, et al. 2001    \\
CD           &  Cox  \& Downes 1996                              & RMG          &  Reif, Mebold, Goss, et al. 1982                   \\
CG           &  Casertano  \& van Gorkom 1991                    & S            &  Sage 1993                                         \\
CS           &  Corbelli \& Salucci 2000                         & Sc           &  Schneider 1989                                    \\
DD           &  Dean  \& Davies 1975                             & SCR          &  Sperandio, Chincarini, Rampazzo, et al. 1995     \\
GDH          &  Garc\'{\i}a-Barreto, Downes  \& Huchtmeier 1994  & Sh           &  Shostak 1978                                      \\
GGK          &  Guhathakurta, vanGorkom, Kotanyi, et al. 1988    & SD           &  Staveley-Smith    \& Davies 1988                  \\
GW           &  Gu\'{e}lin  \& Weliachew 1970                    & SEC          &  Stark, Elmegreen  \& Chance 1987                  \\
H            &  Huchtmeier 1975                                  & SJL          &  Sandqvist, J\"{o}rs\"{a}ter  \& Lindblad 1995     \\
HAS          &  Hoekstra, van Albada  \& Sancisi 2001            & SKB          &  Stark, Knapp, Bally, et al 1986                   \\
HB           &  Helfer  \& Blitz 1995                            & SNK          &  Sorai, Nakai, Kuno, et al 2000                    \\
HBS          &  Heckman, Balick  \& Sullivan 1978                & TM           &  Thornley \& Mundy 1997                            \\
HHG          &  Hewitt, Haynes   \& Giovanelli 1983              & vKS          &  van der Kruit  \& Shostak 1984                    \\
HHM          &  Haynes, Hogg, Maddalena, et al. 1998             & VPD          &  de Vaucouleurs, Pence  \& Davoust 1983           \\
HR           &  Huchtmeier  \& Richter 1986a                     & Wa           &  Warmels 1988a                                     \\
HRb          &  Huchtmeier  \& Richter 1986b                     & Wb           &  Warmels 1988b                                     \\
HRG          &  Hunter, Rubin  \& Gallagher 1986                 & WAD          &  Wevers, Appleton, Davies, et al. 1984             \\
HS           &  Huchtmeier  \& Seiradakis 1985                   & WG           &  Weliachew   \& Gottesman 1973                     \\
HTR          &  Helfer, Thornley, Regan, et al. 2003             & WKA          &  Wevers, van der Kruit  \& Allen 1986              \\
JM           &  J\"{o}rs\"{a}ter  \& van Moorsel 1995            & YXT          &  Young, Xie  \& Tacconi 1995                       \\
KDI          &  Koper, Dame, Israel, et al. 1991                 & ZB           &  van Zee     \& Bryant 1999                       \\
      \hline
\end{tabular}
\end{center}
}
\end{table*}

As was noted above, in investigations of the relationships between the oxygen abundances
and the macroscopic properties of spiral galaxies, the concept of the
characteristic oxygen abundance has been introduced: it is defined as the
oxygen abundance in the disk at a predetermined galactocentric distance.
Following Zaritsky, Kennicutt \& Huchra (1994),  the value of the oxygen abundance at
 $r=0.4R_{\rm 25}$ will be used here as the {\it characteristic} oxygen abundance in a galaxy.
To derive a reliable galaxy luminosity the accurate value of the distance is
necessary. The compilation of the up-to-date distance measurements for our sample
of galaxies is discussed in the Appendix.

\begin{figure}
\centering
\resizebox{1.00\hsize}{!}{\includegraphics[angle=0]{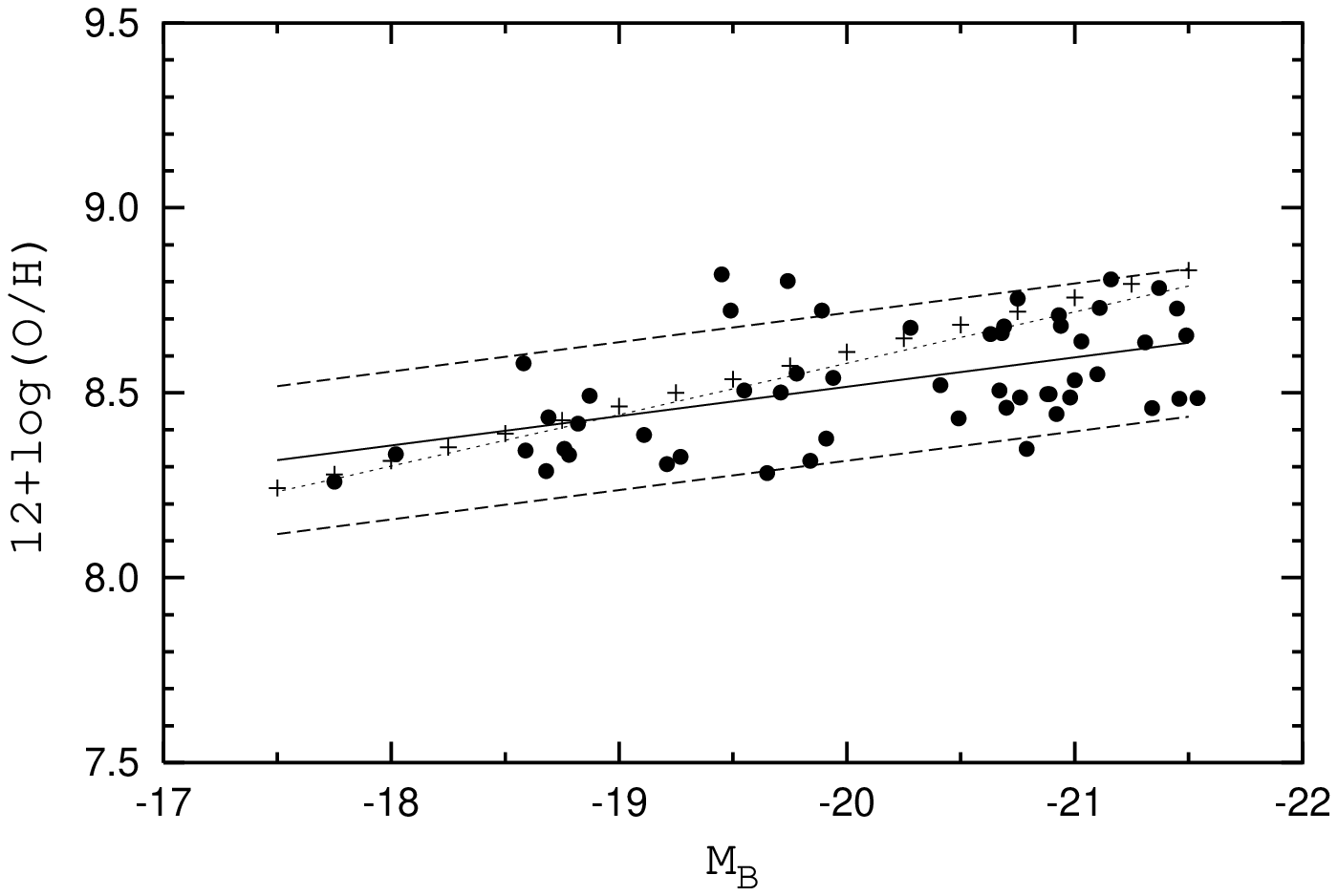}}
\caption{
The characteristic oxygen abundance as a function of absolute blue magnitude M$_B$ for our sample of
spiral galaxies. The solid line is the O/H -- M$_B$ relationship
(best fit derived through the least squares method). The dashed lines
correspond to the lines shifted by --0.2 dex and +0.2 dex along the vertical axes relatively to the
derived O/H -- M$_B$ relationship.
The dotted line is the O/H -- M$_B$ relationship for our sample of irregular
galaxies extrapolated to the luminosity range of spiral galaxies.
The line presented by the plus signs is the extrapolated O/H -- M$_B$ relationship
for irregular galaxies derived by Richer \& McCall (1995).
}
\label{figure:s-l-oh}
\end{figure}

\begin{figure}
\centering
\resizebox{1.00\hsize}{!}{\includegraphics[angle=0]{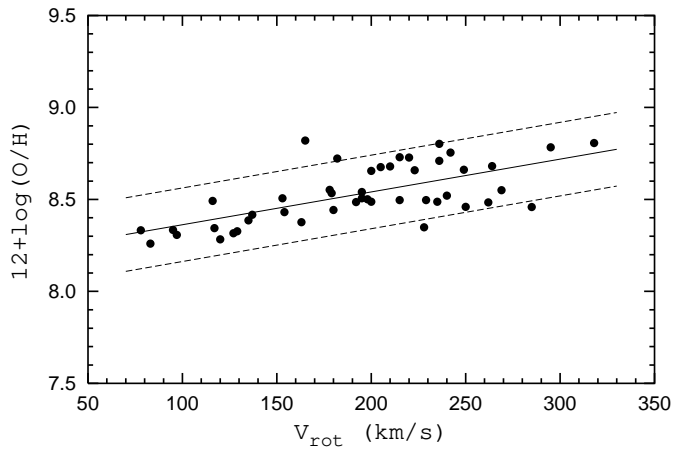}}
\caption{
The characteristic oxygen abundance as a function of rotation velocity for our sample of spiral
galaxies. The solid line is the O/H -- $V_{\rm rot}$ relationship
(best fit derived through the least squares method).
The dashed lines correspond to the lines shifted by --0.2 dex and +0.2 dex relatively to the derived O/H -- $V_{\rm rot}$
relationship.
}
\label{figure:s-v-oh}
\end{figure}

\begin{figure}
\centering
\resizebox{1.00\hsize}{!}{\includegraphics[angle=0]{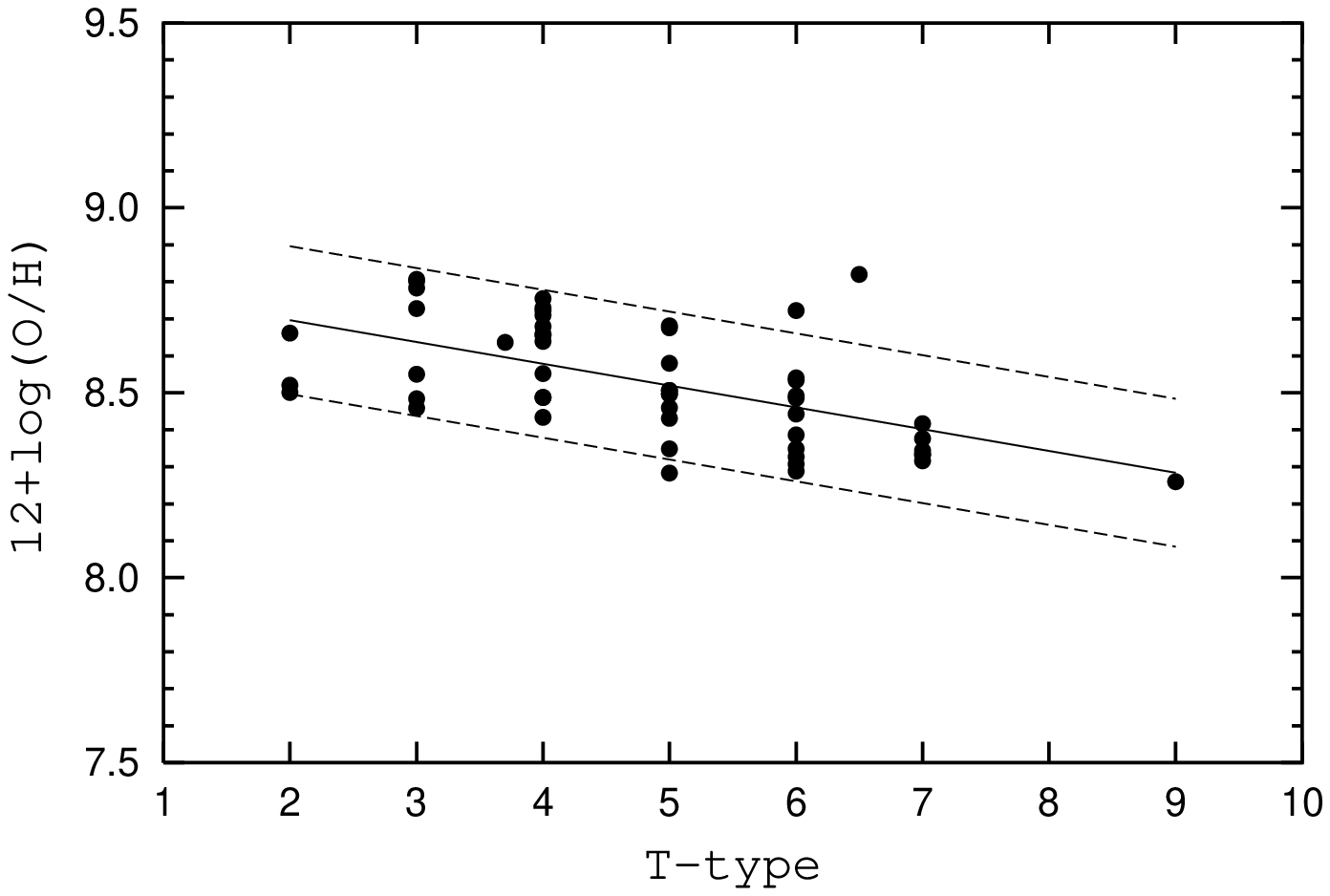}}
\caption{
The characteristic oxygen abundance as a function of Hubble type, expressed in the terms of T-type,
for our sample of spiral galaxies. The solid line is the O/H -- T relationship
derived through the least squares method. The dashed lines
correspond to the lines shifted by --0.2 dex and +0.2 dex relatively to the derived O/H -- T relationship.
}
\label{figure:s-t-oh}
\end{figure}

The characteristic oxygen abundance in spiral galaxies as a function of absolute blue magnitude $M_{\rm B}$ is shown
in Fig. \ref{figure:s-l-oh}. The solid line is the characteristic oxygen abundance -- luminosity relationship
(linear best fit derived through the least squares method):
\begin{equation}
12+\log (O/H) = 6.93\, (\pm 0.37) - 0.079\,(\pm 0.018)\, M_{\rm B}   .
\label{equation:s-ohr-l}
\end{equation}
The dashed lines correspond to this relation shifted by --0.2 dex and +0.2 dex.
Fig. \ref{figure:s-l-oh} clearly demonstrates that there is a correlation between the characteristic oxygen
abundance of a galaxy and its blue luminosity.
The bulk of the galaxies lies within the band defined by the dashed lines ($\pm$ 0.2 dex around
the best fit). The dotted line in Fig. \ref{figure:s-l-oh} is the luminosity --
metallicity relathionship for irregular galaxies (see below) extrapolated to
the luminosity range of spiral galaxies.
The line presented in Fig. \ref{figure:s-l-oh} by the plus signs is the extrapolated 
O/H -- M$_B$ relationship for irregular galaxies derived by Richer \& McCall (1995).
Inspection of Fig. \ref{figure:s-l-oh} shows that the slope of the derived
luminosity -- metallicity relationship for spiral galaxies is lower than the
one for irregular galaxies. 

We have carried out a search in the literature for measurements of rotation velocities for our sample of spiral 
galaxies. Unfortunately, a measurement of the rotation velocity is not available for 7 galaxies of 
our sample. On the contrary, there are two or more measurements of rotation velocities
for a number of galaxies. The more recent value (or a mean value if the
measurements are close to each other) is taken for these galaxies.
The adopted values of the rotation velocity for the galaxies and corresponding reference(s)
are listed in Table \ref{table:Vrot} and Table \ref{table:referV}.

The characteristic oxygen abundance in spiral galaxies as a function of rotation 
velocity $V_{\rm rot}$ is shown in Fig. \ref{figure:s-v-oh}. The solid line is 
the oxygen abundance -- rotation velocity relationship (linear best fit)
\begin{equation}
12+\log (O/H) = 8.18\,(\pm 0.06) + 0.00179\,(\pm0.00031)\, V_{\rm rot}  .
\label{equation:s-v-ohr}
\end{equation}
The dotted lines correspond to this relation shifted by --0.2 dex and +0.2 dex.
Fig.~\ref{figure:s-v-oh} shows that the characteristic oxygen abundance of a galaxy correlates with its
rotation velocity. The deviation of an individual galaxy from the general trend is usually less than 0.2 dex.
The mean value of the scatter in the characteristic oxygen abundances at a given
rotation velocity is 0.12 dex for 46 points.

The characteristic oxygen abundance in spiral galaxies as a function of morphological type, expressed in terms of $T$-type,
is shown in Fig. \ref{figure:s-t-oh}. The solid line is the characteristic oxygen abundance -- $T$ relationship (best fit)
\begin{equation}
12+\log (O/H) = 8.81\,(\pm 0.06) - 0.059\,(\pm 0.011)\, T .
\label{equation:s-t-ohr}
\end{equation}
The dotted lines correspond to this relation shifted by --0.2 dex and +0.2 dex.
Fig. \ref{figure:s-t-oh} shows that the characteristic oxygen abundance of a galaxy correlates with its
morphological type. The deviation of an individual galaxy from the general trend
usually does not exceed the value $\pm$0.2 dex, as in the cases of the O/H -- $M_{\rm B}$ and
the O/H -- $V_{\rm rot}$ diagrams. The mean value of the scatter in the characteristic oxygen abundances at a given
morphological type is 0.13 dex for 53 points.

One can see that the characteristic oxygen abundance correlates well with both the absolute blue luminosity, the 
rotational velocity, and the morphological type; the correlation 
with the rotation velocity perhaps being slightly tighter.

\section{Comparison between O/H-$M_{\rm B}$ relationship for spiral and irregular galaxies}

Zaritsky, Kennicutt \& Huchra (1994) found that the characteristic gas-phase oxygen abundance --
luminosity relation of spiral galaxies extends almost directly to the luminosity -- metallicity
relationship of irregular galaxies. Melbourne \& Salzer (2002) and Lamareille et al. (2004) found 
that the slope of the oxygen abundance -- luminosity relationship for high-luminosity galaxies is steeper 
than when dwarf galaxies are considered alone and may be evidence that the relationship is not linear over
the full luminosity range. Garnett (2002) concluded that the metallicity -- luminosity correlation
shows a roughly uniform growth in the average present-day O/H abundance over 11 mag in absolute blue magnitude
$M_{\rm B}$. Let us compare the derived metallicity -- luminosity relationship for spiral galaxies
with that for irregular galaxies.

The irregular galaxies were selected from the samples of Richer \& McCall (1995) and Pilyugin (2001c).
The luminosities and oxygen abundances for irregular galaxies are taken from these studies and are
listed in Cols. 3 and 4 of Table \ref{table:karlik}.
The values of the gas mass fraction in the irregular galaxies were estimated taking into account the
atomic hydrogen mass only since the molecular hydrogen in dwarf irregular galaxies is only a small
fraction of the total gas mass.
The H\,{\sc i} fluxes were taken mainly from Karachentsev, Makarov \& Huchtmeier (1999).
The derived values of gas mass fraction $\mu$ in irregular galaxies are
reported in Col. 5 of Table \ref{table:karlik}. The rotation velocities for irregular galaxies
taken from the literature, the values of the rotation velocity and corresponding references to the
sources are listed in Cols. 6 and 7 of Table \ref{table:karlik}.

\begin{table*}
\caption[]{\label{table:karlik}
Data for irregular galaxies.
{\it References} --
(a) Lee, McCall, Kingsburgh, et al. 2003;
(b) Pilyugin 2001c;
(c) Karachentsev, Makarov  \& Huchtmeier 1999;
(d) Skillman, Terlevich, Teuben  \& van Woerden 1988;
(e) Hoffman, Salpeter, Farhat, et al. 1996;
(f) C\^{o}t\'{e}, Carignan  \& Freeman 2000;
(g) Kinman  \& Davidson 1981;
(h) Richer  \& McCall 1995;
(i) Wilcots  \& Miller 1998;
(j) Alves  \& Nelson 2000.
 }
\begin{center}
\begin{tabular}{lcccccc} \\ \hline \hline
          &         &            &           &          &           &           \\
galaxy    &  T-type &   $M_{\rm B}$    & 12+logO/H & $\mu$    & $V_{\rm rot}$& references       \\
          &         &            &           &          & km/s      &($M_{\rm B}$, O/H, $\mu$, $V_{\rm rot}$)  \\
          &         &            &           &          &           &           \\   \hline
Sextans B &   10    &  -14.02    &   7.86    &   0.43   &   24      & a, b, c, c   \\
Sextans A &   10    &  -14.04    &   7.71    &   0.60   &   37      & a, b, c, d   \\
GR 8      &   10    &  -12.19    &   7.60    &   0.54   &   13      & a, b, c, e   \\
WLM       &   10    &  -13.92    &   7.78    &   0.44   &   31      & a, b, c, c   \\
UGC 4483  &   10    &  -12.80    &   7.47    &   0.68   &   22      & b, b, c, c   \\
UGC 5423  &   10    &  -12.90    &   7.81    &   0.36   &   27      & b, b, c, c   \\
UGC 6456  &   10    &  -13.24    &   7.71    &   0.68   &   24      & b, b, c, c   \\
Leo A     &   10    &  -11.53    &   7.27    &   0.58   &   16      & a, b, c, c   \\
UGCA 292  &   10    &  -11.43    &   7.22    &   0.83   &   12      & b, b, c, c   \\
DDO 167   &   10    &  -13.30    &   7.81    &   0.50   &   17      & b, b, c, c   \\
SagDIG    &   10    &  -12.10    &   7.48    &   0.51   &   19      & b, b, c, f   \\
A1116+51  &   10    &  -14.99    &   7.76    &   0.69   &   --      & b, b, g, --  \\
A1228+12  &   10    &  -14.57    &   7.79    &   0.46   &   --      & b, b, g, -- \\
A2228-00  &   10    &  -14.57    &   7.79    &   0.65   &   --      & b, b, g, --  \\
ESO 245-G05&  10    &  -15.50    &   7.94    &   0.66   &   48      & b, b, c, i   \\
DDO 53    &   10    &  -13.35    &   7.75    &   0.75   &   --      & b, b, c, --  \\
DDO 190   &   10    &  -15.10    &   7.74    &   0.51   &   --      & b, b, c, --  \\
Holmberg II&  10    &  -15.98    &   7.92    &   0.57   &   34      & a, h, c, c   \\
IC 10     &   10    &  -15.82    &   8.22    &   0.22   &   30      & a, h, c, i   \\
IC 1613   &   10    &  -14.53    &   7.71    &   0.46   &   21      & a, h, c, e   \\
IC 2574   &    9    &  -16.85    &   8.08    &   0.50   &   50      & h, h, c, c   \\
IC 4662   &   10    &  -15.64    &   8.09    &   0.30   &   48      & h, h, c, c   \\
LMC       &    9    &  -17.73    &   8.35    &   0.20   &   72      & h, h, c, j   \\
NGC1560   &    7    &  -16.17    &   8.02    &   0.27   &   59      & h, h, c, c   \\
NGC 2366  &   10    &  -16.28    &   7.92    &   0.48   &   44      & a, h, c, c   \\
NGC 3109  &    9    &  -15.30    &   8.06    &   0.47   &   49      & a, h, c, c   \\
NGC 4214  &   10    &  -17.82    &   8.23    &   0.37   &   --      & h, h, c, --  \\
NGC 5408  &   10    &  -15.60    &   8.01    &   0.35   &   38      & h, h, c, c   \\
NGC 55    &    9    &  -18.07    &   8.34    &   0.20   &   71      & h, h, c, c   \\
SMC       &    9    &  -16.35    &   8.03    &   0.49   &   21      & h, h, c, c   \\
          &         &            &           &          &           &           \\   \hline
\end{tabular}
\end{center}
\end{table*}

\begin{figure}
\centering
\resizebox{1.00\hsize}{!}{\includegraphics[angle=0]{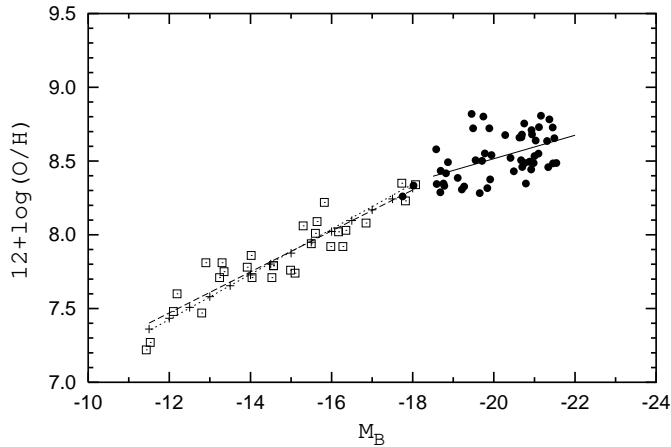}}
\caption{
The characteristic oxygen abundance as a function of absolute blue magnitude
$M_{\rm B}$ for our sample of spiral galaxies (the filled circles). The solid 
line is the best (linear least-squares)  
fit to these data. The open squares are oxygen abundances in
irregular galaxies, the dashed line is the metallicity -- luminosity relationship
for our sample of irregular galaxies. The dotted line is the O/H  -- $M_{\rm B}$ 
relationship for irregular galaxies derived by Lee et al. (2003). The O/H  -- 
$M_{\rm B}$ relationship for irregular galaxies derived by Richer \& McCall 
(1995) is presented by the plus signs.
}
\label{figure:l-oh}
\end{figure}

The oxygen abundances in spiral and irregular galaxies are plotted versus absolute blue magnitude $M_{\rm B}$
in Fig. \ref{figure:l-oh}.
The filled circles in Fig. \ref{figure:l-oh} are spiral galaxies, and the solid line is the O/H -- $M_{\rm B}$
relationship for spiral galaxies, Eq. (\ref{equation:s-ohr-l}).
The open squares in Fig. \ref{figure:l-oh} are irregular galaxies from Table \ref{table:karlik}.
The dashed line is the O/H  -- $M_{\rm B}$ relationship (best fit) for irregular galaxies
\begin{equation}
12+\log (O/H) = 5.80\,(\pm 0.17) - 0.139\,(\pm 0.011)\, M_{\rm B}
\label{equation:i-l-oh}
\end{equation}
The dotted line is the O/H  -- $M_{\rm B}$ relationship for irregular galaxies
\begin{equation}
12+\log (O/H) = 5.59\,(\pm 0.54) - 0.153\,(\pm 0.025)\, M_{\rm B}
\label{equation:lee}
\end{equation}
derived by Lee et al. (2003).
The O/H  -- $M_{\rm B}$ relationship for irregular galaxies
\begin{equation}
12+\log (O/H) = 5.67\,(\pm 0.48) - 0.147\,(\pm 0.029)\, M_{\rm B}
\label{equation:richer}
\end{equation}
derived by Richer \& McCall (1995) is presented by the plus signs.
Inspection of Fig. \ref{figure:l-oh} (as well as the comparison between
Eq. (\ref{equation:i-l-oh}), Eq. (\ref{equation:lee}), and Eq. (\ref{equation:richer}))
shows that the O/H  -- $M_{\rm B}$ relationship for our sample of irregular
galaxies agrees, within the uncertainties, with that from Lee at al. (2003) 
and from Richer \& McCall (1995).
Fig. \ref{figure:l-oh} shows a familiar oxygen abundance -- luminosity correlation for late-type galaxies,
an increase in metallicity with luminosity over the full range of absolute blue magnitude, $M_{\rm B}$
from $\sim -22$ to $\sim -11$. Comparison between Eq. (\ref{equation:s-ohr-l}) and
Eq. (\ref{equation:i-l-oh}) shows that the O/H -- $M_{\rm B}$ relationships for
spiral and irregular galaxies have slightly different slopes.

\begin{figure}
\centering
\resizebox{1.00\hsize}{!}{\includegraphics[angle=0]{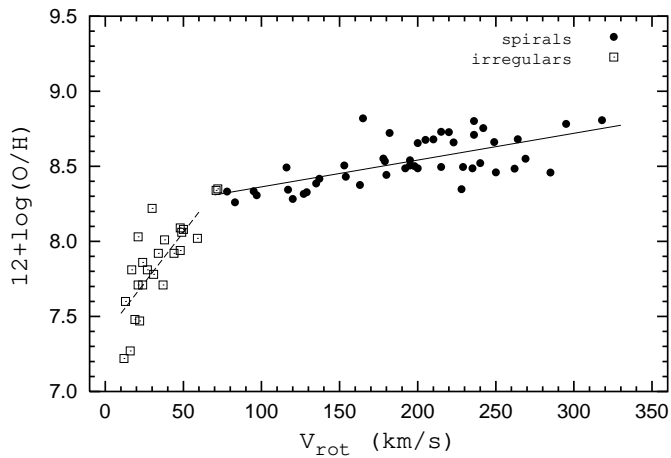}}
\caption{
The characteristic oxygen abundance as a function of rotation velocity 
V$_{rot}$ for our sample of spiral galaxies (the filled circles). The solid 
line is the best (linear least-squares) fit to these data. The open squares are 
oxygen abundances in irregular galaxies, the dashed line is the linear 
least-squares fit to these data. 
}
\label{figure:v-oh}
\end{figure}

\begin{figure}
\centering
\resizebox{1.00\hsize}{!}{\includegraphics[angle=0]{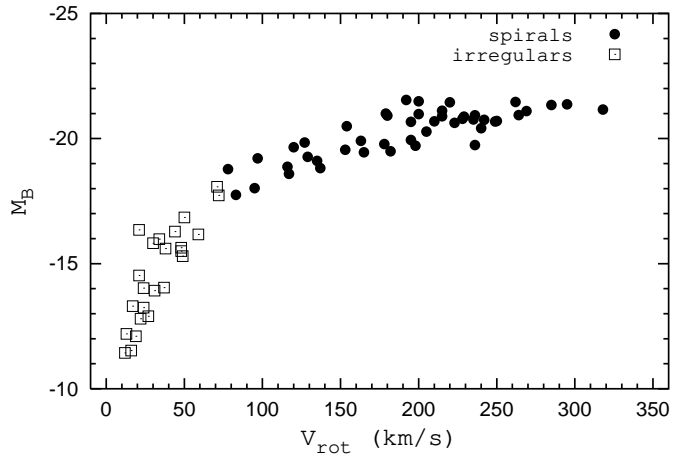}}
\caption{
The absolute blue magnitude $M_{\rm B}$ as a function of the rotation velocity.
The filled circles are spiral galaxies, the open squares are irregular galaxies.
}
\label{figure:v-l}
\end{figure}

Fig. \ref{figure:v-oh} shows the oxygen abundance in spiral (filled circles)
and irregular (open squares) galaxies as a function of rotation velocity
$V_{\rm rot}$. The most prominent feature is the bend in the O/H -- $V_{\rm rot}$
relation. Garnett (2002) also found that the correlation between O/H and
$V_{\rm rot}$ does not increase steadily but rather turns over for
rotation speeds greater than 125 kms$^{-1}$. It is worth noting that the presence of the bend
in the O/H -- $V_{\rm rot}$ relation does not necessary imply the existence of a bend in
the O/H -- $M_{\rm B}$ trend.
Fig. \ref{figure:v-l} shows the absolute blue magnitude $M_{\rm B}$
as a function of the rotation velocity $V_{\rm rot}$ for spiral (filled circles)
and irregular (open squares) galaxies. Inspection of Fig. \ref{figure:v-l}
shows that the correlation between $M_{\rm B}$ and $V_{\rm rot}$ is not linear 
but rather shows a bend. This bend in the O/H -- $V_{\rm rot}$ trend would thus 
occur even if the increase in oxygen abundance with luminosity can be described 
by a single linear function over the full magnitude range.

\section{The effective oxygen yields}

The observed oxygen abundance in a galaxy is defined not only by the astration level
but also by the mass exchange between a galaxy and its environment.
The latter can alter the relation between oxygen abundance and gas
mass fraction; it mimics the variation in the oxygen yield.
To investigate the possibility of a varying yield, it is
useful to define the ``effective'' oxygen yield, $y_{\rm eff}$, as the yield 
that would be deduced if a system was assumed to behave as in the simplest model 
of chemical evolution (Edmunds 1990; Vila-Costa \& Edmunds 1992)
\begin{equation}
y_{\rm eff} = \frac{z_O}{\ln (\frac{1}{\mu})}   .
\label{equation:yeff}
\end{equation}
Since we used the oxygen abundances given by number relative to hydrogen, while the value of z$_O$ in
Eq. (\ref{equation:yeff}) is given in units of mass fraction, the conversion for oxygen
from Garnett et al. (1997) and Garnett (2002)
\begin{equation}
z_{O} = 12 \frac{O}{H}
\end{equation}
was adopted.

\subsection{The gas mass fractions}

\begin{figure}
\centering
\resizebox{1.00\hsize}{!}{\includegraphics[angle=0]{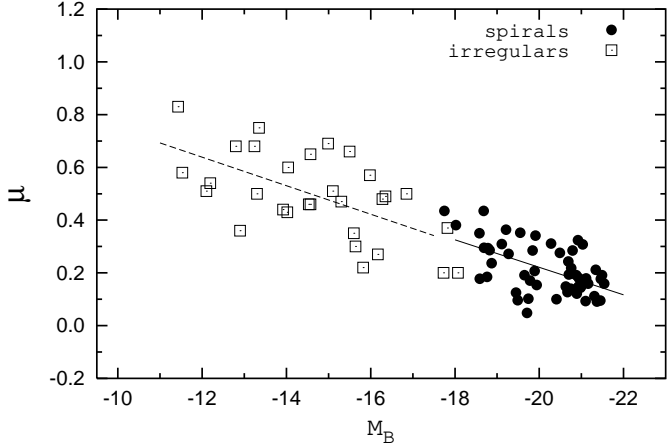}}
\caption{
Gas mass fraction as a function of the absolute blue magnitude M$_B$.
The filled circles are spiral galaxies, the open squares are irregular galaxies.
The solid line is the $\mu$ -- M$_B$ relationship for spiral galaxies,
the dashed line is the $\mu$ -- M$_B$ relationship for irregular galaxies.
}
\label{figure:l-mu}
\end{figure}

The gas fraction $\mu$ in a galaxy is estimated using the following standard relation:
\begin{equation}
\mu  = \frac{M_{H_I} + M_{H_2}}{M_{H_I} + M_{H_2} + k\,L_B}   ,
\label{equation:mu}
\end{equation}
where $M_{\rm H_I}$ is the mass of atomic hydrogen in the galaxy, $M_{\rm H_2}$ is the mass of molecular hydrogen
in the galaxy (both $M_{\rm H_I}$ and $M_{\rm H_2}$ are corrected for the helium and heavy element contribution),
L$_B$ is the blue luminosity of the galaxy, and $k$ is the mass-to-luminosity ratio. The mass of atomic hydrogen
is derived in a standard way using the measured H\,{\sc i} flux taken from literature and the adopted distance
(Table \ref{table:distance}). If the mass of atomic hydrogen instead of the H\,{\sc i} flux measurement is reported
in literature, this value is rescaled to the adopted distance. The atomic hydrogen masses expressed in
terms of $M_{\rm H_I}$/L$_B$ and corresponding references to the H\,{\sc i} flux measurements are listed
in Table \ref{table:Vrot} and Table \ref{table:referV}.

The mass of molecular hydrogen can only be estimated with indirect methods. The commonly accepted method
is the use of the CO line flux and $X$ conversion factor between the flux in the CO line and the
amount of molecular hydrogen. The conversion factor $X$ = N(H$_2$)/I(CO) strongly depends on the physical
properties of the interstellar medium which are known to vary from galaxy to galaxy. The best-estimated
values of the conversion factor  for a sample of well-studied nearby galaxies span the range
$0.6 \leq X \leq 10 \times 10^{20}$ mol cm$^{-2}$ (K km s$^{-1}$)$^{-1}$ (Boselli, Lequeux \&
Gavazzi 2002). The high values of the conversion factor correspond to low-luminosity irregular
galaxies, and the low values of the conversion factor are found in spiral galaxies.
Then the value of $X =  1 \times 10^{20}$ mol cm$^{-2}$ (K km s$^{-1}$)$^{-1}$ (including helium
and heavy elements contribution) is adopted here for spiral galaxies.  The molecular gas content
is derived with this conversion factor  using the measured CO flux from literature and distance
(Table \ref{table:distance}). 
If the mass of molecular hydrogen instead of the CO flux measurement is reported
in the literature, this value is rescaled to the adopted distance and adopted conversion factor. The molecular
hydrogen masses expressed in terms of $M_{\rm H_2}$/L$_B$ and corresponding references to the CO flux
measurements are given in Table \ref{table:Vrot} and Table \ref{table:referV}.

The mass of the stellar component of the galaxy is estimated by converting the measured luminosity to mass
via the mass-to-luminosity ratio. It is difficult to get a reliable estimation of the mass-to-luminosity
ratio for individual galaxies. The $\mu$ and $\mu$ -- dependent values will be used here only for examination
of the general trends of these values with luminosity and rotation velocity but not for examination of
individual galaxies. A similar investigation was carried out by Garnett (2002). He considered the impact of the
choice of the mass-to-luminosity ratio on the trends by comparing the trends derived with the color-based
mass-to-luminosity ratio and constant mass-to-luminosity ratio. Garnett (2002) has found that the trends are
fairly robust against the choice of mass-to-luminosity ratio. Based on this conclusion, a constant
value of the mass-to-luminosity ratio $k=1.5$ is adopted here for all spiral galaxies, and $k=1$ is adopted for
all irregular galaxies.

The derived values of the gas mass fraction in galaxies $\mu$ are listed in Table \ref{table:Vrot} (Col. 3).
Unfortunately, for several spiral galaxies on our list no measurement of the CO flux is 
available. For these galaxies, the gas mass fraction is based on the atomic hydrogen mass only and
is a lower limit. Taking into account that the average value is $M_{\rm H_2}$/$M_{\rm H_I}$ = 0.14
(Boselli, Lequeux \& Gavazzi 2002), one can hope that the use of a lower limit instead of the estimated
value of gas fraction for several galaxies is quite acceptable.

The gas mass fraction in spiral galaxies as a function of absolute blue
magnitude $M_{\rm B}$ is shown by the filled circles in Fig. \ref{figure:l-mu}. The solid line
is the gas mass fraction -- luminosity relationship (best fit determined via
the least squares method) for spiral galaxies
\begin{equation}
\mu = 1.264\,(\pm 0.220)  + 0.0522\,(\pm 0.0109)\, M_{\rm B}   .
\label{equation:s-lmu}
\end{equation}
The open squares in Fig. \ref{figure:l-mu} are irregular galaxies.
The dashed line is the gas fraction  -- luminosity relationship (best fit) for irregular galaxies
\begin{equation}
\mu = 1.289\,(\pm 0.198)  + 0.0542\,(\pm 0.0134)\, M_{\rm B} .
\label{equation:i-lmu}
\end{equation}
Comparison between Eq. (\ref{equation:s-lmu}) and Eq. (\ref{equation:i-lmu}) shows that
the $\mu$  -- $M_{\rm B}$ relationship for spiral galaxies coincides remarkably well with
the $\mu$  -- $M_{\rm B}$ relationship for irregular galaxies. It can also be seen in Fig. \ref{figure:l-mu}
that the gas mass fraction  -- luminosity relation for spiral galaxies is a direct 
continuation of the gas mass fraction  -- luminosity relationship for irregular galaxies.

\subsection{The effective oxygen yields}

\begin{figure}
\centering
\resizebox{1.00\hsize}{!}{\includegraphics[angle=0]{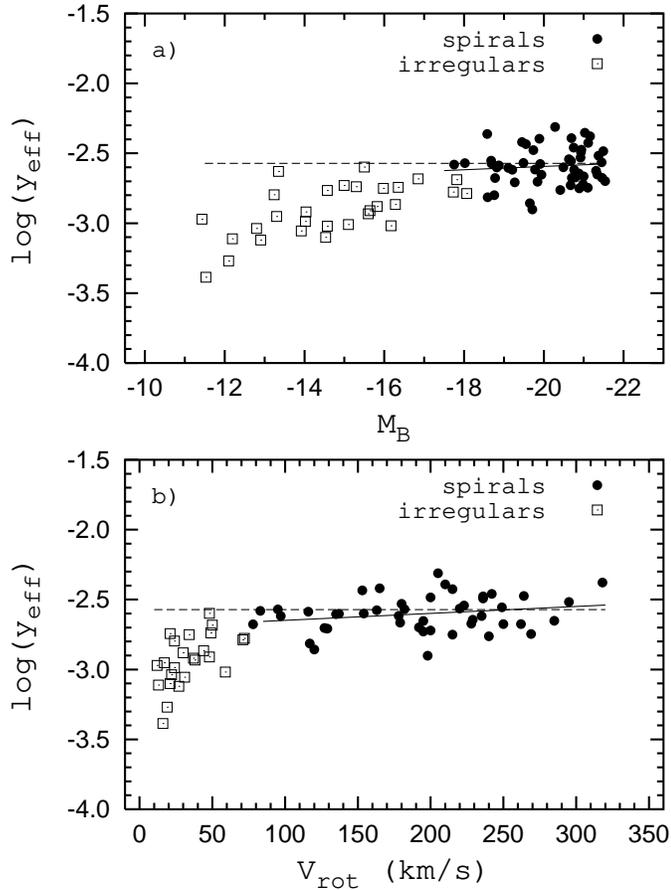}}
\caption{
{\bf (a)} Effective oxygen yield as a function of the absolute blue magnitude 
$M_{\rm B}$. The filled circles are spiral galaxies, the open circles are 
irregular galaxies. The solid line is the $y_{\rm eff}$ -- $M_{\rm B}$ relation 
for spiral galaxies (best fit). The dashed line corresponds to the mean 
$y_{\rm eff}$ = 0.00268 for spiral galaxies.
{\bf (b)} Effective oxygen yield as a function of the rotation velocity.
Symbols are the same as in panel {\bf (a)}.
}
\label{figure:lv-y}
\end{figure}

The derived effective oxygen yields are presented as a function of absolute blue magnitude $M_{\rm B}$ in
Fig. \ref{figure:lv-y}a and as a function of rotation velocity $V_{\rm rot}$ in Fig. \ref{figure:lv-y}b.

Inspection of Figs. \ref{figure:lv-y}a,b shows that there is only a hint of trends of the effective
oxygen yield with luminosity and rotation velocity.
The formal best fit to the log($y_{\rm eff}$) -- $M_{\rm B}$ relation (solid line in Fig. \ref{figure:lv-y}a) is
\begin{equation}
\log y_{\rm eff} = -2.834\,(\pm 0.371) - 0.0120\,(\pm 0.0183)\, M_{\rm B}    .
\label{equation:s-ly}
\end{equation}
The mean value of the deviation of individual galaxies from the general trend
is 0.13 dex. The formal best fit (least squares method) to data for spiral
galaxies, the log($y_{\rm eff}$) -- $V_{\rm rot}$ relation (solid line in
Fig. \ref{figure:lv-y}b) is
\begin{equation}
\log y_{\rm eff} = -2.698\,(\pm 0.065) + 0.000496\,(\pm 0.000321)\, V_{\rm rot}  .
\label{equation:s-vy}
\end{equation}
The mean value of the deviation of individual galaxies from the general trend is
0.12 dex. The mean value of the effective oxygen yield in spiral galaxies is
$y_{\rm eff}$ = 0.00268. The dashed line in Figs. \ref{figure:lv-y}a,b
corresponds to the mean value of $y_{\rm eff}$. Taking into account the large
scatter of points in the log($y_{\rm eff}$) -- $M_{\rm B}$ and the
log($y_{\rm eff}$) -- $V_{\rm rot}$ diagrams, it is difficult to establish
firmly how real are the trends of the effective oxygen yield in spiral galaxies
with luminosity and rotation velocity. More accurate data are needed to confirm
or reject these trends. Here we can only conclude that the effective oxygen
yield in spiral galaxies remains approximatively constant.

\begin{figure}
\centering
\resizebox{1.00\hsize}{!}{\includegraphics[angle=0]{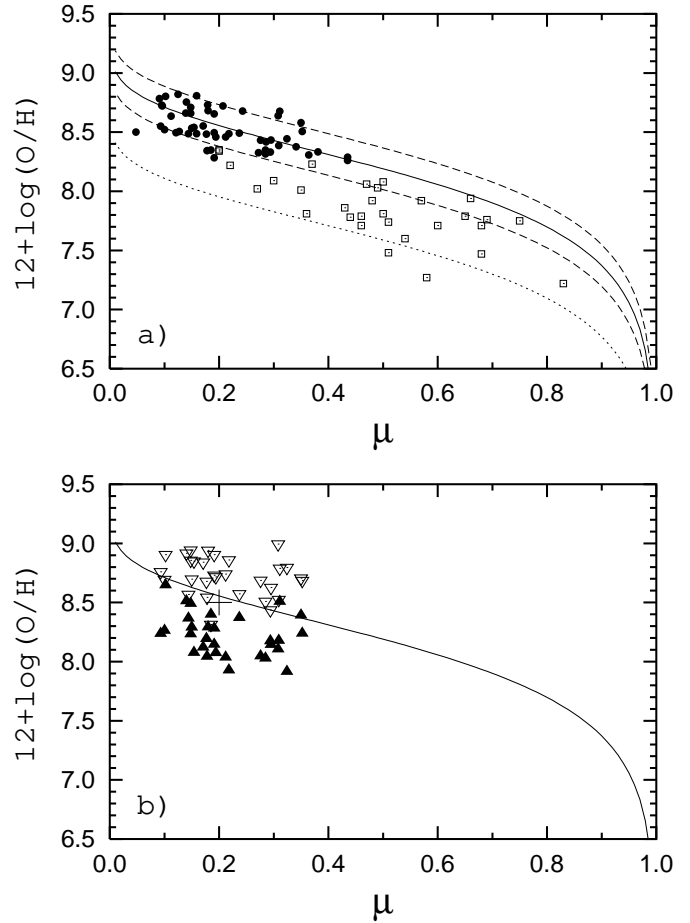}}
\caption{
{\bf (a)}
The positions of spiral (filled circles) and irregular (open squares) galaxies in
the $\mu$ -- O/H diagram together with the predictions of the closed-box models.
The solid curve is the prediction of the closed-box model with oxygen yield $y_0$ = 0.00268
(the mean value of the effective oxygen yields derived for spiral galaxies).
The dashed curves correspond to the predictions of the closed-box models with
oxygen yields $y=1.5\times y_0$ and $y=y_0$/1.5.
The dotted curve is the prediction of the closed-box models with
oxygen yields $y=y_0$/4.
{\bf (b)}
The oxygen abundances extrapolated to zero radius (open triangles) and
extrapolated
to the isophotal radius $R_{\rm 25}$ (filled triangles) in spiral galaxies
with well defined radial oxygen abundance gradients together with the
$\mu$ - O/H track
predicted by the closed-box model with oxygen yield $y_0$ = 0.00268.
The large plus sign is  the position of the solar vicinity region.
}
\label{figure:mu-oh}
\end{figure}

It is widely accepted that the mass exchange (gas infall and/or galactic wind)
between a galaxy and its environment plays a major role in the evolution of
galaxies. Gas infall and galactic winds produce a shift of the position of a
galaxy in the $\mu$ -- O/H diagram towards lower oxygen abundances
compared to the predictions of the closed-box models (see e.g. Mouhcine \&
Contini 2002). The location of spiral and irregular galaxies in the
$\mu$ -- O/H diagram together with the predictions of the closed-box model
\begin{equation}
z_O = y_O \ln (\frac{1}{\mu})
\end{equation}
are presented in Fig. \ref{figure:mu-oh}a.

The solid line in Fig. \ref{figure:mu-oh}a is the prediction of the closed-box model for the evolution
of galaxies with oxygen yield $y_0$ = $y_{\rm eff}$ = 0.00268.
As can be seen in Fig. \ref{figure:mu-oh}a, the positions of spiral galaxies in the $\mu$ -- O/H diagram
are close to the track predicted by the closed-box model. The mean value of the O/H deviation of the individual
galaxies from the closed-box model is 0.13 dex.
The dashed lines in Fig. \ref{figure:mu-oh}a correspond to the predictions of the closed-box models with
oxygen yields $y=1.5\times y_0$ and $y=y_0$/1.5. The majority of spiral galaxies is located within the
band defined by dashed lines.

It is well known that the closed-box model predicts many more low-metallicity stars than are observed
in the solar neighbourhood, the so called ``G--dwarf'' paradox. 
Various versions of the infall model, in which an
infall of gas onto the disk takes place for a long time, have been suggested (Tosi 1988a,b; Pagel \&
Tautvai\^{s}ien\.{e} 1995; Pilyugin \& Edmunds 1996a,b; Chiappini, Matteucci \& Romano 2001;
among many others) to explain the observed metallicity function. An infall model has also been
applied to other spiral galaxies (D\'{\i}az \& Tosi 1986;
Moll\'{a}, Ferrini \& D\'{\i}az  1996,1997; Pilyugin et al. 2002; Mouhcine \& Contini 2002;
Chiappini, Romano \& Matteucci 2003).
It is thus generally accepted that gas infall can play an important role in the chemical evolution of
disks of spiral galaxies. Therefore, the fact that the positions of spiral galaxies are close to the
position of the track of the closed-box model may appear surprising. It has been shown (Pilyugin \&
Ferrini 1998) that the present-day position of a system in the $\mu$ -- O/H diagram is governed by
its evolution in the recent past and is, in fact, independent of its evolution on long timescales.
Therefore, the fact that the present-day position of spiral galaxies is close to the position of
the closed-box model is not in conflict with the model, in which an infall of gas onto disk takes
place for a long time (which is necessary to satisfy the observed abundance distribution function and
the age -- metallicity relation in the solar neighbourhood) since these observational data reflect the
evolution of a system in the distant past. Of course, the simple model 
of chemical evolution of galaxies can be used only as the first-order approximation. To establish a more accurate relation between the present-day values of oxygen abundance 
and gas mass fraction in a galaxy, an appropriate model of chemical evolution should be 
developed (e.g. Pagel \& Tautvai\^{s}ien\.{e} 1995).

Inspection of Figs. \ref{figure:lv-y}a,b shows that the values of the effective oxygen yield in irregular
galaxies are lower than that in spiral galaxies. The effective oxygen yield in irregular galaxies
shows a clear trend with luminosity and rotation velocity.
The positions of irregular galaxies in the $\mu$ -- O/H diagram are systematically shifted towards
lower oxygen abundances compared to the predictions of the closed-box model with oxygen yield $y_0$
typical for spiral galaxies. This suggests that the mass exchange between a galaxy and its environment
(galactic winds) plays an important role in the chemical evolution of irregular galaxies.
The dotted line in Fig. \ref{figure:mu-oh}a corresponds to the closed-box models with
oxygen yield $y=y_0$/4. Only two irregular galaxies from our sample are located
below the dotted line. This implies that irregular galaxies lose only a moderate part of their 
manufactured heavy elements. Two facts are important:
{\it i)}  the smooth variation in gas mass fraction in spiral and irregular galaxies
with luminosity over the full luminosity range,
{\it ii)}  the growth of effective oxygen yield with increasing luminosity from
$M_{\rm B} \sim$ --11 to --18 which remains approximately constant for more luminous
galaxies. These two facts taken together allow us to suggest that
the difference in the slopes of the O/H -- $M_{\rm B}$ relationships for spiral
and irregular galaxies is real and that the variation of the effective oxygen yield
with luminosity is responsible for the bend in the luminosity -- metallicity relationship.

How realistic is the value of the mean oxygen yield derived here for spiral galaxies?
Fig. \ref{figure:mu-oh}b shows the oxygen abundances extrapolated to zero radius in spiral galaxies
together with the $\mu$ - O/H track
predicted by the closed-box model with the derived oxygen yield.
One can expect that the central oxygen abundances in the most evolved spiral galaxies
correspond to the oxygen abundance when the system had exhausted the gas. As can be seen
in Fig. \ref{figure:mu-oh}b, the upper limit of the central oxygen abundances in spiral
galaxies is close to the oxygen abundances predicted by the model with derived oxygen yield
when the gas mass fraction is close to zero. The present-day oxygen abundance at the solar 
galactocentric distance in our Galaxy is about 12 + log(O/H) = 8.50 as traced by H\,{\sc ii} 
regions (Rodr\'{\i}guez 1999; Caplan et al. 2000;  Deharveng et al. 2000;
Pilyugin, Ferrini \& Shkvarun 2003) and derived from the interstellar absorption
lines towards the stars (Meyer, Jura \& Cardelli 1998; Sofia \& Meyer 2001),
and for the present-day gas mass fraction, 0.15 -- 0.20 appears to be a reasonable value
(Malinie et al. 1993). As shown in Fig. \ref{figure:mu-oh}b, the position of the solar
vicinity (large plus sign) is close to the $\mu$ - O/H track predicted by the
closed-box model with derived oxygen yield. Thus, the data in Fig. \ref{figure:mu-oh}b imply
that the value of the oxygen abundance derived here is quite realistic.

It should be noted that the value of the oxygen yield is derived here on the basis of gas-phase
oxygen abundances. Some fraction of the oxygen is locked into dust grains in H\,{\sc ii} regions.
Meyer, Jura \& Cardelli (1998) obtained a limit to the dust-phase oxygen abundance in the
interstellar medium in the vicinity of the Sun. Assuming various mixtures of oxygen-bearing
grain compounds, they found that it is difficult to increase the oxygen dust fraction beyond
$\sim$ 0.14 dex, simply because the requisite metals are far less abundant than oxygen.
Esteban et al. (1998) found that the fraction of the dust-phase oxygen abundance in the
Orion nebula is about 0.1 dex. Thus, the true value of the oxygen yield is slightly higher
than the value obtained here and is around $y =$ 0.003.

\section{Comparison with previous studies}

How reliable are the results obtained here? How significant is the difference
between the P-based and the $R_{\rm 23}$-based O/H -- M$_{B}$ relationships?
The validity of the oxygen abundances obtained here can be tested by a 
comparison with the T$_{e}$-based oxygen abundances determined recently
in the disk of the galaxy NGC 5457 by Kennicutt, Bresolin \& Garnett (2003).
It is worth noting that the comparison of oxygen abundances derived through
the P--method and through the T$_e$--method has already been made 
for the galaxy NGC 5457 (Pilyugin 2001b), as well as for
the Milky Way (Pilyugin, Ferrini \& Shkvarun 2003), and for
a sample of star-forming emission-line galaxies from the KPNO International
Spectroscopic Survey (KISS) (Melbourne et al. 2004). A good agreement has
been found in all cases. Recently Kennicutt, Bresolin \& Garnett (2003)
obtained new spectra for a number of H\,{\sc ii} regions in the disk
of the galaxy NGC 5457 and determined the T$_e$-based abundances.
They compared their temperature-based abundances with those derived for the
same H\,{\sc ii} regions using different calibrations. According to their
Fig. 14, the validity of the P-based abundances is questionable.
The data from Kennicutt, Bresolin \& Garnett (2003) have been used here
to test the validity of the P--calibration as well as other popular
calibrations.  To enlarge the comparison we have also included direct
abundance measurements of H\,{\sc ii} regions in this galaxy from the literature
(the compilation of the spectral data was performed by Pilyugin 2001b)
as well as direct abundance measurements of H\,{\sc ii} regions in other
galaxies (compilation from Pilyugin 2001a).

\begin{figure}
\centering
\resizebox{0.98\hsize}{!}{\includegraphics[angle=0]{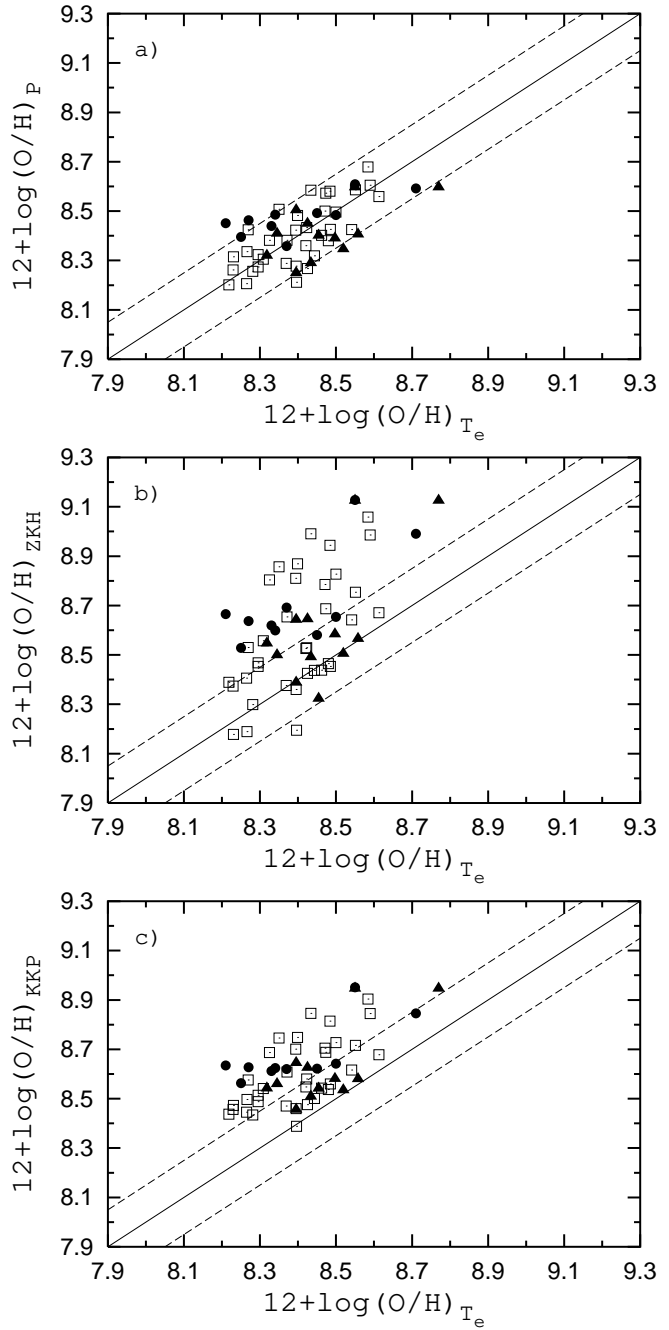}}
\caption{Comparison of temperature-based oxygen abundances O/H$_{T_e}$ with
those derived with different calibrations for high-metallicity H\,{\sc ii}
regions.
{\bf (a)} The O/H$_{P}$ -- O/H$_{Te}$ diagram. The filled circles represent
the O/H$_{Te}$ abundances for H\,{\sc ii} regions in the galaxy NGC 5457
from Kennicutt, Bresolin \& Garnett (2003), the filled triangles are
the O/H$_{Te}$ abundances for H\,{\sc ii} regions in this galaxy
from other authors. The open squares are
the O/H$_{Te}$ abundances for H\,{\sc ii} regions in other galaxies.
The O/H$_P$ abundances are derived through the high-metallicity P--calibration.
The solid line is the equal-value line, the dashed lines are shifted by +0.15 dex
and --0.15 dex.
{\bf (b)} The O/H$_{ZKH}$ -- O/H$_{Te}$ diagram for the same H\,{\sc ii} regions.
The O/H$_{ZKH}$ abundances are derived using the high-metallicity calibration
from Zaritsky, Kennicutt \& Huchra (1994).
{\bf (c)} The O/H$_{KKP}$ -- O/H$_{Te}$ diagram for the same H\,{\sc ii} regions.
The O/H$_{KKP}$ abundances are derived using the high-metallicity calibration
from Kobulnicky, Kennicutt \& Pizagno (1999).
}
\label{figure:zz}
\end{figure}

Fig. \ref{figure:zz}a shows the O/H$_{P}$ -- O/H$_{Te}$ diagram (the analog
of Fig.14 from Kennicutt, Bresolin \& Garnett 2003).
The filled circles are direct abundance measurements of H\,{\sc ii} regions
in the galaxy NGC 5457 with 12+logO/H$_{Te}$ $>$ 8.2 from Kennicutt, Bresolin \& Garnett (2003).
The filled triangles are direct abundance measurements of H\,{\sc ii} regions
in this galaxy from other authors (compilation from Pilyugin 2001b)
The open squares are direct abundance measurements of H\,{\sc ii} regions
in other galaxies (compilation from Pilyugin 2001a). Fig. \ref{figure:zz}a shows
that there is a satisfactory agreement between O/H$_{T_{e}}$ and O/H$_{P}$
abundances. The origin of the disagreement between Kennicutt et al's and our
conclusions is evident. The relationship between oxygen
abundance and strong oxygen line intensities is double-valued with two distincts parts
usually known as the ``lower'' and ``upper'' branches of the $R_{\rm 23}$ -- O/H
relationship. The high-metallicity P--calibration is valid only for
H\,{\sc ii} regions which belong to the upper branch, i.e. with 12+log(O/H) higher
than $\sim 8.2$. The unjustified use of the high-metallicity P--calibration
in the determination of the oxygen abundance in low-metallicity H\,{\sc ii}
regions results in overestimated P-based oxygen abundances (Pilyugin 2003a).
Our O/H$_{P}$ -- O/H$_{Te}$ diagram (Fig. \ref{figure:zz}a) shows only the
high-metallicity H\,{\sc ii} regions with 12+logO/H$_{Te} >$ 8.2.
On the contrary, Fig.14 of Kennicutt, Bresolin \& Garnett (2003) also includes 
low-metallicity H\,{\sc ii} regions, with 12+log(O/H) lower than $\sim 8.2$.
They do not belong to the upper branch, and, as a consequence, the high-metallicity
P--calibration cannot be used in the determination of the oxygen abundance
in these H\,{\sc ii} regions. This is the reason why the low-metallicity 
H\,{\sc ii} regions in the galaxy NGC 5457 are not presented in our 
Fig. \ref{figure:zz}a and are excluded from the analysis above (see Sect. 2).

\begin{figure}
\centering
\resizebox{1.00\hsize}{!}{\includegraphics[angle=0]{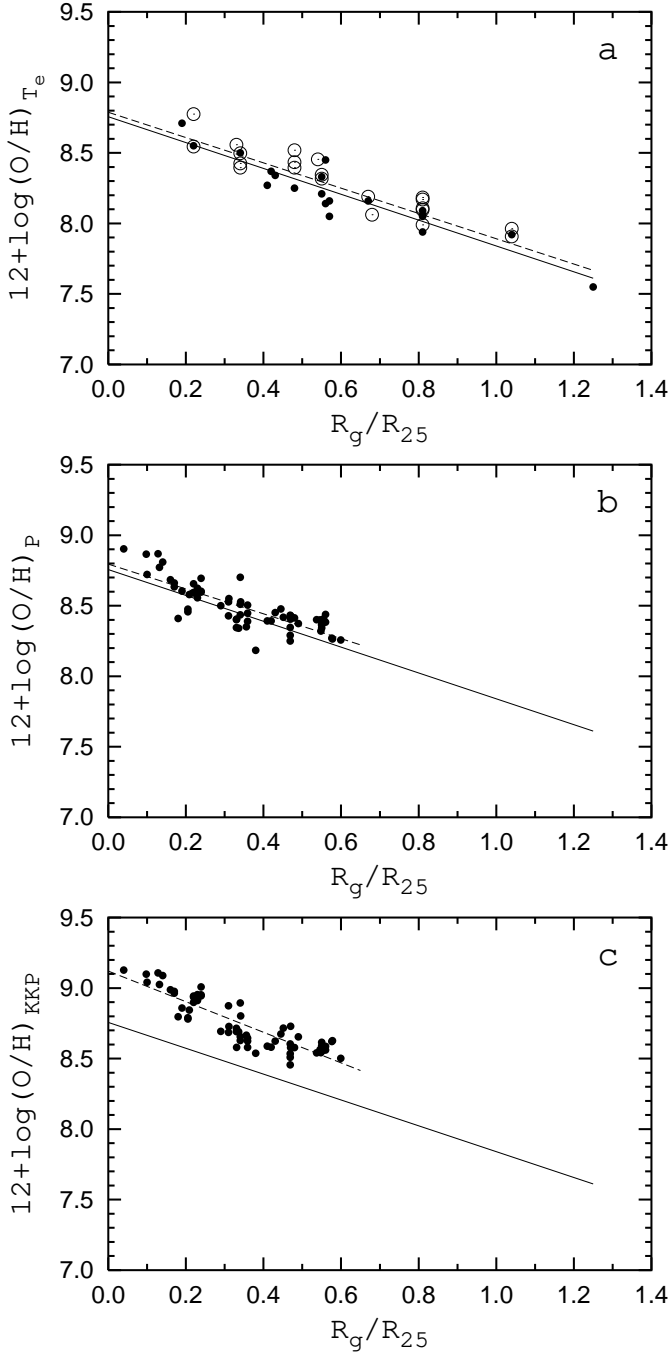}}
\caption{
The oxygen abundance versus galactocentric distance for the galaxy NGC 5457.
{\bf (a)} The filled circles are temperature-based abundances from Kennicutt,
Bresolin \& Garnett (2003). The solid line is the least squares fit to these
data. The open circles are temperature-based abundances from other authors
(compilation from Pilyugin 2001b). The dashed line is the least squares fit to
all the temperature-based abundances.
{\bf (b)} The filled circles are oxygen abundances derived through the
high-metallicity P--calibration for a large sample of H\,{\sc ii} regions
with $R_g < 0.60 R_{\rm 25}$. The best fit to these data is presented by the dashed
line. The solid line is the same as in {\bf (a)}.
{\bf (c)} The filled circles are abundances derived using the calibration
from Kobulnicky, Kennicutt \& Pizagno (1999) for the same H\,{\sc ii} regions.
The least squares fit to these data is presented by the dashed line.
}
\label{figure:m101}
\end{figure}

\begin{figure}
\centering
\resizebox{1.00\hsize}{!}{\includegraphics[angle=0]{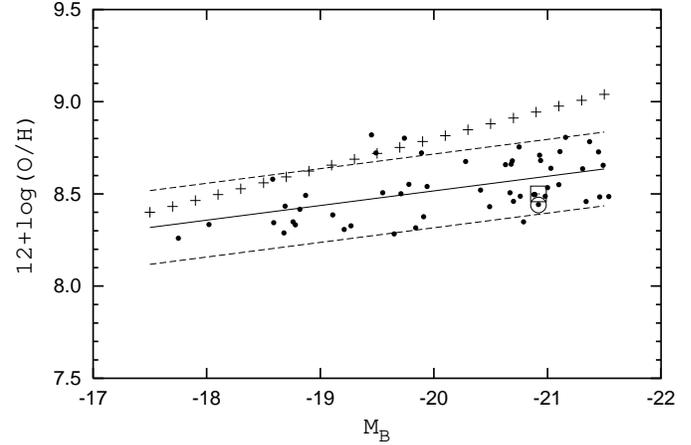}}
\caption{
The characteristic oxygen abundance as a function of absolute blue magnitude 
M$_B$ for our sample of spiral galaxies (the filled circles). The solid line is 
the O/H -- M$_B$ relationship (best fit derived through the least squares method). 
The dashed lines correspond to the lines shifted by --0.2 dex and +0.2 dex along 
the vertical axes relatively to the derived O/H -- M$_B$ relationship.
The oxygen abundances in the galaxy NGC 5457 are presented by the open circle
(O/H at $R=0.4R_{\rm 25}$) and open square (O/H at the disk half-light radius).
The O/H -- M$_B$ relationship from Garnett (2002) is shown by plus signs.
}
\label{figure:garnet}
\end{figure}

Fig. \ref{figure:zz}b shows the O/H$_{ZKH}$ versus (O/H)$_{T_{e}}$ diagram
for the same sample of H\,{\sc ii} regions as in Fig. \ref{figure:zz}a.
The O/H$_{ZKH}$ abundances are derived using the high-metallicity calibration
from Zaritsky, Kennicutt \& Huchra (1994). Zaritsky et al.'s calibration is
an average of the three calibrations by Edmunds \& Pagel (1994), McCall et al.
(1985), and Dopita \& Evans (1986), hence this calibration can be considered
as ``representative''  for one-dimensional calibrations.
Fig. \ref{figure:zz}c shows the O/H$_{KKP}$ versus (O/H)$_{T_{e}}$ diagram
for the same sample of H\,{\sc ii} regions as in Fig. \ref{figure:zz}a.
The O/H$_{KKP}$ abundances are derived using the two-dimensional
high-metallicity calibration from Kobulnicky, Kennicutt \& Pizagno (1999).
Fig. \ref{figure:zz}a-c shows that the O/H$_{P}$ abundances agree significantly
better with the O/H$_{T_{e}}$ abundances than do the (O/H)$_{ZKH}$ and
(O/H)$_{KKP}$ abundances.

We assume that the temperature-based abundance is the ``true'' abundance
and that the difference between, e.g. the P-based abundance and the T$_e$-based 
abundance is due only to the error attached to the P-based abundance. 
However, the O/H$_{T_{e}}$
abundances determined by Kennicutt, Bresolin \& Garnett (2003) in H\,{\sc ii}
regions of the galaxy NGC 5457 with close galactocentric distances
($0.55R_{\rm 25} \leq R_g \leq 0.57R_{\rm 25}$) show a large scatter, from
12+log(O/H)$_{T_{e}}$ = 8.05 for H\,{\sc ii} region H71 to
12+log(O/H)$_{T_{e}}$ = 8.45 for H\,{\sc ii} region H128.
This may be evidence for large uncertainty in O/H$_{T_{e}}$ abundances.
If this is the case, the uncertainty in the O/H$_{T_{e}}$ abundances can make an
appreciable contribution to the disagreement between the P-based and the
$T_{e}$-based abundances.

It is worth noting that the direct comparison
of temperature-based oxygen abundances O/H$_{Te}$ with those derived with
different calibrations is affected by the selection criteria. Indeed, in the high
metallicity range, the weak temperature-sensitive line O[III]$\lambda$4363
is usually measured in high-excitation H\,{\sc ii} regions. It is known that
the disagreement between temperature-based abundances and those derived with
one-dimensional $R_{\rm 23}$ calibrations is lowest for high-excitation H\,{\sc ii}
regions and is larger for low-excitation H\,{\sc ii} regions, while the P--based
abundances agree with the T$_{e}$--based abundances both for high-- and
low--excitation H\,{\sc ii} regions. Since the two-dimensional calibration of
Kobulnicky, Kennicutt \& Pizagno (1999) converts, in fact, into one-dimensional
calibration at high metallicities, one can expect that the mean value of the 
disagreement between O/H$_{KKP}$ and (O/H)$_{T_{e}}$ abundances is larger than
for the sample of high-excitation H\,{\sc ii} regions in
Fig. \ref{figure:zz}c. This statement can be checked through comparing 
the radial gradients traced by oxygen abundances derived using different methods. 
This also provides additional verification of the validity of oxygen abundances 
derived through different methods.

The O/H$_{T_e}$ abundances versus galactocentric distance for the galaxy NGC 5457
are shown in Fig. \ref{figure:m101}a. The small filled circles are the
data from Kennicutt, Bresolin \& Garnett (2003). The best fit to these data
\begin{eqnarray}
12+\log(O/H)_{Te, KBG}  =  \nonumber  \\
8.76\,(\pm 0.06) - 0.90\,(\pm 0.08) \times (R/R_{25})
\label{equation:gr03}
\end{eqnarray}
is shown by the solid line. The open circles are O/H$_{T_e}$ abundances from other 
authors (compilation from Pilyugin 2001b). The best fit to all data
\begin{eqnarray}
12+\log(O/H)_{Te, all}  = \nonumber  \\
8.79\,(\pm 0.04) - 0.90\,(\pm 0.06) \times (R/R_{25})
\label{equation:teall}
\end{eqnarray}
is shown by the dashed line. Fig. \ref{figure:m101}a, as well as
the comparison between Eq. (\ref{equation:gr03}) and Eq. (\ref{equation:teall}), 
shows that the O/H$_{T_e}$ abundances from Kennicutt, Bresolin \& Garnett (2003)
are slightly shifted towards lower abundances compared to those from other authors.

Fig. \ref{figure:m101}b shows the oxygen abundances derived here using the
P--calibration for a large (65) sample of H\,{\sc ii} regions with
$R_g > 0.6R_{\rm 25}$. The least squares fit to these data
\begin{eqnarray}
12+\log(O/H)_{P}  = \nonumber  \\
8.79\,(\pm 0.03) - 0.88\,(\pm 0.08) \times (R/R_{25})
\label{equation:here}
\end{eqnarray}
is shown in Fig. \ref{figure:m101}b by the dashed line. The O/H$_{Te}$ -- R$_g$
relationship of Kennicutt, Bresolin \& Garnett (2003) is shown by the solid
line. Fig. \ref{figure:m101}b, as well as the comparison between
Eq. (\ref{equation:gr03}), Eq. (\ref{equation:teall}) and Eq. (\ref{equation:here}), 
shows that there is a very good agreement, both in the zero point and slope,
between the radial O/H$_{T_{e}}$ abundance gradient and the O/H$_{P}$
abundance gradient in the disk of the galaxy NGC 5457.

Fig. \ref{figure:m101}c shows the radial distribution of the O/H$_{KKP}$
abundances, derived using the high-metallicity calibration from Kobulnicky,
Kennicutt \& Pizagno (1999), for the sample of H\,{\sc ii} regions from
Fig. \ref{figure:m101}b.  The least squares fit to these data
\begin{eqnarray}
12+\log(O/H)_{KKP}  =
\nonumber  \\
9.12\,(\pm 0.03) - 1.08\,(\pm 0.07) \times (R/R_{25})
\label{equation:kkp}
\end{eqnarray}
is shown in Fig. \ref{figure:m101}c by the dashed line. The O/H$_{Te}$ -- R$_g$
relationship of Kennicutt, Bresolin \& Garnett (2003) is shown by the solid
line. Fig. \ref{figure:m101}c, as well as the comparison between
Eq. (\ref{equation:gr03}), Eq. (\ref{equation:teall}) and Eq. (\ref{equation:kkp}), 
clearly shows that the central oxygen abundance in NGC 5457, derived using the 
calibration from Kobulnicky, Kennicutt \& Pizagno (1999),
is overestimated (by a factor $\sim $ 2) in comparison with the one derived
with the T$_e$ and P--methods. The slopes of the (O/H)$_{T_{e}}$-based  and
the (O/H)$_{R_{23}}$-based gradients are significantly different.

According to Garnett et al. (1997), the radial oxygen abundance gradient within the 
disk of NGC 5457 with oxygen abundances derived through the $R_{\rm 23}$--calibration is
\begin{equation}
12+\log(O/H)_{R_{23}}  = 9.13 - 1.36 \times (R/R_{25}) .
\label{equation:gr97}
\end{equation}
Comparison between Eq. (\ref{equation:gr03}), Eq. (\ref{equation:teall}) and
Eq. (\ref{equation:gr97}) clearly shows that the central oxygen abundance in
NGC 5457, derived using the $R_{\rm 23}$ calibration,
is overestimated (by a factor $\sim $ 2) in comparison with the one derived
with the T$_e$ and P--methods. The slopes of the (O/H)$_{T_{e}}$-based  and
the (O/H)$_{R_{23}}$-based gradients are significantly different.

There is thus a very good agreement, both in the zero point and slope,
between the radial O/H$_{T_{e}}$ abundance gradient and the O/H$_{P}$
abundance gradient within the disk of the galaxy NGC 5457.
For radial distributions of abundances derived with other calibrations, both the
central oxygen abundances and the slopes do not agree with (O/H)$_{T_{e}}$-based
values. This strongly supports our claim that the O/H$_{P}$ abundances are 
significantly more realistic than abundances derived with other calibrations.

The temperature-based characteristic oxygen abundance in the galaxy NGC 5457
can also be used as the "Rosetta Stone" to test the validity of the
O/H -- M$_B$ relationship obtained
here. The characteristic oxygen abundance as a function of absolute blue
magnitude M$_B$ for our sample of spiral galaxies is presented in
Fig. \ref{figure:garnet}. The positions of individual galaxies (points) and
the O/H -- M$_B$ relationship (solid line) are the same as in
in Fig. \ref{figure:s-l-oh}. The dashed lines
correspond to the lines shifted by --0.2 dex and +0.2 dex along the vertical 
axes relatively to the derived O/H -- M$_B$ relationship.
The oxygen abundances in the galaxy NGC 5457 are presented by the open circle
(O/H at $R=0.4R_{\rm 25}$) and open square (O/H at the disk half-light radius).
The line shown by plus signs in Fig. \ref{figure:garnet} is the luminosity --
metallicity relathionship for spiral galaxies obtained by Garnett (2002).
Fig. \ref{figure:garnet} shows that there is a large difference between the
luminosity -- metallicity relationship derived here and that from Garnett (2002).
The main reason of this discrepancy is due to the fact that our luminosity --
metallicity relationship is based on the oxygen abundances derived through
the P--calibration while Garnett has used oxygen abundances derived
through the $R_{\rm 23}$--calibrations.
The recent results on the oxygen abundance determination in the disk of
the well-studied spiral galaxy NGC 5457 (M 101) provide a possibility to check
the credibility of these relationships. Kennicutt, Bresolin \& Garnett (2003)
derived the radial distribution of the oxygen abundance in the disk of
NGC 5457, traced by the H\,{\sc ii} regions with oxygen abundances derived
with the classical T$_e$ -- method.
The characteristic oxygen abundance (as adopted here the oxygen abundance
at $R=0.4R_{\rm 25}$) in the galaxy NGC 5457
is presented by the open circle in Fig. \ref{figure:garnet}.
It is evident that the T$_e$-based characteristic oxygen abundances in the
galaxy NGC 5457 are in conflict with Garnett's luminosity -- metallicity
relationship but agree with our luminosity -- metallicity relationship.
It is not suprising since it was shown above that there is
a very good agreement, both in the zero point and slope, between 
our P-based oxygen abundance distribution, Eq. (\ref{equation:here}), 
and the T$_e$-based oxygen abundance distribution, Eq. (\ref{equation:gr03}), 
of Kennicutt, Bresolin \& Garnett (2003), while the $R_{\rm 23}$-based oxygen 
abundance distribution, Eq. (\ref{equation:gr97}), of Garnett et al. (1997) is 
quite different. It should be noted that Garnett (2002) used the oxygen abundances 
at the disk half-light radius to built the luminosity -- metallicity relationship.
Can this make a significant contribution to the deviation of the characteristic
oxygen abundance in the galaxy NGC 5457 from the luminosity -- metallicity relationships 
based on oxygen abundances at the disk half-light radius? Based on 
Garnett's conclusion: ``for comparison, I include results
based on using the value of O/H at one disk scale length; this change has
little effect on the overall results'', one can expect that the difference
in the choice of the ``characteristic'' radius plays a minor role. For
verification and illustration, the T$_e$-based oxygen abundance in the galaxy
NGC 5457 at the disk half-light radius (equal to 1.685 times the exponential
scale length) is obtained and presented in Fig. \ref{figure:garnet}
by the open square. Fig. \ref{figure:garnet} shows that using the characteristic 
oxygen abundance at $R=0.4_{\rm 25}$ in the galaxy
NGC 5457 instead of the value of O/H at the disk half-light radius has
little effect on the overall result.
Thus, the T$_e$-based characteristic oxygen abundances in the galaxy NGC 5457
are in conflict with the luminosity -- metallicity relationship derived by
Garnett (2002) with the $R_{\rm 23}$-based oxygen abundances, but agree with our
luminosity -- metallicity relationship derived with the P-based oxygen abundances.

\section{Conclusions}

We performed a compilation of more than 1000 published spectra of H\,{\sc ii} regions
in 54 spiral galaxies. The oxygen and nitrogen abundances in each H\,{\sc ii} region
were recomputed in a homogeneous way, using the P--method. The radial distributions
of the oxygen abundances, nitrogen abundances, and nitrogen-to-oxygen abundance ratios
were derived. The parameters of the radial distributions (the extrapolated central
intersect value and the gradient) are listed.

The correlations between oxygen abundance and macroscopic properties of galaxies 
are examined. The oxygen abundance in the disk of a galaxy at $r=0.4R_{\rm 25}$, 
where $R_{\rm 25}$ is the isophotal
radius, is used as a characteristic (or representative) oxygen abundance for spiral galaxies.
We found that the oxygen abundance in spiral galaxies  correlates with its luminosity,
rotation velocity, and morphological type: the correlation 
with the rotation velocity may be slightly tighter. 

There is a significant difference between the luminosity -- metallicity relationship
derived here and that based on the oxygen abundances derived through the 
$R_{\rm 23}$--calibrations. The T$_e$-based characteristic oxygen abundances in the
galaxy NGC 5457 (Kennicutt, Bresolin \& Garnett 2003) agree with our
luminosity -- metallicity relationship derived with the P-based oxygen abundances
but is in conflict with the luminosity -- metallicity relationship derived by
Garnett (2002) with the $R_{\rm 23}$-based oxygen abundances.

The derived luminosity -- metallicity relation for spiral galaxies is compared
to that for irregular galaxies. We found that the slope of the oxygen
abundance  -- luminosity relationship for spirals is slightly more shallow
than the one for irregular galaxies.

The effective oxygen yields were derived for spiral and irregular galaxies.
We found that the effective oxygen yield increases with increasing
luminosity from $M_{\rm B} \sim -11$ to $M_{\rm B} \sim -18$ (or with
increasing rotation velocity from $V_{\rm rot} \sim 10$ km s$^{-1}$ to 
$V_{\rm rot} \sim 100$ km s$^{-1}$) and then remains approximately constant.
Irregular galaxies from our sample have effective oxygen yields
lowered by a factor of 3 at maximum, i.e. irregular galaxies usually retain 
at least 1/3 of the oxygen they manufactured during their evolution.
From the comparison between the effective oxygen yields for spiral and irregular
galaxies (with $R_{\rm 23}$-based oxygen abundances in spiral galaxies),
Garnett (2002) found however that an irregular galaxy can lose up to
90\% $\div$ 95\% of the oxygen manufactured.

\begin{acknowledgements}
We thank Prof. B.E.J.~Pagel for a careful reading of the manuscript and useful 
comments and suggestions which helped to improve the paper.
We thank the anonymous referee for helpful comments.
This study was supported by the sabbatical grant SAB2001-0165 of the Spanish
Ministerio de Educaci\'{o}n, Cultura y Deporte (L.S.P.) and by the Ukrainian 
Fund of Fundamental Investigation, grant No 02.07/00132 (L.S.P.). 
\end{acknowledgements}

\appendix  

\section{Distances and luminosities}

To derive a reliable galaxy luminosity, an accurate value of the distance is necessary.
Over the past decade, great progress has been made in accurate distance measurements
for galaxies using many different methods. Therefore,
we have performed a thorough literature search for distance information to determine the
most reliable, up-to-date distances for our sample.

The Cepheid period--luminosity relation remains the most useful method for 
estimating the distance of galaxies. Another efficient tool is the luminosity of 
the tip of the red giant branch (TRGB) stars.
One of the most promising distance indicators is believed to be the peak brightness
of type Ia supernovae (SNe Ia). The supernovae of type II (SNe II) are also
reliable distance indicators; the method of distance determination is based
on models of the expanding photospheres of SNe II (EPM). The planetary nebula
luminosity function (PNLF) is a reliable distance indicator for galaxies
potentially as far away as $\sim$ 25 Mpc. Surface brightness fluctuations (SBF)
have emerged as a reliable distance indicator. The classic method of distance
measurements via  the luminosity of the brighest blue and red supergiants
(BBSG and BRSG) has been widely used over last decade. The Tully-Fisher (TF) relation,
or ``luminosity--rotation velocity'' relation, is the most commonly applied distance
indicator for spiral galaxies at the present time. There are different versions of the
TF relation: the luminosity in different bands can be used in the TF relation,
and the CO linewidth instead of the H\,{\sc i} linewidth can be used as the tracer of rotation
velocity.

\begin{table*}
\caption[]{\label{table:distance}
The general properties of spiral galaxies from our sample}
{\footnotesize
\begin{center}
\begin{tabular}{lclrrrl} \\ \hline \hline
               &        &            &         &           &            &                                          \\
galaxy         & T-Type & morphology & B$_T^0$ & logL$_B$  & distance   &  references for                          \\
               &        &            &        &(L$_{\odot}$)&    Mpc     & distance values                         \\  
               &        &            &         &           &            &                                          \\  \hline
               &        &            &         &           &            &                                          \\
NGC224 = M31   &  3.0   &  SA(s)b    &  3.36   &   10.63   &  0.78      & FMG01, JPN03, MK86, JTB03, TDB01, SS94   \\
NGC253         &  5.0   &  SAB(s)c   &  7.09   &   10.55   &  3.94      & KSD03, TS00                              \\
NGC300         &  7.0   &  SA(s)d    &  8.49   &    9.40   &  2.00      & FMG01, SMJ96                             \\
NGC598 = M33   &  6.0   &  SA(s)cd   &  5.75   &    9.74   &  0.84      & FMG01, TS00, MCM00, LKS02, KKL02         \\
NGC628 = M74   &  5.0   &  SA(s)c    &  9.76   &   10.01   &  7.28      & SD96, SKT96, HFS97                       \\
NGC753         &  4.0   &  SAB(rs)bc & 12.35   &   10.64   & 49.30      & R02                                      \\
NGC925         &  7.0   &  SAB(s)d   &  9.97   &   10.13   &  9.16      & FMG01, SD98, TPE02                       \\
NGC1058        &  5.0   &  SA(rs)c   & 11.55   &    9.62   & 10.60      & SKE94, TPE02                             \\
NGC1068 = M77  &  3.0   &(R)SA(rs)b  &  9.47   &   10.77   & 15.30      & SS94, TS00, HFS97                        \\
NGC1232        &  5.0   &  SAB(rs)c  & 10.38   &   10.55   & 17.90      & R02, TPE02                               \\
NGC1365        &  3.0   &(R)SBb(s)c  &  9.93   &   10.73   & 17.95      & FMG01, SS94                              \\
NGC2403        &  6.0   &  SAB(s)cd  &  8.43   &    9.84   &  3.22      & FMG01, CFJ02                             \\
NGC2442        &  3.7   &  SAB(s)bc  & 10.36   &   10.72   & 21.58      & RKS01, TDB01                             \\
NGC2541        &  6.0   &  SA(s)cd   & 11.57   &    9.66   & 11.22      & FMG01, FHS97                             \\
NGC2805        &  7.0   &  SAB(rs)d  & 11.17   &    9.70   &  9.77      & TDB01                                    \\
NGC2835        &  5.0   &  SAB(rs)c  & 10.31   &   10.05   &  9.80      & R02, TDB01                               \\
NGC2841        &  3.0   &  SA(r)b    &  9.58   &   10.66   & 14.06      & MSB01, HFS97                             \\
NGC2903        &  4.0   &  SB(s)d    &  9.11   &   10.44   &  8.87      & DK00, TPE02, TS00                        \\
NGC2997        &  5.0   &  SA(s)c    &  9.34   &   10.46   & 10.05      & R02, SPU94, T88, V79                     \\
NGC3031 = M81  &  2.0   &  SA(s)ab   &  7.39   &   10.36   &  3.63      & FMG01, JCB89, JTB03, TDB01, TPE02        \\
NGC3184        &  6.0   &  SAB(rs)cd & 10.34   &   10.15   & 11.10      & HFS97, LFL02, V79                        \\
NGC3198        &  5.0   &  SB(rs)c   & 10.21   &   10.39   & 13.80      & FMG01, TPE02                             \\
NGC3344        &  4.0   &(R)SAB(r)bc & 10.50   &    9.67   &  6.90      & VBA00                                    \\
NGC3351 = M95  &  3.0   &  SB(r)b    & 10.26   &   10.09   & 10.00      & CFJ02, FMG01, TPE02                      \\
NGC3521        &  4.0   & SAB(rs)bc  &  9.29   &   10.49   & 10.21      & HFS97, RST00, TS00                       \\
NGC3621        &  7.0   &  SA(s)d    &  9.20   &   10.16   &  6.64      & FMG01, TPE02                             \\
NGC4254 = M99  &  5.0   &  SA(s)c    & 10.10   &   10.57   & 16.14      & R02, SS97, SSS02                         \\
NGC4258 = M106 &  4.0   &  SAB(s)bc  &  8.53   &   10.58   &  7.98      & CFJ02, CMM02, FMG01, HMG99, NFS01        \\
NGC4303 = M61  &  4.0   &  SAB(rs)bc & 10.12   &   10.10   &  9.55      & SPU94, SS97, TS00                        \\
NGC4321 = M100 &  4.0   &  SAB(s)bc  &  9.98   &   10.56   & 15.21      & FMG01, R02, SKE94, SSS02, TPE02, TS00    \\
NGC4395        &  9.0   &  SA(s)m    & 10.57   &    9.29   &  4.61      & HFS97, KMS03, KD98                       \\
NGC4501 = M88  &  3.0   &  SA(rs)b   &  9.86   &   10.74   & 17.58      & R02, SS97, SSS02                         \\
NGC4559        &  6.0   &  SAB(rs)cd &  9.76   &    9.90   &  6.40      & HFS97, RST00                             \\
NGC4571        &  6.5   &  SA(r)d    & 11.73   &    9.97   & 17.22      & PMR92, SS97, SSS02, TGD00                \\
NGC4651        &  5.0   &  SA(rs)c   & 11.04   &   10.47   & 22.28      & SS97, SSS02, TPE02                       \\
NGC4654        &  6.0   &  SAB(rs)cd & 10.75   &   10.17   & 13.74      & SS97, SSS02                              \\
NGC4689        &  4.0   &  SA(rs)bc  & 11.39   &    9.99   & 15.00      & SS97, SSS02                              \\
NGC4713        &  7.0   &  SAB(rs)d  & 11.85   &    9.72   & 13.61      & SSS02                                    \\
NGC4725        &  2.0   &  SAB(r)ab  &  9.78   &   10.46   & 12.36      & JTB03, FMG01, TDB01, TPE02               \\
NGC4736 = M94  &  2.0   & (R)SA(r)ab &  8.75   &   10.08   &  4.93      & KMS03, SS94, TDB01, TPE02, TS00          \\
NGC5033        &  5.0   &  SA(s)c    & 10.21   &   10.51   & 15.82      & RST00, SFE01, TPE02                      \\
NGC5055 = M63  &  4.0   &  SA(rs)bc  &  9.03   &   10.47   &  8.81      & RST00, TPE02, TS00                       \\
NGC5068        &  6.0   &  SB(s)d    & 10.09   &    9.70   &  5.90      & SFE01                                    \\
NGC5194 = M51  &  4.0   &  SA(s)bc   &  8.67   &   10.49   &  7.64      & CFJ02, FCJ97, SS94, TDB01, TS00          \\
NGC5236 = M83  &  5.0   &  SAB(s)c   &  7.98   &   10.30   &  4.49      & ESK96, SKE94, TTS03                      \\
NGC5457 = M101 &  6.0   &  SAB(rs)cd &  8.21   &   10.56   &  6.70      & DK00, FCJ97, FMG01, JPJ00, SKE94         \\
NGC6384        &  4.0   &  SAB(r)bc  & 10.60   &   10.79   & 26.14      & P94, PSS00, S97, TPE02                   \\
NGC6744        &  4.0   &  SAB(r)bc  &  8.82   &   10.60   &  9.33      & SFE01, TPE02                             \\
NGC6946        &  6.0   &  SAB(rs)cd &  7.78   &   10.59   &  5.70      & ESK96, SKE94, SKT97, SS94, TPE02, TS00   \\
NGC7331        &  3.0   &  SA(s)b    &  9.38   &   10.78   & 14.72      & FMG01, JTB03, SKE94, TDB01, TPE02        \\
NGC7793        &  7.0   &  SA(s)d    &  9.37   &    9.63   &  3.91      & KSD03, D98                               \\
IC342          &  6.0   &  SAB(rs)cd &  6.04   &   10.81   &  3.28      & KT93, SCH02, SS94, TS00                  \\
IC5201         &  6.0   &  SB(s)cd   & 11.00   &    9.87   & 11.00      & T88                                      \\
               &        &            &         &           &            &                                          \\  \hline
\end{tabular}
\end{center}
}
\end{table*}

\begin{table*}
\caption[]{\label{table:referd}
List of references to Table \ref{table:distance}}
\begin{center}
\begin{tabular}{lll} \hline \hline
             &                                                   &         \\
Abbreviation & Reference                                         & Method  \\
             &                                                   &         \\
 \hline
             &                                                   &         \\
             &                                                   &         \\
CFJ02        &  Ciardullo, Feldmeier, Jacoby, et al. 2002        & PNLF    \\
CMM02        &  Caputo, Marconi  \& Musella 2002                 & Cepheid \\
D98          &  Davidge 1998                                     & BRSG    \\
DK00         &  Drozdovsky \& Karachentsev 2000                  & TRGB    \\
ESK96        &  Eastman, Schmidt  \& Kirshner 1996               & EPM     \\
FCJ97        &  Feldmeier, Ciardullo  \& Jacoby  1997            & PNLF    \\
FMG01        &  Freedman, Madorw, Gibson, et al.  2001           & Cepheid \\
HFS97        &  Ho, Filippenko  \& Sargent 1997                  & H$_0$   \\
HMG99        &  Herrnstein, Moran, Greenhil, et al. 1999         & geometric \\
JCB89        &  Jacoby, Ciardullo, Booth, et al. 1989            & PNLF    \\
JPJ00        &  Jurcevic, Pierce  \& Jacoby 2000                 & RSGV    \\
JPN03        &  Joshi, Pandey, Narasimha, et al. 2003            & RSGV    \\
JTB03        &  Jensen, Tonry, Barris, et al.  2003              & SBF     \\
KD98         &  Karachentsev  \& Drozdovsky  1998                & BBSG    \\
KKL02        &  Kim, Kim, Lee, et al. 2002                       & TRGB    \\
KMS03        &  Karachentsev, Makarov, Sharina, et al. 2003a     & TRGB    \\
KSD03        &  Karachentsev, Sharina, Dolphin, et al. 2003b     & TRGB    \\
KT93         &  Karachentsev  \& Tikhonov 1993                   & BRGB    \\
LFL02        &  Leonard, Filippenko, Li, et al.  2002b           & EPM     \\
LKS02        &  Lee, Kim, Sarajedini, et al. 2002                & Cepheid \\
MCM00        &  Magrini, Corradi, Mampaso, et al. 2000           & PNLF    \\
MK86         &  Mould  \& Kristian  1986                         & TRGB    \\
MSB01        &  Macri, Stetson, Bothun, et al.  2001             & Cepheid \\
NFS01        &  Newman, Ferrarese, Stetson, et al.  2001         & Cepheid \\
P94          &  Pierce  1994                                     & SNIa    \\
PMR92        &  Pierce, McClure  \& Racine  1992                 & BRSG    \\
PSS00        &  Parodi, Saha, Sandage, et al. 2000               & SNIa    \\
R02          &  Russel 2002                                      & TF      \\
RKS01        &  Ryder, Koribalski, Staveley-Smith, et al.  2001  & H$_0$   \\
RST00        &  Rothberg, Saunders, Tully, et al. 2000           & H$_0$   \\
S97          &  Shanks  1997                                     & SNIa    \\
SCH02        &  Saha, Claver  \& Hoessel 2002                    & Cepheid \\
SD96         &  Sohn  \& Davidge 1996                            & BRSG    \\
SD98         &  Sohn  \& Davidge 1998                            & BRSG    \\
SFE01        &  Shapley, Fabbiano  \& Eskridge  2001             & H$_0$   \\
SKE94        &  Schmidt, Kirshner, Eastman, et.al.  1994         & EPM     \\
SKT96        &  Sharina, Karachentsev  \& Tikhonov  1996         & BBSG    \\
SKT97        &  Sharina, Karachentsev  \& Tikhonov  1997         & BBSG    \\
SMJ96        &  Soffner, M\'{e}ndez, Jacoby, et al. 1996         & BBSG    \\
SPU94        &  Saikia, Pedlar, Unger, et al. 1994               & H$_0$   \\
SS94         &  Sch\"{o}niger  \& Sofue  1994                    & TF      \\
SS97         &  Sch\"{o}niger  \& Sofue  1997                    & TF      \\
SSS02        &  Solanes, Sanchis, Salvador, et.al.  2002         & TF      \\
T88          &  Tully  1988                                      & H$_0$   \\
TDB01        &  Tonry, Dressler, Blakeslee, et al.  2001         & SBF     \\
TGD00        &  Tikhonov, Galazutdinova  \& Drozdovsky  2000     & BSG     \\
TPE02        &  Terry, Paturel  \& Ekholm  2002                  & sosie(TF) \\
TS00         &  Takamiya  \& Sofue  2000                         & TF      \\
TTS03        &  Thim, Tammann, Saha, et al.  2003                & TF      \\
V79          &  de Vaucouleurs  1979                             &tertiary \\
VBA00        &  Verdes-Montenegro, Bosma  \& Athanassoula  2000  & H$_0$   \\
             &                                                   &         \\
             &                                                   &         \\
             &                                                   &         \\
 \hline
\end{tabular}
\end{center}
\end{table*}

The important point is that the TRGB, SNIa, PNLF, SBF, BRSG, and TF methods
request a calibration. The {\it Hubble Space Telescope} Key Project on the
Extragalactic Distance Scale has provided Cepheid distances to more than 20
galaxies, in addition to the existing ground-based Cepheid distances to about
10 galaxies.  This data set constitutes a solid basis for calibrating other
distance indicators (Ferrarese et al. 2000). It should be
noted, however, that some uncertainty in the Cepheid distances still remains,
due to uncertainties in the slope and zero point of the Cepheid period --
luminosity relation (see discussion in Ferrarese et al. 2000; Freedman et
al. 2001; among others).
Therefore, the slightly different values of the Cepheid distance to a given
galaxy have been derived by different authors using the same observational data.
As a result, there is some disagreement in the literature on the magnitudes of
other standard candles. For example, two recent calibrations of the peak absolute
magnitude of SNe Ia, based on the SNe Ia events in galaxies with known Cepheid
distances result in slightly different magnitudes:
$M_{\rm B} = -19.56$ (Parodi et al. 2000) and $M_{\rm B} = -19.32$
(Gibson \& Stetson 2001). The method based on modeling the expanding
photospheres of SNe II is a "physical method" and is hence completely independent
of the calibration of the local distance scale. The EPM distances  agree with
the Cepheid distances within around 10\% (Eastman, Schmidt \& Kirshner 1996;
Leonard et al. 2002a). Since systematic uncertainty clearly remains at the 10\%
level for Cepheid and at the 10--20\% level for EPM, it is not clear whether the
observed discrepancy is significant (Leonard et al. 2002a).

The most important point to take into account when comparing different methods of
distance determination is their accuracy. The {\it relative} Cepheid distances
are determined to $\sim \pm 5$\% (Freedman et al. 2001). A typical accuracy of the
TRGB distances is $\sim 12$\% (Karachentsev et al. 2003a, b). According to Parodi
et al. (2000), the SNe Ia are the best standard candles known so far, with a luminosity
scatter of less than 0.25 mag.  The EPM distances agree with the Cepheid distances
within $\sim 10$\% (Eastman, Schmidt \& Kirshner 1996; Leonard et al. 2002a).
There is also a good agreement between PNLF and Cepheid distances over two
orders of magnitude in the derived distances (Feldmeier, Ciardullo  \& Jacoby 1997).
The SBF distance measurements rely on the empirical calibration of absolute fluctuation
amplitudes. A relative accuracy of $\sim 10$\% in extragalactic SBF distances can be
reached, provided that the galaxy color is well known (Jensen et al. 2003).
Distance moduli derived with the BBSG and BRSG method have typical errors of 0.4 --
0.5 mag (Drozdovsky \& Karachentsev 2000). Solanes et al. (2002) have examined
the TF distance moduli of spiral galaxies in the region of the Virgo cluster taken from
eight published datasets and have found a dispersion of 0.41 mag.
Tutui \& Sofue (1997, 1999) have analysed H\,{\sc i} and CO TF methods. They concluded that
the errors in the H\,{\sc i} TF distances can be especially large in the case of interacting
galaxies since the H\,{\sc i} linewidth may be strongly disturbed by galaxy -- galaxy interaction
and may not reflect correctly the rotation velocity of the galaxy.
The method of ``sosie galaxies'' (look-alike galaxies) is a particular application of the TF
method which avoids some practical problems (Paturel 1984; Terry, Paturel  \& Ekholm
2002). The distances derived with the ``sosie galaxies'' method appear to be more reliable
than those obtained with the direct TF relation.

Typically, two or more distance estimations for a galaxy from our sample are available
in the literature. Examination of available distances confirms that the
Cepheid, EPM, SBF, SNIa, PNLF, and TRGB methods provide accurate distance determinations.
Estimations of distances with two or more of these methods exist for 14 galaxies from
our sample. The deviation of individual distance estimations from the mean value is usually
less than 10\%. These methods will be refered to as ``high-precision'' methods below.
The agreement between the TF (and BBSG, BRSG) distances and distances derived with the
high-precision methods is appreciably worse. Moreover, the TF distances derived by
different authors for the same galaxy can show a significant scatter. Taking these
facts into account, the adopted value of the distance for a galaxy is chosen in the following
way. If the distance derived with the ``high-precision'' method(s) is available, this
distance is adopted. If only the TF and/or BBSG, BRSG distances are available, the mean
value is adopted. If a galaxy is a member of compact group, the information on the
distances to other members of the group is used to check the distance.
We have to adopt the redshift (H$_0$) distances for three galaxies (NGC3344, NGC5068, and
IC5201) from our sample due to the lack of other reliable distance estimations.

Here are some comments on individual galaxies.

{\it NGC628. --} There is no distance to NGC628 determined with one of the ``high-precision''
methods. However, the available distances to NGC628 derived with the method of brightest stars
seem to be quite reliable. Firstly, the distances derived from two different investigations are in
agreement: 7.24 Mpc (Sohn \& Davidge 1996) compared to 7.32 Mpc (Sharina, Karachentsev  \& Tikhonov 1996).
Secondly, the distance moduli estimated by Sharina, Karachentsev  \& Tikhonov (1996) for four galaxies in
the group of NGC628 have a mean value of 7.80 Mpc with a small scatter. As noted by Sharina,
Karachentsev  \& Tikhonov (1996), if a group contains a sufficient number of irregular galaxies, the
``brightness stars'' method makes it possible to measure the distance to the group with almost the same accuracy
as with the Cepheid method.

{\it NGC753. --} The galaxy NGC753 belongs to the Pisces filament. The TF distances of several dozens
of galaxies of this filament lie between $\sim$ 50 Mpc and $\sim$ 80 Mpc (Tully \& Pierce 2000).
The TF distance (49.30 Mpc) of NGC753 from Russel (2002) is adopted here.

{\it NGC1068. --}
Two values of TF distances to NGC1068 show an appreciable discrepancy: 12.45 Mpc (Sch\"{o}niger \& Sofue 1994)
compared with 18.10 Mpc (Takamiya \& Sofue 2000). The galaxy NGC1068 is a member of the LGG73 group (Garcia 1993).
A distance of 16.07 Mpc to the galaxy NGC1055 (also a member of LGG73 group) was estimated by
Paturel et al. (1998) using the method of ``sosies galaxies''. The mean TF distance of 15.30 Mpc is
confirmed by the Paturel et al.'s data and is adopted here.

{\it NGC2442. --}
There is no reliable distance determination for NGC2442. This galaxy is a member of the LGG147 group (Garcia 1993).
The distance to another member (NGC2434) of this group has been determined with the SBF method and is equal to 21.58 Mpc
(Tonry et al. 2001). This value is adopted as distance to NGC2442.

{\it NGC2805. --}
There is no reliable distance determination for NGC2805. This galaxy is a member of the LGG173 group (Garcia 1993).
The distance to another member (NGC2880) of this group has been determined with the SBF method and is equal to 9.77 Mpc
(Tonry et al. 2001). This value is adopted as distance to NGC2805.

{\it NGC2835. --}
Different distance values to NGC2835 were used recently: 10.76 Mpc (Elmegreen \& Salzer 1999) and
6.03 Mpc (Larsen \& Richtler 2000). The TF distance to NGC2835 is 9.80 Mpc (Russel 2002). The galaxy NGC2835
is a member of the LGG172 group (Garcia 1993). The distance to NGC2784 (a member of the LGG172 group)
has been determined with the SBF method and is equal to 9.82 Mpc (Tonry et al. 2001). This value is adopted as
distance to NGC2835.

{\it Virgo galaxies. --}
We chose  individual distance estimations to Virgo galaxies in the same way as in the case of
field galaxies. If the distance estimation with one of the ``high-precision'' methods is available,
this distance is adopted. Otherwise the distance is derived by averaging the available TF distances.
The distances of Virgo galaxies NGC4254, NGC4501, NGC4571, NGC4651, NGC4654, NGC4689,
and NGC4713 were taken from Solanes et al. (2002). These distances were obtained by averaging distance
moduli based upon the TF relationship taken from eight published datasets (Ekholm et al. 2000;
Fouqu\'{e} et al. 1990; Federspiel, Tammann  \& Sandage 1998; Gavazzi et al. 1999; Kraan-Korteweg,
Cameron  \& Tammann 1988; Mould, Aaronson  \& Huchra 1980; Pierce \& Tully 1988; Yasuda, Fukugita \&
Okamura 1997). The distances from these works have been previously homogenized. It seems quite impossible to
check the distance of individual Virgo cluster galaxies by comparing their value with the "group"
distance. Indeed, the galaxies NGC4321, NGC4651, and NGC4689 belong to the LGG289 group (Garcia 1993). Their
TF distances range from 15.00 Mpc for NGC4689 to 22.28 Mpc for NGC4651. The distances for nine galaxies
of the LGG289 group have been determined with the SBF method (Tonry et al. 2001), these distances lie between
15.00 Mpc (NGC4564) and 22.42 Mpc (NGC4365). Thus, both the TF distances and the SBF distances
of galaxies of LGG289 group are spread over more than 7 Mpc. There are two estimations of the distance to the
Virgo cluster galaxy NGC4571 with the ``brightest supergiant stars'' method. The mean BRSG distance value
of 17.15 Mpc is in agreement with Solanes et al.'s TF distance of 17.22 Mpc.

The adopted values of distance and the corresponding references to sources of data are presented in
Table \ref{table:distance} (Cols. 6 and 7). The list of references to Table \ref{table:distance} is given
in Table \ref{table:referd}. Using the adopted distances and the total face-on blue magnitudes corrected
for galactic and internal absorption $B_{\rm T}^0$, the luminosities of galaxies were determined.
The $B_{\rm T}^0$ values are taken from the Third Reference Catalog of bright Galaxies (de Vaucouleurs et al
1991, RC3) and are listed in Col. 4 of Table \ref{table:distance}, the derived luminosities are given
in Col. 5 of Table \ref{table:distance}.


\begin{thebibliography}{99}
\bibitem{01}
         Alloin, D., Collin-Souffrin, S., Joly, M. \& Vigroux, L.
         1979, A\&A, 78, 200
\bibitem{02}
         Alloin, D., Edmunds, M.G., Lindblad, P.O. \& Pagel, B.E.J. 1981, A\&A, 101, 377
\bibitem{001}
         Alves, D.R. \& Nelson, C.A. 2000, ApJ, 542, 789
\bibitem{002}
         Amram, P., Marcelin, M., Balkowski, C., Cayatte, V., Sullivan, W.T. \& Le Coarer, E. 1994, A\&AS, 103, 5
\bibitem{003}
         Appleton, P.N., Davies, R.D. \& Stephenson, R.J. 1981, MNRAS, 195, 327
\bibitem{004}
         Bajaja, E. \& Martin, M.C. 1985, AJ, 90, 1783
\bibitem{005}
         Bajaja, E., Wielebinski, R., Reuter, H.-P., Harnett, J.I. \& Hummel, E. 1995, A\&AS, 114, 147
\bibitem{006}
         Becker, R., Mebold, U., Reif, K. \& van Woerden, H. 1988, A\&A, 203, 21
\bibitem{007}
         Begeman, K.G., Broeils, A.H. \& Sanders, R.H. 1991, MNRAS, 249, 523
\bibitem{03}
         Blair, W.P., Kirshner, R.P. \& Chevalier, R.A. 1982, ApJ, 254, 50
\bibitem{008}
         Bosma, A. 1981, AJ, 86, 1791
\bibitem{009}
         Bosma, A., Casini, C., Heidmann, J., van der Hulst, J.M. \& van Woerden, H. 1980, A\&A, 89, 345
\bibitem{010}
         Bosma, A., van der Hulst, J.M.  \& Sullivan, W.T. 1977, A\&A, 57, 373
\bibitem{011}
         Bravo-Alfaro, H., Szomoru, A., Cayatte, V., Balkowski, C. \& Sancisi, R. 1997, A\&AS, 126, 537
\bibitem{04}
         Bresolin, F. \& Kennicutt, R.C. 2002, ApJ, 572, 838
\bibitem{05}
         Bresolin, F., Kennicutt, R.C. \& Garnett, D.R. 1999, ApJ, 510, 104
\bibitem{012}
         Broeils, A.H. \& Rhee, M.-H. 1997, A\&A, 324, 877
\bibitem{013}
         Buta, R. 1988, ApJS, 66, 233
\bibitem{014}
         Caplan, J., Deharveng, L., Pe\~{n}a, M., Costero, R. \& Blondel, C.
         2000, MNRAS, 311, 328
\bibitem{015}
         Caputo, F., Marconi, M. \& Musella, I. 2002, ApJ, 566, 833
\bibitem{016}
         Carignan, C., Charbonneau, P., Boulanger, F. \& Viallefond, F. 1990, A\&A, 234, 43
\bibitem{017}
         Carignan, C. \& Puche, D. 1990, AJ, 100, 394
\bibitem{018}
         Carollo, C.M. \& Lilly, S.J. 2001, ApJL, 548, L153
\bibitem{06}
         Castellanos, M. 2000, Ph.D., Madrid
\bibitem{019}
         Castellanos, M., D\'{\i}az, A.I. \& Terlevich, E. 2002, MNRAS, 329, 315
\bibitem{020}
         Casertano, S. \& van Gorkom, J.H. 1991, AJ, 101, 1231
\bibitem{021}
         Chiappini, C., Matteucci, F. \& Romano, D. 2001, ApJ, 554, 1044
\bibitem{022}
         Chiappini, C., Romano, D. \& Matteucci, F. 2003, MNRAS, 339, 63
\bibitem{07}
         Christensen, T., Petersen, L. \& Gammelgaard, P. 1997, A\&A, 322, 41
\bibitem{023}
         Ciardullo, R., Feldmeier, J.J., Jacoby, G.H., et al. 2002, ApJ, 577, 31
\bibitem{08}
         Consid{\` e}re, S., Coziol, R., Contini, T. \& Davoust, E.  2000, A\&A, 356, 89
\bibitem{024}
         Corbelli, E. \& Salucci, P. 2000, MNRAS, 311, 441
\bibitem{025}
         C\^{o}t\'{e}, S., Carignan, C. \& Freeman, K.C. 2000, AJ, 120, 3027
\bibitem{026}
         Cox, P. \& Downes, D. 1996, ApJ, 473, 219
\bibitem{027}
         Davidge, T.J. 1998, ApJ, 497, 650
\bibitem{028}
         Dean, J.F. \& Davies, R.D. 1975, MNRAS, 170, 503
\bibitem{09}
         Deharveng, L., Caplan, J., Lequeux, J., Azzopardi, M., Breysacher, J.,
         Tarenghi, M. \& Westerlund, B. 1988, A\&AS, 73, 407
\bibitem{029}
         Deharveng, L., Pe\~{n}a, M., Caplan, J. \& Costero, R.
         2000, MNRAS, 311, 329
\bibitem{030}
         de Blok, W.J.G. \& Bosma, A. 2002, A\&A, 385, 816
\bibitem{10}
         Dennefeld, M. \& Kunth, D. 1981, AJ, 86, 989
\bibitem{031}
         de Vaucouleurs, G. 1979, ApJ, 227, 729
\bibitem{032}
         de Vaucouleurs, G., de Vaucouleurs, A., Corvin, H.G., Buta,
         R.J., Paturel, J. \& Fouque, P. 1991, Third Reference Catalog
         of bright Galaxies(New York: Springer) (RC3)
\bibitem{033}
         de Vaucouleurs, G., Pence, W.D. \& Davoust, E. 1983, ApJS, 53, 17
\bibitem{12}
         D\'{\i}az, A.I., Terlevich, E., Pagel, B.E.J., V\'{\i}lchez, J.M. \& Edmunds, M.G.
         1987, MNRAS, 226, 19
\bibitem{13}
         D\'{\i}az, A.I., Terlevich, E., V\'{\i}lchez, J.M., Pagel, B.E.J. \& Edmunds M.G.
         1991, MNRAS, 253, 245
\bibitem{034}
         Diaz, A.I. \& Tosi, M. 1986, A\&A, 158, 60
\bibitem{14}
         d'Odorico, S., Rosa, M. \& Wampler, E.J. 1983, A\&A, 53, 97
\bibitem{15}
         Dopita, M.A. \& Evans, I.N. 1986, ApJ, 307, 431
\bibitem{035}
         Drozdovsky, I.O. \& Karachentsev, I.D. 2000, A\&AS, 142, 425
\bibitem{16}
         Dufour, R.J., Talbot, R.J., Jensen, E.B. \& Shields G.A. 1980, ApJ, 236, 119
\bibitem{036}
         Eastman, R.G., Schmidt, B.P. \& Kirshner, R. 1996, ApJ, 466, 911
\bibitem{aaa}
         Edmunds, M.G. 1990, MNRAS, 246, 678
\bibitem{17}
         Edmunds, M.G. \& Pagel, B.E.J. 1984, MNRAS, 211, 507
\bibitem{037}
         Ekholm, T., Lanoix, P., Teerikorpi, P., Fouqu\'{e}, P. \& Paturel, G. 2000, A\&A, 355, 835
\bibitem{038}
         Elmegreen, D.M. \& Salzer, J.J. 1999, AJ, 117, 764
\bibitem{039}
         Esteban, C., Peimbert, M., Torres-Peimbert, S. \& Escalante, V. 1998, MNRAS, 295, 401
\bibitem{040}
         Federspiel, M., Tammann, G.A. \& Sandage, A. 1998, ApJ, 495, 115
\bibitem{041}
         Feldmeier, J.J., Ciardullo, R. \& Jacoby, G.H.  1997, ApJ, 479, 231
\bibitem{18}
         Ferguson, A.M.N., Gallagher, J.S. \& Wyse R.F.G. 1998, AJ, 116, 673
\bibitem{042}
         Ferrarese, L., Ford, H.C., Huchra, J., et al. 2000, ApJS, 128, 431
\bibitem{19}
         Fierro, J., Torres-Peinmert, S. \& Peimbert, M. 1986, PASP, 98, 1032
\bibitem{043}
         Fouqu\'{e}, P., Bottinelli, L., Gouguenheim, L. \& Paturel, G. 1990, ApJ, 349, 1
\bibitem{044}
         Freedman, W.L., Madore, B.F., Gibson, B.K., et al. 2001, ApJ, 553, 47
\bibitem{045}
         Garcia, A.M. 1993, A\&AS, 100, 47
\bibitem{046}
         Garc\'{\i}a-Barreto, J.A., Downes, D. \& Huchtmeier, W.K. 1994, A\&A, 288, 705
\bibitem{20}
         Garnett, D.R. 1992, AJ, 103, 1330
\bibitem{047}
         Garnett, D.R. 2002, ApJ, 581, 1019
\bibitem{abc}
         Garnett, D.R., Kennicutt, R.C. \& Bresolin, F. 2004, ApJ, 607, L21
\bibitem{21}
         Garnett, D.R., Odewanh, S.C. \& Skillman, E.D. 1992, AJ, 104, 1714
\bibitem{048}
         Garnett, D.R. \& Shields, G.A. 1987, ApJ, 317, 82
\bibitem{23}
         Garnett, D.R., Shields, G.A., Peimbert, M., Torres-Peimbert, S.,
         Skillman, E.D., Dufour, R.J., Terlevich, E. \& Terlevich, R.J. 1999, ApJ, 513, 168
\bibitem{049}
         Garnett, D.R., Shields, G.A., Skillman, E.D., Sagan, S.P. \& Dufour, R.J. 1997, ApJ, 489, 63
\bibitem{050}
         Gavazzi, G., Boselli, A., Scodeggio, M., Pierini, D. \& Belsole, E. 1999, MNRAS, 304, 595
\bibitem{051}
         Gibson, B.K. \& Stetson, P.B. 2001, ApJL, 547, L103
\bibitem{052}
         Gu\'{e}lin, M. \& Weliachew, L. 1970, A\&A, 7, 141
\bibitem{053}
         Guhathakurta, P., van Gorkom, J.H., Kotanyi, C.G. \& Balkowski, C. 1988, AJ, 96, 851
\bibitem{054}
         Haynes, M.P., Hogg, D.E., Maddalena, R.J., Roberts, M.S. \& van Zee, L. 1998, AJ, 115, 62
\bibitem{055}
         Heckman, T.M., Balick, B. \& Sullivan, W.T. 1978, ApJ, 224, 745
\bibitem{056}
         Helfer, T.T. \& Blitz, L. 1995, ApJ, 450, 90
\bibitem{057}
         Helfer, T.T., Thornley, M.D., Regan, M.W., et al. 2003, ApJS, 145, 259
\bibitem{25}
         Henry, R.B.C., Balkowski, C., Cayatte, V., Edmunds, M.G. \& Pagel, B.E.J.
         1996, MNRAS, 283, 635
\bibitem{26}
         Henry, R.B.C., Pagel, B.E.J. \& Chincarini, G.L. 1994, MNRAS, 266, 421
\bibitem{27}
         Henry, R.B.C., Pagel, B.E.J., Lasseter, D.F. \& Chincarini G.L.
         1992, MNRAS, 258, 321
\bibitem{058}
         Herrnstein, J.R., Moran, J.M., Greenhill, L.J., et al. 1999, Nature, 400, 539
\bibitem{059}
         Hewitt, J.N., Haynes, M.P. \& Giovanelli, R. 1983, AJ, 88, 272
\bibitem{060}
         Ho, L.C., Filippenko, A.V. \& Sargent, W.L.W. 1997, ApJS, 112, 315
\bibitem{061}
         Hoekstra, H., van Albada, T.S. \& Sancisi, R. 2001, MNRAS, 323, 453
\bibitem{062}
         Hoffman, G.L., Salpeter, E.E., Farhat, B., Roos, T., Williams, H. \& Helow, G.
	 1996, ApJS, 105, 269
\bibitem{063}
         Huchtmeier, W.K. 1975, A\&A, 45, 259
\bibitem{064}
         Huchtmeier, W.K. \& Richter, O.G. 1986a, A\&AS, 63, 323
\bibitem{065}
         Huchtmeier, W.K. \& Richter, O.G. 1986b, A\&AS, 64, 111
\bibitem{066}
         Huchtmeier, W.K. \& Seiradakis, J.H. 1985, A\&A, 143, 216
\bibitem{067}
         Hunter, D.A., Rubin, V.C. \& Gallagher, J.S. 1986, AJ, 91, 1086
\bibitem{068}
         Jacoby, G.H., Ciardullo, R., Booth, J. \& Ford, H.C.  1989, ApJ, 344, 704
\bibitem{069}
         Jensen, J.B., Tonry, J.L., Barris, B.J., et al. 2003, ApJ, 583, 712
\bibitem{070}
         J\"{o}rs\"{a}ter, S. \& van Moorsel, G.A. 1995, AJ, 110, 2037
\bibitem{071}
         Joshi, Y.C., Pandey, A.K., Narasimha, D., Sagar, R. \& Giraud-H\'{e}raud, Y.
         2003, A\&A, 402, 113
\bibitem{072}
         Jurcevic, J.S., Pierce, M.J. \& Jacoby, G.H. 2000, MNRAS, 313, 868
\bibitem{073}
         Karachentsev, I.D. \& Drozdovsky, I.O. 1998, A\&AS, 131, 1
\bibitem{074}
         Karachentsev, I.D., Makarov, D.I. \& Huchtmeier, W.K. 1999, A\&AS, 139, 97
\bibitem{075}
         Karachentsev, I.D., Makarov, D.I., Sharina, M.E., et al. 2003a, A\&A, 398, 479
\bibitem{076}
         Karachentsev, I.D., Sharina, M.E., Dolphin, A.E., et al. 2003b, A\&A, 398, 467
\bibitem{077}
         Karachentsev, I.D. \& Tikhonov, N.A. 1993, A\&AS, 100, 227
\bibitem{078}
         Kenney, J.D. \& Young, J.S. 1988, ApJS, 66, 261
\bibitem{079}
         Kennicutt, R.C., Bresolin, F. \& Garnett, D.R. 2003, ApJ, 591, 801
\bibitem{28}
         Kennicutt, R.C. \& Garnett, D.R. 1996, ApJ, 456, 504
\bibitem{30}
         Kewley, L.J. \& Dopita, M.A. 2002, ApJS, 142, 35
\bibitem{080}
         Kim, M., Kim, E., Lee, M.G., Sarajedini, A. \& Geisler, D. 2002, AJ, 123, 244
\bibitem{081}
         Kinkel, U. \& Rosa, M.R. 1994, A\&A, 282, L37
\bibitem{082}
         Kinman, T.D. \& Davidson, K. 1981, ApJ, 243, 127
\bibitem{32}
         Kobulnicky, H.A., Kennicutt, R.C. \& Pizano J.L. 1999, ApJ, 514, 544
\bibitem{083}
         Koper, E., Dame, T.M., Israel, F.P. \& Thaddeus, P. 1991, ApJ, 383, L11
\bibitem{084}
         Kraan-Korteweg, R.C., Cameron, L.M. \& Tammann, G.A. 1988, ApJ, 331, 620
\bibitem{085}
         Krumm, N. \& Salpeter, E.E. 1979, AJ, 84, 1138
\bibitem{086}
         Kuno, N. \& Nakai, N. 1997, PASJ, 49, 279
\bibitem{33}
         Kwitter, K.B. \& Aller, L.H. 1981, MNRAS, 195, 939
\bibitem{500}
         Lamareille, F., Mouhcine, M., Contini, T., Lewis, I., Maddox, S.  2004, MNRAS, 350, 398
\bibitem{087}
         Larsen, S.S. \& Richtler, T. 2000, A\&A, 354, 836
\bibitem{088}
         Lee, M.G., Kim, M., Sarajedini, A., Geisler, D. \& Gieren, W. 2002, ApJ, 565, 959
\bibitem{000}
         Lee, H., McCall, M.L., Kingsburgh, R.L., Ross, R., Stevenson, C.C. 
         2003, AJ, 125, 146
\bibitem{089}
         Leonard, D.C., Filippenko, A.V., Gates, E.L., et al. 2002a, PASP, 114, 35
\bibitem{090}
         Leonard, D.C., Filippenko, A.V., Li, W., et al. 2002b, AJ, 124, 2490
\bibitem{091}
         Lequeux, J., Peimbert, M., Rayo, J.F., Serrano, A. \& Torres-Peimbert, S.
         1979, A\&A, 80, 155
\bibitem{092}
         Macri, L.M., Stetson, P.B., Bothun, G.D., et al. 2001, ApJ, 559, 243
\bibitem{093}
         Magrini, L., Corradi, R.L.M., Mampaso, A. \& Perinotto, M. 2000, A\&A, 355, 713
\bibitem{094}
         Malinie, G., Hartmann, D.H., Cayton, D.D. \& Mathews, G.J. 1993, ApJ, 413, 633
\bibitem{34}
         McCall, M.L., Rybski, P.M. \& Shields G.A. 1985, ApJS, 57, 1
\bibitem{35}
         McGaugh, S.S. 1991, ApJ, 380, 140
\bibitem{095}
         Melbourne, J. \& Salzer, J.J. 2002, AJ, 123, 2302
\bibitem{}
         Melbourne, J., Phillips, A., Salzer, J.J., Gronwall, C. 
         \& Sarajedini, V.L.  2004, AJ, 127, 686
\bibitem{096}
         Meyer, D.M., Jura, M. \& Cardelli, J.A. 1998, ApJ, 493, 222
\bibitem{097}
         Milliard, B. \& Marcelin, M. 1981, A\&A, 95, 59
\bibitem{098}
         Moll\'{a}, M., Ferrini, F. \& D\'{\i}az, A.I. 1996, ApJ, 466, 668
\bibitem{099}
         Moll\'{a}, M., Ferrini, F. \& D\'{\i}az, A.I. 1997, ApJ, 475, 519
\bibitem{0991}
         Mouhcine, M. \& Contini, T. 2002, A\&A, 389, 106
\bibitem{100}
         Mould, J., Aaronson, M. \& Huchra, J. 1980, ApJ, 238, 458
\bibitem{101}
         Mould, J. \& Kristian, J. 1986, ApJ, 305, 591
\bibitem{102}
         Newman, J.A., Ferrarese, L., Stetson, P.B., et al. 2001, ApJ, 553, 562
\bibitem{103}
         Nishiyama, K. \& Nakai, N. 2001, PASJ, 53, 713
\bibitem{104}
         Nishiyama, K., Nakai, N.  \& Kuno, N. 2001, PASJ, 53, 757
\bibitem{105}
         Ondrechen, M.P. \& van der Hulst, J.M. 1989, ApJ, 342, 29
\bibitem{36}
         Oey, M.S. \& Kennicutt, R.C. 1993, ApJ, 411, 137
\bibitem{106a}
         Pagel, B.E.J. 2003, in {\it CNO in the Universe}, Eds. C. Charbonnel,
         D. Schaerer \& G. Meynet, ASP Conference Series, v. 304, p. 187
\bibitem{106b}
         Pagel, B.E.J., Edmunds, M.G., Blackwell, D.E., Chun, M.S. \& Smith, G.
         1979, MNRAS, 189, 95
\bibitem{38}
         Pagel, B.E.J., Simonson, E.A., Terlevich, R.J. \& Edmunds, M.G.
         1992, MNRAS, 255, 325
\bibitem{107}
         Pagel, B.E.J. \& Tautvai\^{s}ien\.{e}, G. 1995, MNRAS, 276, 505
\bibitem{108}
         Parodi, B.R., Saha, A., Sandage, A. \& Tammann, G.A. 2000, ApJ, 540, 634
\bibitem{109}
         Paturel, G. 1984, ApJ, 282, 382
\bibitem{110}
         Paturel, G., Lanoix, P., Teerikorpi, P., et al. 1998, A\&A, 339, 671
\bibitem{39}
         Peimbert, M. 1970, PASP, 82, 636
\bibitem{111}
         Peterson, C.J. 1978, ApJ, 226, 75
\bibitem{112}
         Pierce, M.J. 1994, ApJ, 430, 53
\bibitem{113}
         Pierce, M.J., McClure, R.D. \& Racine, R. 1992, ApJ, 393, 523
\bibitem{114}
         Pierce, M.J. \& Tully, R.B. 1988, ApJ, 330, 579
\bibitem{115}
         Pilyugin, L.S., 2000, A\&A, 362, 325
\bibitem{116}
         Pilyugin, L.S., 2001a, A\&A, 369, 594
\bibitem{117}
         Pilyugin, L.S., 2001b, A\&A, 373, 56
\bibitem{118}
         Pilyugin, L.S., 2001c, A\&A, 374, 412
\bibitem{44}
         Pilyugin, L.S., 2003a, A\&A, 397, 109
\bibitem{119}
         Pilyugin, L.S., 2003b, A\&A, 399, 1003
\bibitem{120}
         Pilyugin, L.S. \& Edmunds, M.G. 1996a, A\&A, 313, 792
\bibitem{121}
         Pilyugin, L.S. \& Edmunds, M.G. 1996b, A\&A, 313, 803
\bibitem{122a}
         Pilyugin, L.S. \& Ferrini, F. 1998, A\&A, 336, 103
\bibitem{122b}
         Pilyugin, L.S. \& Ferrini, F. 2000a, A\&A, 354, 874
\bibitem{122c}
         Pilyugin, L.S. \& Ferrini, F. 2000b, A\&A, 358, 72
\bibitem{123}
         Pilyugin, L.S., Ferrini, F., Shkvarun, R.V., 2003, A\&A, 401, 557
\bibitem{124}
         Pilyugin, L.S., Moll\'{a}, M., Ferrini, F. \& V\'{\i}lchez, J.M. 2002, A\&A, 383, 14
\bibitem{46}
         Pilyugin, L.S., Thuan, T.X. \& V\'{\i}lchez, J.M. 2003, A\&A, 397, 487
\bibitem{126}
         Puche, D., Carignan, C. \& Bosma, A.  1990, AJ, 100, 1468
\bibitem{127}
         Puche, D., Carignan, C. \& van Gorkom, J.H. 1991, AJ, 101, 456
\bibitem{47}
         Rayo, J.F., Peimbert, M. \& Torres-Peimbert, S. 1982, ApJ, 255, 1
\bibitem{128}
         Reakes, M. 1979, MNRAS, 187, 525
\bibitem{129}
         Richer, M.G. \&  McCall, M.L. 1995, ApJ, 445, 642
\bibitem{130}
         Reif, K., Mebold, U., Goss, W.M., van Woerden, H.  \& Siegman, B. 1982, A\&AS, 50, 451
\bibitem {131}
         Rodr\'{\i}guez, M. 1999, A\&A, 351, 1075
\bibitem{132}
         Rogstad, D.H., Crutcher, R.M.  \& Chu, K. 1979, ApJ, 229, 509
\bibitem{133}
         Rothberg, B., Saunders, W., Tully, R.B.  \& Witchalls, P.L. 2000, ApJ, 533, 781
\bibitem{134}
         Rots, A.H. 1975, A\&A, 45, 43
\bibitem{135}
         Rots, A.H. 1980, A\&AS, 41, 189
\bibitem{48}
         Roy, J.-R.  \& Walsh, J.R. 1997, MNRAS, 288, 715
\bibitem{136}
         Rubin, V.C., Ford, W.K.  \& Thonnard, N. 1980, ApJ, 238, 471
\bibitem{137}
         Russell, D.G.  2002, ApJ, 565, 681
\bibitem{49}
         Ryder, S.D. 1995, ApJ, 444, 610
\bibitem{138}
         Ryder, S.D., Koribalski, B., Staveley-Smith, L., et al. 2001, ApJ, 555, 232
\bibitem{139}
         Sage, L.J. 1993, A\&A, 272, 123
\bibitem{140}
         Saha, A., Claver, J.  \& Hoessel, J.G. 2002, AJ, 124, 839
\bibitem{141}
         Saikia, D.J., Pedlar, A., Unger, S.W.  \& Axon, D.J., 1994, MNRAS, 270, 46
\bibitem{142}
         Sandqvist, A., J\"{o}rs\"{a}ter, S.  \& Lindblad, P.O. 1995, A\&A, 295, 585
\bibitem{143}
         Schmidt, B.P., Kirshner, R.P., Eastman, R.G., et al. 1994, ApJ, 432, 42
\bibitem{144}
         Schneider, S.E. 1989, ApJ, 343, 94
\bibitem{145}
         Sch\"{o}niger, F.  \& Sofue, Y. 1994, A\&A, 283, 21
\bibitem{146}
         Sch\"{o}niger, F., \& Sofue, Y. 1997, A\&A, 323, 14
\bibitem{50}
         Searle, L. 1971, ApJ, 168, 327
\bibitem{147}
         Shanks, T. 1997, MNRAS, 290, L77
\bibitem{148}
         Shapley, A., Fabbiano, G.  \& Eskridge, P.B. 2001, ApJS, 137, 139
\bibitem{149}
         Sharina, M.E., Karachentsev, I.D.  \& Tikhonov, N.A. 1996, A\&AS, 119, 499
\bibitem{150}
         Sharina, M.E., Karachentsev, I.D.  \& Tikhonov, N.A. 1997, PAZh, 23, 373
\bibitem{51}
         Shields, G.A.  \& Searle, L. 1978, ApJ, 222, 821
\bibitem{52}
         Shields, G.A., Skillman, E.D.  \& Kennicutt, R.C. 1991, ApJ, 371, 82
\bibitem{151}
         Shostak, G.S. 1978, A\&A, 68, 321
\bibitem{152}
         Skillman, E.D., Kennicutt, R.C.  \& Hodge, P.W. 1989, ApJ, 347, 875
\bibitem{53}
         Skillman, E.D., Kennicutt, R.C., Shields, G.A.  \& Zaritsky, D.
         1996, ApJ, 462, 147
\bibitem{153}
         Skillman, E.D., Terlevich, R., Teuben, P.J.  \& van Woerden, H. A\&A, 198, 33
\bibitem{154}
         Smith, H.E. 1975, ApJ, 199, 591
\bibitem{155}
         Soffner, T., M\'{e}ndez, R.H., Jacoby, G.H., et al. 1996, A\&A, 306, 9
\bibitem{156}
         Sofia, U.J.  \& Meyer, D.M. 2001, ApJL, 554, L221
\bibitem{157}
         Sohn, Y.-J.  \& Davidge, T.J. 1996, AJ, 111, 2280
\bibitem{158}
         Sohn, Y.-J.  \& Davidge, T.J. 1998, AJ, 115, 130
\bibitem{159}
         Solanes, J.M., Sanchis, T., Salvador-Sol\'{e}, E., Giovanelli, R.  \& Haynes, M.P. 2002, AJ, 124, 2440
\bibitem{160}
         Sorai, K., Nakai, N., Kuno, N., Nishiyama, K.  \& Hasegawa, T. 2000, PASJ, 52, 785
\bibitem{161}
         Sperandio, M., Chincarini, G., Rampazzo, R.  \& de Souza, R. 1995, A\&AS, 110, 279
\bibitem{162}
         Stark, A.A., Elmegreen, B.G.  \& Chance, D. 1987, ApJ, 322, 64
\bibitem{163}
         Stark, A.A., Knapp, G.R., Bally, J., et al. 1986, ApJ, 310, 660
\bibitem{164}
         Staveley-Smith, L.  \& Davies, R.D. 1988, MNRAS, 231, 833
\bibitem{55}
         Stauffer, J.R.  \& Bothun, G.D. 1984, AJ, 89, 1702
\bibitem{165}
         Takamiya, T.  \& Sofue, Y. 2000, ApJ, 534, 670
\bibitem{166}
         Terry, J.N., Paturel, G.  \& Ekholm, T. 2002, A\&A, 393, 57
\bibitem{167}
         Thim, F., Tammann, G.A., Saha, A., et al. 2003, ApJ, 590, 256
\bibitem{168}
         Thornley, M.D.  \& Mundy, L.G. 1997, ApJ, 484, 202
\bibitem{169}
         Tikhonov, N.A., Galazutdinova, O.A.  \& Drozdovsky, I.O. 2000, Afz, 43, 367
\bibitem{170}
         Tonry, J.L., Dressler, A., Blakeslee, J.P., et al. 2001, ApJ, 546, 681
\bibitem{56}
         Torres-Peimbert, S., Peimbert, M.  \& Fierro, J. 1989, ApJ, 345, 186
\bibitem{171}
         Tosi, M. 1988a, A\&A, 197, 33
\bibitem{172}
         Tosi, M. 1988b, A\&A, 197, 47
\bibitem{173}
         Tully, R.B. 1988, Nearby Galaxies Catalogue (Cambridge: Cambridge Univ. Press)
\bibitem{174}
         Tully, R.B.  \& Pierce, M.J., 2000, ApJ, 533, 744
\bibitem{175}
         Tutui, Y.  \& Sofue, Y. 1997, A\&A, 326, 915
\bibitem{176}
         Tutui, Y.  \& Sofue, Y. 1999, A\&A, 351, 467
\bibitem{177}
         van Albada, T.S., Bahcall, J.N., Begeman, K.  \& Sanscisi, R. 1985, ApJ, 295, 305
\bibitem{178}
         van Albada, G.D.  \& Shane, W.W. 1975, A\&A, 42, 433
\bibitem{179}
         van der Kruit, P.C.  \& Shostak, C.S. 1984, A\&A, 134, 258
\bibitem{180}
         van Zee, L.  \& Bryant, J. 1999, AJ, 118, 2172
\bibitem{57}
         van Zee, L., Salzer, J.J., Haynes, M.P., O'Donoghue, A.A.  \& Balonek T.J.
         1998, AJ, 116, 2805
\bibitem{181}
         Verdes-Montenegro, L., Bosma, A.  \& Athanassoula, E. 2000, A\&A, 356, 827
\bibitem{182}
         Vila-Costas, M.B.  \& Edmunds, M.G. 1992, MNRAS, 259, 121
\bibitem{59}
         V\'{\i}lchez, J.M., Edmunds, M.G.  \& Pagel, B.E.J. 1988, PASP, 100, 1428
\bibitem{60}
         V\'{\i}lchez, J.M., Pagel, B.E.J., D\'{\i}az, A.I., Terlevich, E.  \& Edmunds M.G.
         1988, MNRAS, 235, 633
\bibitem{183}
         Warmels, R.H. 1988a, A\&AS, 72, 19
\bibitem{184}
         Warmels, R.H. 1988b, A\&AS, 72, 57
\bibitem{61}
         Webster, B.L.  \& Smith, M.G. 1983, MNRAS, 204, 743
\bibitem{185}
         Weliachew, L. \& Gottesman, S.T.  1973, A\&A, 24, 59
\bibitem{186}
         Wevers, B.M.H.R., Appleton, P.N., Davies, R.D.   \& Hart, L. 1984, A\&A, 140, 125
\bibitem{187}
         Wevers, B.M.H.R., van der Kruit, P.C.   \& Allen, R.J. 1986, A\&AS, 66, 505
\bibitem{188}
         Wilcots, E.M.  \& Miller, B.W. 1998, AJ, 116, 2363
\bibitem{189}
         Yasuda, N., Fukugita, M.  \& Okamura, S. 1997, ApJS, 108, 417
\bibitem{190}
         Young, J.S., Xie, S., Tacconi, L., et al. 1995, ApJS, 98, 219
\bibitem{191}
         Zaritsky, D., Kennicutt, R.C.  \& Huchra, J.P. 1994, ApJ, 420, 87
\end{thebibliography}
\end{document}